\documentclass[aps,showpacs,preprint,epsf,epsfig, nofootinbib]{revtex4-2}
\usepackage{ifxetex,ifluatex}
\if\ifxetex T\else\ifluatex T\else F\fi\fi T%
\usepackage{fontspec}
\usepackage{tikz-feynman}
\usepackage{tikz}
\usepackage{mcite}
\usepackage[compat=1.1.0]{tikz-feynman}
\else
\usepackage[T1]{fontenc}
\usepackage[utf8]{inputenc}
\usepackage{lmodern}
\fi
\usepackage{mathrsfs}
\usepackage[utf8]{inputenc}
\usepackage{dcolumn}
\usepackage{bm}
\usepackage{hyperref}
\usepackage{epsfig, graphics, color, appendix}
\usepackage{multirow}
\usepackage{tabularx}
\usepackage{capt-of}
\usepackage{caption}
\usepackage{multirow}
\usepackage[countmax]{subfloat}
\usepackage{float}
\usepackage{placeins}
\usepackage[utf8]{inputenc}
\usepackage{graphicx}
\renewcommand{\baselinestretch}{1.10}
\usepackage{threeparttable}
\usepackage{amssymb}
\usepackage{array}
\usepackage{amsmath}
\usepackage{bbold}
\usepackage{mathtools}
\usepackage{subcaption}
\graphicspath{{Img/}}
\usepackage[english]{babel}
\begin{document}
	\count\footins=850
	\def\be{\begin{eqnarray}}
		\def\en{\end{eqnarray}}
	\def\non{\nonumber}
	\def\q2{q^2}
	\def\obt{\frac{1}{2}}
	\def\tbt{\frac{3}{2}}
	\def\obtp{\frac{1^{\prime}}{2}}
	\def\la{\langle}
	\def\ra{\rangle}
	\def\t{\times}
	\def\e{\epsilon}
	\def\r{\gamma}
	\def\w{\omega} 
	\def\u{\mu}
	\def\v{\nu}
	\def\lbd{\lambda}
	\def\ve{\varepsilon}
	\def\p{{\prime}}
	\def\pp{{\prime\prime}}
	\def\nc{N_c^{\rm eff}}
	\def\hep{\hat{\varepsilon}}
	\def\a{{\cal A}}
	\def\B{{\cal B}}
	\def\C{{\cal C}}
	\def \d {{\rm d}}
	\def\E{{\cal E}}
	\def\P{{\cal P}}
	\def\tt{{\cal T}}
	\def\qb{{\bf q}_\bot}
	\def\kb{{\bf k}_\bot}
	\def\pb{{\bf p}_\bot}
	\def\up{\uparrow}
	\def\dw{\downarrow}
	\def\vma{{_{V-A}}}
	\def\vpa{{_{V+A}}}
	\def\smp{{_{S-P}}}
	\def\spp{{_{S+P}}}
	\def\J{{J/\psi}}
	\def\ov{\overline}
	\def\Lqcd{{\Lambda_{\rm QCD}}}
	\def\pr{{Phys. Rev.}~}
	\def\prl{{ Phys. Rev. Lett.}~}
	\def\pl{{ Phys. Lett.}~}
	\def\np{{ Nucl. Phys.}~}
	\def\zp{{ Z. Phys.}~}
	\def\plus{\texttt{+}}
	\def\minus{\texttt{-}}
	\renewcommand{\baselinestretch}{1.10}
	\long\def\symbolfootnote[#1]#2{\begingroup%
		\def\thefootnote{\fnsymbol{footnote}}\footnote[#1]{#2}\endgroup}
	\def\lsim{ {\ \lower-1.2pt\vbox{\hbox{\rlap{$<$}\lower5pt\vbox{\hbox{$\sim$}
			}}}\ } }
	\def\gsim{ {\ \lower-1.2pt\vbox{\hbox{\rlap{$>$}\lower5pt\vbox{\hbox{$\sim$}
			}}}\ } }
	\font\el=cmbx10 scaled \magstep2{\obeylines\hfill \today}
	
	\vskip 1.5 cm
	\centerline{\large\bf \bm{$B_c$} to \textit{A} Transition Form Factors and Semileptonic Decays}
	\centerline{\large\bf in Self-consistent Covariant Light-front Approach}

	\small
	\vskip 1.0 cm
	
	\centerline{\bf Avijit Hazra$^{\dagger}$\footnote[1]{\href{mailto:hazra\_avijit@outlook.com}{hazra\_avijit@outlook.com}}, \bf Thejus Mary S.$^{\dagger}$\footnote[2]{\href{mailto:thejusmarys@gmail.com}{thejusmarys@gmail.com}}, \bf Neelesh Sharma$^{\ddagger}$\footnote[3]{\href{mailto:nishu.vats@gmail.com}{nishu.vats@gmail.com}}, Rohit Dhir$^{\dagger}$\footnote[4]{Corresponding author: \href{mailto:dhir.rohit@gmail.com}{dhir.rohit@gmail.com}}}
	
	\medskip
	\centerline{\it $^{\dagger}$Department of Physics and Nanotechnology,}
	\centerline{\it SRM Institute of Science and Technology, Kattankulathur 603203, India.}
	\centerline{$^{\ddagger}$ \it Paradigm of Science Cultivation and Ingenious, Kangra 176032, India.}
	\bigskip
	\bigskip
	\begin{center}
		{\large \bf Abstract}
	\end{center}	
	We present a comprehensive analysis of the semileptonic weak decays of $B_c$ meson decaying to axial-vector ($A$) mesons for bottom-conserving and bottom-changing decay modes. We employ self-consistent covariant light-front quark model (CLF QM) that uses type-II correspondence to eliminate inconsistencies in the traditional type-I CLF QM. As a fresh attempt, we test the self-consistency in CLF QM through type-II correspondence for $B_c \to A$ meson transition form factors. We establish that in type-II correspondence the form factors for longitudinal and transverse polarization states are numerically equal and are free from zero-mode contributions, which confirms the self-consistency of type-II correspondence for $B_c \to A$ transition form factors. Furthermore, we ascertain that the problems of inconsistency and violation of covariance of CLF QM within the type-I correspondence are resolved in type-II correspondence for $B_c \to A$ transitions. We thoroughly investigate the effects of self-consistency between type-I and type-II schemes using a comparative analysis. In this investigation, we employ a direct calculation approach to determine the form factors within the space-like region. Subsequently, we employ a vector meson dominance (VMD)-inspired three-parameter form to extrapolate these form factors into the physically accessible region. This enables us to study the $\q2$ dependence of the form factors in weak hadronic currents for the whole accessible kinematic range for both bottom-conserving as well as bottom-changing transitions. In addition, we extend our analysis to predict the branching ratios of the semileptonic weak decays of $B_c$ meson involving axial-vector meson in the final state to quantify the effects of self-consistency in these decays that were not studied before. We evaluate the lepton mass effect on these branching ratios and various other important physical observables, such as forward-backward asymmetries, lepton-side convexity parameter, asymmetry parameter, and longitudinal polarization asymmetries and fractions. Finally, we obtain the lepton flavor universality ratios for various decays.
	\newpage
	\section{Introduction}
	The CDF collaboration discovered $B_c$ meson ground state at Fermilab in 1998 \cite{CDF:1998axz, CDF:1998ihx}. Thereafter, considerable experimental progress has been made in $B_c$ meson studies, including spectroscopy, lifetime, decay channels, \textit{etc}. In the past few decades, the study of $B_c$ meson decays has garnered significant interest because of its unique characteristic of containing two distinct heavy open flavors. This flavor asymmetry in $B_c$ prevents it from strong (electromagnetic) annihilation into gluons (photons), however, it can decay weakly. As a result, the $B_c$ state is much more stable than heavy quarkonia, with a relatively longer life span of $(0.5134 \pm 0.011 \pm 0.0057)$~ps~\cite{LHCb:2014glo}. The LHCb reported latest precise measurement of $B_c$ meson mass as $(6274.47 \pm 0.27 \pm 0.17)$~MeV \cite{LHCb:2020ayi}. Thus, the $B_c$ system is an exceptional tool for exploring decay properties in both strong and weak interactions, as well as probing heavy quark dynamics.
	
	The modern \textit{B} factories and collaborations such as LHCb, CMS, ATLAS, and CDF have explored and planned \cite{Berezhnoy:2019yei, CDF:2016hra, Chen:2018obq} experimental study of $B_c$ meson, where the substantial volume of data on $B_c$ will yield valuable insights into the understanding of heavy flavor dynamics. In this regard, significant advancements have been achieved through the measurement of the relative $B_c/B$ production fraction, $B_c^{+}-B_c^{-}$ production asymmetry, and properties and decays of $B_c$ system \cite{LHCb:2012ihf, LHCb:2013xlg, LHCb:2013vrl, LHCb:2013hwj, Berezhnoy:2019yei, LHCb:2019bem, CMS:2019uhm, LHCb:2017lpu, ATLAS:2015jep, LHCb:2015azd, LHCb:2014mvo, ATLAS:2014lga, LHCb:2014rck, ATLAS:2019jpi, LHCb:2019tea, CDF:2005yjh, CDF:2007umr, D0:2008bqs, D0:2008thm, LHCb:2016vni, LHCb:2016djy, LHCb:2017ogk, LHCb:2017rqe, LHCb:2022ioi, LHCb:2021tdf}. Recently, LHCb and the ATLAS collaboration have investigated $B_c^{+}$ decays into two charm mesons \cite{LHCb:2021azb, ATLAS:2022aiy}.  ATLAS reported the ratio of branching fractions, Br$(B_c^{+}\to J/\psi D_s^{*+})$/Br$(B_c^{+}\to J/\psi D_s^{+})=1.93\pm 0.26$, and measured the transverse polarization fraction for $B_c^{+}\to J/\psi D_s^{*+}$ \cite{ATLAS:2022aiy}. Additionally, the LHCb collaboration announced the first observation of the decay $B_c^{+} \rightarrow \chi_{c2} \pi^{+}$, involving a \textit{p}-wave meson in the final state, and measured Br$(B_c^{+} \rightarrow \chi_{c2} \pi^{+})$/Br$(B_c^{+}\to J/\psi \pi^{+})=0.37 \pm 0.06 \pm 0.02 \pm 0.01$ \cite{LHCb:2023fqn}. Furthermore, evidence for the decay mode $B_c^{+} \rightarrow \chi_{c0}(\to K^{+}K^{-}) \pi^{+}$ has also been reported by LHCb \cite{LHCb:2016utz}. While a number of observations have been reported for bottom-changing transitions of $B_c$, the $B_c^{+} \rightarrow B_s^0 \pi^{+}$ mode is the only bottom-conserving transition seen experimentally. Moreover, the $B_c^{+} \rightarrow B_s^0 \pi^{+}$ decay is expected to have a large branching ratio due to the Cabibbo-Kobayashi-Maskawa (CKM) favored, $c\to s$, transition \cite{LHCb:2013xlg}. Lately, LHCb reported the first direct measurement of the ratio of branching fractions of $B_c^{+} \rightarrow B_s^0 \pi^{+}$ and $B_c^{+} \rightarrow J/\psi \pi^{+}$ \cite{LHCb:2022htj}. Among the semileptonic decays, only $B_c^{+}\to J / \psi(1S)\mu^{+}\nu_{\mu}$ and $B_c^{+}\to J / \psi(1S)\tau^{+}\nu_{\tau}$ were seen by CDF and LHCb collaborations \cite{LHCb:2017vlu, LHCb:2014rck}, however, their branching ratios are yet to be measured.
	
	Concurrently, investigating the semileptonic decays of $B_{(s)}$ mesons into excited charm mesons such as $D_{(s)}^{**}$ offers valuable insights into the properties and behavior of these excited states \cite{Chen:2016spr, Atoui:2013ksa, Gan:2014jxa, Huang:2004et, Zhao:2006at}. Similarly, semileptonic $B_c$ decay into $B^{**}$ and $B_s^{**}$ mesons can significantly contribute to understanding the spectroscopy and mixing phenomena of $B_{(s)}$ excitations \cite{CMS:2017hfy, LHCb:2016dxl, D0:2017qqm, Chen:2016spr, ParticleDataGroup:2022pth, ATLAS:2018udc, D0:2016mwd, Wang:2016wkj, Chen:2016mqt}. The excited states $D_{(s)}^{**} ~(B_{(s)}^{**})$ are referred to as \textit{p}-wave, \textit{i.e.}, scalar ($S,~J^P = 0^+$), axial-vector ($A,~1^+$), and tensor ($T,~2^+$) mesons, for example, $D_{(s)0}, D_{(s)1}, D_{(s)2}$ ($B_{(s)0}, B_{(s)1}, B_{(s)2}$), respectively\footnote{The orbitally excited charm and bottom mesons (and their charge conjugates) have a valence quark composition of $c\bar{q}$ and $b\bar{q}$, respectively, where $q = u, d, s$ \cite{ParticleDataGroup:2022pth}. Among p-wave mesons, physical axial-vector states exist in two types, \textit{i.e.}, $A$ and $A'$, which are mixtures of ${}^3P_1$ ($1^{++}$) and ${}^1P_1$ ($1^{+-}$) states. Detailed spectroscopic information, including masses and mixing schemes, can be found in Appendix~\ref{axial_mixing}.}. These studies encourage experimental efforts for the semileptonic decay of $B_c$ to orbitally excited mesons. The analysis of semileptonic weak decays of the $B_c$ meson holds significant importance in experimental studies due to its simple theoretical description via tree-level processes in the Standard Model (SM). Additionally, these decays act as probes for strong and weak interaction interference, phases of the CKM matrix elements, \textit{CP}-violation within and beyond the SM, and examine the lepton flavor universality (LFU) in the electroweak interactions. It is important to note that the $B_c$ meson decays are categorized by $c$-quark ($c\to d/s$) and $b$-quark ($b\to u/c$) transitions and are given by following selection rules \cite{Fayyazuddin:2012qfa}: i) bottom-conserving: CKM-enhanced, $c\to s$, ($\Delta b = 0, \Delta C =-1, \Delta S = -1$) and CKM-suppressed, $c\to d$, ($\Delta b = 0, \Delta C =-1, \Delta S = 0$); ii) bottom-changing: CKM-enhanced, $b\to c$, ($\Delta b = -1, \Delta C =-1, \Delta S = 0$) and CKM-suppressed, $b\to u$, ($\Delta b = -1, \Delta C =0, \Delta S = 0$).
	
	Theoretically, numerical estimates of semileptonic $B_c$ decays involving axial-vectors have been reported based on a variety of models, namely, the CLF QM, relativistic quark model (RQM), nonrelativistic quark model (NRQM), relativistic constituent quark model (RCQM), light cone QCD sum rules (LCQSR), QCD motivated relativistic quark model (QRQM), perturbative QCD (pQCD), Bethe-Salpeter (BS) approach, and three-point QCD sum rules (QCDSR) \cite{Zhang:2023ypl, Wang:2009mi, Ivanov:2005fd, Ivanov:2006ni, Hernandez:2006gt, Akan:2022vtf, Ebert:2010zu, Colangelo:2022awx, Rui:2018kqr, Shi:2016gqt, Wang:2011jt, Azizi:2009ny, Khosravi:2015tea}. Thus far, the main focus of such theoretical studies has been on $B_c$ meson decaying to charmonium states. However, comprehensive investigations involving bottom-conserving and bottom-changing semileptonic decays of $B_c$ (except for decays to charmonia) are scarce.
	
	Exclusive semileptonic decays can be characterized by their dependence on weak transition form factors. Knowledge of these form factors of heavy meson transitions is crucial for precise phenomenological predictions of helicity amplitude, branching ratios, and other observables in various decay processes. In this regard, the traditional light-front quark model (LF QM), also known as standard light-front (SLF), offers a conceptually straightforward and phenomenologically reliable framework for calculating decay constants and form factors. The SLF quark model (SLF QM) is a relativistic quark model based on the light-front (LF) formalism \cite{Dirac:1949cp} and LF quantization of QCD \cite{Brodsky:1997de}. It provides a full relativistic treatment of spin through the Melosh rotation \cite{Terentev:1976jk, Berestetsky:1977zk}. However, a key limitation of LF QM lies in its lack of manifest Lorentz covariance within the evaluated matrix elements. Moreover, it is unable to explicitly determine the zero-mode contributions. To address persisting inconsistencies and ambiguities of LF QM approach~\cite{Terentev:1976jk, Chung:1988mu, Jaus:1989au, Jaus:1991cy, Cheng:1997au, Carbonell:1998rj, Jaus:1999zv, Karmanov:1991fv, Karmanov:1994ck, Karmanov:1996un, Choi:1998nf}, Jaus \cite{Jaus:1999zv} put forth the manifestly CLF approach over the SLF QM. The CLF is a quantum field theory framework aiming to reconcile the simplicity of LF quantization with Lorentz covariance. It incorporates zero-mode contributions and employs covariant vertex functions with specific asymmetry properties to eliminate spurious LF dependencies. The CLF QM specifically focuses on mesons as quark-antiquark bound states and combines LF quantization with a manifestly covariant formalism. The hadronic bound state, represented by a covariant vertex function that includes quark interactions, serves as a building block for calculating hadronic matrix elements \cite{Jaus:1999zv}. Furthermore, the CLF formalism describes particle dynamics using LF coordinates, ensuring consistency with special relativity. Unlike standard non-covariant light-front dynamics defined on the plane $x^0+x^3=0$, the covariant approach utilizes a general plane $\w\cdot x=x^+=x^0+x^3=0$, where $\w$ is a null four-vector $(\w^{\u}\w_{\u}=0)$\footnote{By setting $\w=(1,0,0,-1)$, the SLF dynamics on $x^0+x^3=0$ is recovered.}. The formalism's covariance stems from the invariance of the $\w\cdot x=0$ plane under Lorentz transformations of $\w$ and $x$, necessitating $\w$'s variation across reference frames~\cite{Carbonell:1998rj}. This separation of coordinate and dynamical transformations offers advantages, including an explicit parametrization of wave functions (WFs) and off-shell amplitudes' dependence on the LF plane's orientation via $\w$. While this approach \cite{Jaus:1999zv} ensured the covariant results and provided a systematic calculation of zero-mode contributions, it could not fully account for the effects of zero-modes \cite{Cheng:2003sm, Choi:2013mda}. However, these zero-mode contributions are crucial for preserving symmetries and ensuring consistencies in LF quantization. Thus, neglecting zero-mode contributions can lead to inconsistencies in CLF results for the physical quantities $\cal Q$, such as, decay constants and form factors. For example, $[{\cal Q}]^{\lambda=0}_{CLF}\neq [{\cal Q}]^{\lambda=\pm}_{CLF}$, \textit{i.e.}, the numerical evaluation of $\cal Q$ derived from longitudinal ($\lambda=0$) and transverse ($\lambda=\pm$) polarization states are not consistent with each other. 
	
	In this paper, to calculate the required transition form factors, we employ the CLF approach, which provides full relativistic treatment of hadronic quantities, \textit{e.g.}, decay constant and form factors \cite{Jaus:1999zv, Cheng:2003sm}. Aforementioned, the CLF QM is manifestly covariant and self-consistent and offers a systematic method for investigating zero-mode effects. The resulting outcomes are ensured to be covariant once the spurious contributions related to the light-like four-vector $\omega^\mu=(0, 2, 0_{\bot})$ are nullified by the inclusion of zero-mode contributions (for details see \cite{Jaus:1999zv, Cheng:2003sm}). The matrix elements in CLF QM contains inadvertent $\omega$-dependencies\footnote{In order to segregate its $\omega$-dependent components, three coefficients, namely, $A^{(i)}_{j}$, $B^{(i)}_{j}$, and $C^{(i)}_{j}$ functions, are introduced \cite{Jaus:1999zv}. Details are discussed in Section~\ref{methodology}.}. The primary $\omega$-dependences are associated with the $C^{(i)}_{j}$ functions and can be eradicated by zero-mode contributions. However, non-zero spurious contributions associated with $B^{(i)}_{j}$ functions still retain some residual $\omega$-dependences \cite{Jaus:1999zv, Jaus:2002sv, Chang:2018zjq, Chang:2019mmh}. These contributions persist within the established correspondence between the BS approach and LF QM called the type-I scheme, leading to a violation of covariance in the results of CLF QM \cite{Choi:2013mda}. In addition, Cheng \textit{et al.} \cite{Cheng:2003sm} have identified inconsistencies in CLF results for vector decay constant derived from $\lambda=0$ and $\lambda=\pm$ polarization states because of an additional contribution characterized by the $B^{(i)}_{j}$ functions. This implies that type-I CLF QM is not self-consistent, and the replacement of $M$ with $M_0$ only in the $D$ factor does not guarantee consistency within LF QM \cite{Choi:2013mda}, where $M$ and $M_0$ are physical and kinetic invariant masses of mesons, respectively\footnote{The $D$ factors account for the model-specific characteristics of the spin-$1$ meson and appear in the denominator of the associated vertex operator \cite{Jaus:1999zv, Cheng:2003sm}. These factors are given in Section~\ref{methodology}.}. Furthermore, in both SLF and CLF QMs, transition form factors can be derived from various combinations of weak currents ($ J^\mu $), where $ \mu = (+, -, \bot) $. Typically, these form factors are extracted from the plus component ($J^+ \equiv J^0 + J^3$), which is considered the “good” component of the weak current \cite{Frankfurt:1993ut}. For transitions involving only pseudoscalar mesons, zero-mode contributions can be avoided by calculating the hadronic form factors from the plus components of the respective currents. This approach is effective because, in a purely transverse frame where $ q^+ = q^0 + q^3= 0 $, only diagonal matrix elements are required, thus simplifying the calculation and potentially circumventing complications from zero modes \cite{Cheng:2003sm}. This is particularly relevant for LF formalism results related to pseudoscalar transitions, where the integrands are independent of the physical mass $M$. Consequently, the traditional type-I scheme, which restricts $M \to M_0$ within the D-factor, may result in an incomplete representation. However, this method of eliminating zero-mode contributions by selecting the plus component ($ J^+ $) does not universally apply to composite spin-1 systems \cite{Jaus:1999zv, Jaus:2002sv}. Therefore, in order to revive self-consistency in CLF QM, the type-II correspondence has been proposed in \cite{Choi:2013mda, Chang:2018zjq, Chang:2019mmh}. The crucial aspect of the type-II correspondence is to apply the replacement of $M$ with $M_0$ to all physical mass terms involved in the calculation of physical quantities. Therefore, we exploit type-II correspondence in CLF QM in our work to obtain the transition form factors.
	
	In this work, we calculate the axial-vector meson emitting transition form factors of $B_c$ for bottom-conserving and bottom-changing transitions using type-II correspondence in the CLF QM \cite{Choi:2013mda}. It should be noted that type-II correspondence in CLF QM (type-II CLF QM) is termed as self-consistent CLF QM on account of the challenges associated with type-I correspondence \cite{Jaus:1999zv, Choi:2013mda, Cheng:2003sm}. Qin Chang \textit{et al.} \cite{Chang:2018zjq, Chang:2019mmh} have investigated the issues of self-consistency and covariance in the CLF QM within the type-I scheme (hereafter termed as type-I CLF QM). It has been found that, analogous to the decay constants of vector (\textit{V}) and axial-vector mesons, the form factors also encounter problems in terms of self-consistency, particularly with regard to the $B^{(i)}_{j}$ functions \cite{Chang:2019mmh}. The authors have examined the form factors for pseudoscalar (\textit{P}) to vector transitions, focusing on both type-I and type-II correspondence schemes in the CLF QM; however, they have restricted their study to bottom-changing $P \to V$ transitions only. Furthermore, it has been shown in \cite{Chang:2019mmh} that the self-consistency issues in $a_-(q^2)$ and $f(q^2)$ form factors of $P \to V$ transitions can be addressed by adopting the type-II correspondence scheme, as the contributions associated with $B^{(i)}_{j}$ functions numerically diminish when the meson mass $M \to M_0$. The zero-mode contributions to these form factors are present in theory for both type-I as well as type-II schemes; however, they are numerically negligible in the type-II scheme. Thus, the self-consistency and covariance challenges in the CLF QM can be understood as stemming from the same source and thus addressed simultaneously by employing the type-II correspondence scheme. It is worth remarking that although self-consistency effects on the form factors involving vector mesons in the final state have been studied in type-II CLF QM, their corresponding effects have not yet been quantified for semileptonic or nonleptonic decays. On the other hand, despite a reasonable number of investigations on $B_c$ transition form factors to the ground and excited states of charmonia, studies involving bottom-conserving decays of $B_c$ meson are sparse within type-I CLF QM.
	
	We should emphasize that similar to vector mesons, axial-vector states also encounter issues concerning self-consistency and covariance of the matrix element. However, these issues so far have not been studied on form factors for axial-vector meson emitting transitions, particularly for $B_c \to A$ form factors\footnote{Note that A represents both types of axial-vector mesons.} for both bottom-conserving and bottom-changing decays in type-II correspondence. Furthermore, it is necessary to conduct these analyses to validate the recent advancements in the CLF approach and to assert the challenges, specifically, associated with understanding of $\q2$ dependence formulation of bottom-conserving as well as bottom-changing $B_c$ meson transitions and the mixing of axial-vector states. The main goal of the present work is to confirm and test the findings of the type-II correspondence for axial-vector meson emitting form factors of $B_c$ meson, and consequently, to quantitatively observe their implications on semileptonic decays. Therefore, we incorporate the suggested improvements by using type-II CLF QM for the first time in a comprehensive analysis of $B_c \to A$ form factors. Furthermore, we believe that the perceptible impact of such effects can only be weighed by studying weak decays. Thus, we also present an analysis of semileptonic $B_c\to A \ell\nu_\ell$ decays in the current work, for both bottom-conserving and bottom-changing modes. We study the self-consistency effects, in corresponding $c_{-}(\q2)$ and $l(\q2)$ form factors involving axial-vector mesons, through numerical evaluation of these form factors for $\lambda=0$ and $\lambda=\pm$ polarization states\footnote{The form factors $l(\q2)$, $q(\q2)$, $c_{+}(\q2)$, and $c_{-}(\q2)$ are related to CLF form factors $A(\q2)$, $V_{0}(\q2)$, $V_{1}(\q2)$, $V_{2}(\q2)$, and $V_{3}(\q2)$, as shown in Sec.~\ref{methodology}.}. It should be noted that in CLF QM the form factors are directly calculable only in the space-like region ($\q2 \leq 0$). However, the physical decay processes occur exclusively in the time-like region ($\q2 \geq 0$). Consequently, an essential step involves the extrapolation of these form factors into the physically relevant region using an appropriate $\q2$ formulation. In this work, we adopt a three-parameter function inspired by VMD to perform this extrapolation. In contrast to the vector meson system, it is noteworthy that self-consistency effects have a substantial numerical influence on $B_c \to A$ transitions and subsequently lead to prominent impact on the decay rates. We further investigate the implication of $\q2$ dependence for $B_c\to A$ form factors over the entire momentum range, \textit{i.e.}, $0 \leqslant \q2\leqslant \q2_{max}$, where $\q2_{max}=(M_{B_c}-M_{A})^2$. We will particularly focus on characterizing the $\q2$ dependence of $B_c\to B_{1}/ B_{s1}$ and $ B_c\to D_{1}/ \chi_{c1}$ form factors, utilizing the VMD-like three-parameter fit within available $\q2$ range. We plot the type-II $B_c\to A$ form factors to analyze their behavior with respect to available $\q2$. Moreover, we use these form factors to predict the helicity amplitudes and consequently the branching ratios of the semileptonic decays. In addition, we calculated various physical observables important for experimental as well as theoretical understanding, such as $\la A_{FB}\ra$, $\la C^\ell_{F}\ra$, $\la P^\ell_{L}\ra$, $\la F_{L}\ra$, and $\la \alpha^{*}\ra$ and plot their $\q2$ dependence. Further, our investigation includes the calculation of LFU ratios pertaining to $B_c\to B_{1}/ B_{s1}/D_{1}/ \chi_{c1}$ semileptonic decay modes.
	
	The paper is organized as follows: The form factors of $B_c$ meson transitions are calculated using the methodology given in Sec.~\ref{methodology}. We discuss the $\q2$ dependence of the form factors and its implications in Sec.~\ref{sec:ffq2}. Further, in Sec.~\ref{sec:drha}, we provide the expressions for decay rates, helicity amplitudes, and other observables for $B_c$ semileptonic decays. Following this, we present our numerical results and discussions in Sec.~\ref{sec:NR&D}. Finally, we summarize our findings in Sec.~\ref{sec:summary}. Appendices~\ref{axial_mixing},~\ref{cov_ml}, \ref{FF_space-like}, and~\ref{Br_unmix}, include the details of the mixing between the axial-vector mesons, the zero-mode contributions, and covariance of matrix element, space-like transition form factor, and the unmixed branching ratios, respectively.
	\section{Self-consistent covariant light-front approach}
	In this work, we focus on the proposed self-consistency in the CLF QM; therefore, we summarize the formulation of the theoretical framework and adopt the notations of \cite{Chang:2019mmh} (more details can be found in Refs.~\cite{Jaus:1999zv, Jaus:2002sv, Cheng:2003sm, Choi:2013mda, Chang:2018zjq, Chang:2019mmh, Chang:2020wvs}). The SLF framework derives physical quantities from the plus component of current matrix elements by constraining meson constituent quarks to be on-shell. This framework, however, neglects zero-mode effects, leading to non-covariant matrix elements \cite{Cheng:2003sm}. Jaus proposed a CLF method \cite{Jaus:1999zv} to address zero-mode contributions using covariant Feynman momentum loop integrals with off-shell constituent quarks. This approach employs phenomenological vertex functions similar to those in the SLF framework. The covariant approach represents hadronic matrix elements of one-body currents as one-loop diagrams, evaluable using standard space-time formalism. In practice, LF matrix elements can be obtained through LF decomposition of the loop momentum and integration over the minus component ($k_1^{\p-}$) using contour methods. The integration over the negative component of loop momentum defines the corresponding LF vertex functions reflecting the transition from CLF to LF approach. Proper $k_1^{\p-}$ integration eliminates the spurious contributions proportional to the light-like four-vector $\omega^\mu=(0, 2, 0_{\bot})$, thereby  effectively handling zero-mode contributions. As an initial step, we examine decay and transition amplitudes described by one-loop diagrams, as illustrated in Figure~\ref{fig:feynman}, in the context of decay constants and form factors for ground-state $s$-wave mesons and low-lying $p$-wave mesons.
	\label{methodology}
	\subsection{Preliminaries}
\begin{figure}[t!]
		\centerline{
			{\epsfxsize2.1 in \epsffile{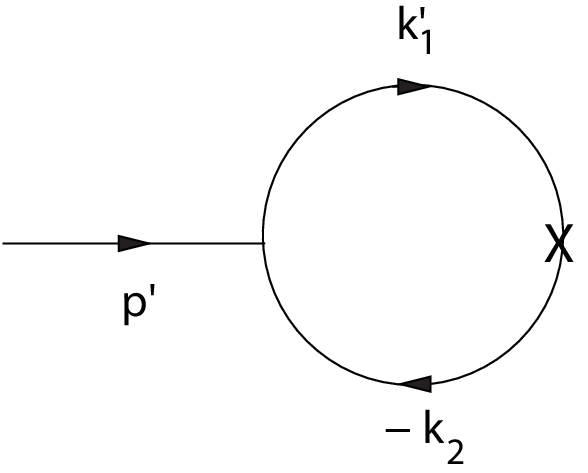}}
			\hspace{1cm}
			{\epsfxsize3 in \epsffile{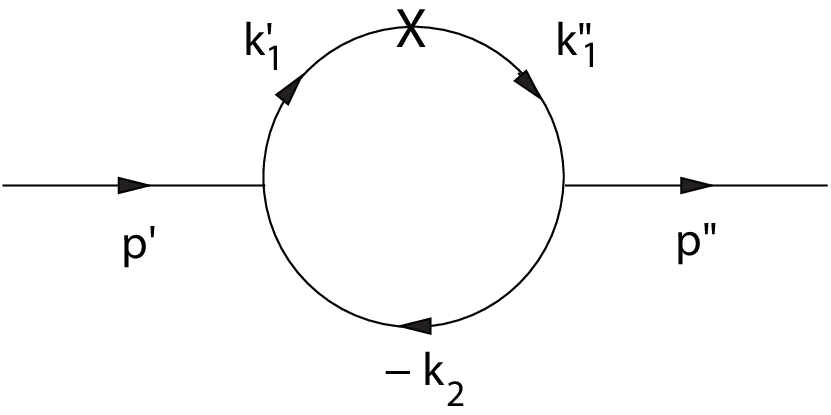}}}
		\centerline{\,\,\,\,\,(a)\hspace{6.2cm}(b)} \vskip0.2cm
		\caption{Feynman loop diagrams for (a) meson decay and (b) meson
			transition amplitudes, where $X$ denotes the corresponding $\gamma^\mu-\gamma^\mu\gamma_5$ current vertex.}\label{fig:feynman}
	\end{figure}

	For a meson transition, we define the momentum of the initial and final mesons as $p^{\p}$ and $p^{\p\p}$, respectively, as shown in Figure~\ref{fig:feynman}. 
	Furthermore, the meson momenta can be expressed in terms of the off-shell quark ($q_1^{\p}$), and an antiquark ($\bar{q}_2$), momenta, \textit{i.e}, $p^{\p(\p\p)}=k_1^{\p(\p\p)}+k_2$, where $k_1^{\p(\p\p)}$ and $k_2$ represent quarks' momenta. We denote $m_1^{\p(\p\p)}$ and $m_2$ as the masses of the incoming (outgoing) quark and the spectator quark, respectively. In addition, quarks' momenta can be expressed in terms of the internal LF variables $(x,\kb^{\p})$ as 
	\begin{align}\label{eq:momk1}
		k_1^{\p+}=xp^{\p+}\,,\quad\, \mathbf{k}_{1\bot}^{\p}=x\pb^{\p}+\kb^{\p} \,,\qquad  k_2^+=\bar{x}p^{\p+} \,,\quad\, \mathbf{k}_{2\bot}=\bar{x}\pb^{\p}-\kb^{\p}\,,
	\end{align}
	where $x$ and $\bar{x}$ are the longitudinal momentum fractions of the quark and antiquark, respectively, such that $x+\bar{x}=1$. The meson momentum is defined as $p^{\p}=(p^{\p+}, p^{\p-}, \pb^{\p})$ with $p^{\p\pm}=p^{\p0}\pm p^{\p3}$, such that $p^{\p 2}=p^{\p+}p^{\p-}-\pb^{\p2}= M^{\p 2}$, while the transverse quark and meson momenta are given by $\kb^{\p}=(k^{\p x}\,,k^{\p y})$ and $\pb^{\p}=(p^{\p x}\,,p^{\p y})$, respectively. Furthermore, the kinetic invariant mass squared, $M^{\p 2}_0$, of the incoming $q_1^{\p}\bar{q}_2$ system, and other useful quantities, such as the energy of $i$th quark, $E_i$, are defined in terms of internal LF variables \cite{Chang:2019mmh, Chang:2020wvs} as follows:
	\begin{equation}
		\label{eq:kinec_mass}
		M^{\p 2}_0 =(E_1^{\p}+E_2)^2=\frac{k^{2}_\bot+m_1^{\p 2}}
		{x}+\frac{k^{2}_{\bot}+m_2^2}{\bar{x}},
	\end{equation} 
	\begin{equation}
		\hat M_0^{\p}=\sqrt{M_0^{\p 2}-(m_1^{\p}-m_2)^2}, \quad\quad
		E^{\p}_i = \sqrt{m^{\p 2}_i+k^{\p2}_\bot+k^{\p2}_z},
	\end{equation} 
	\begin{eqnarray} \label{eq:int_kz}
		k_z^{\p}=(x-\frac{1}{2})M_0^{\p}+\frac{m_2^2-m_1^{\p2}}{2 M_0^{\p}}.
	\end{eqnarray}
	Conventionally, a meson bound-state ($q_1^{\p}, \bar{q}_2$) can be represented as 
	\begin{eqnarray}\label{eq:Fockexp}
		|M(p, ^{2S+1}L_J, J_z)\ra
		=\int\{d^3\tilde{k}_1\}\{d^3\tilde{k}_2\} ~2(2\pi)^3 \delta^3 ({\tilde{p}-\tilde{k}_1-\tilde{k}_2})~\nonumber \\
		\t\sum_{h_1,h_2}\Psi^{JJ_z}_{LS}(\tilde k_1,\tilde k_2,h_1,h_2)
		|q_1^{\p}(k_1,h_1) \bar q_2(k_2,h_2)\ra,
	\end{eqnarray}
	where \textit{L} and \textit{J} are orbital angular and total spin quantum numbers, respectively \cite{Cheng:2003sm}. Further, $\tilde{p}=(p^{\p +},\pb^{\p})$, and $\tilde{k}_{1,2}=(k_{1,2}^{\p +},\mathbf{k}^{\p}_{1,2\bot})$ represent the on-mass-shell LF momenta, and $\{d^3\tilde{k}\} \equiv {1\over 2(2\pi)^3}dk^{\p +}d^2\kb^{\p}$. The wave function $\Psi^{JJ_z}_{LS}(\tilde k_1,\tilde k_2,h_1,h_2)$, which describes the distribution of momentum in space for $^{2S+1}L_J$ meson, can be written as
	\begin{equation}
		\Psi^{JJ_z}_{LS}(\tilde k_1,\tilde k_2,h_1,h_2)
		= \frac{1}{\sqrt N_c}\la L S; L_z S_z|L S;J J_z\ra
		R^{SS_z}_{h_1h_2}(x,\kb^{\p})~ \psi_{LL_z}(x, \kb^{\p}).
	\end{equation}
	The radial wave function $\psi_{LL_z}(x,\kb^{\p})$ characterizes how the constituent quarks' momenta are distributed in a bound state that possesses orbital angular momentum $L$, and the corresponding Clebsch-Gordan coefficient is represented by $\la LS; L_z S_z|L S; J J_z\ra$ \cite{Cheng:2003sm}. The spin-orbit LF wave function ($R^{SS_Z}_{h_1h_2}$), which can be obtained by Melosh transformation, represents the definite spin state ($S, S_Z$) corresponding to the LF helicity ($h_1,h_2$) eigenstates. For more details on the polarization states and normalization procedure see \cite{Cheng:2003sm}. It is noteworthy that the LF wave functions possess a remarkable property of being frame-independent \cite{Brodsky:1998hn, Brodsky:1981fz, Brodsky:2007hb}. This characteristic makes LF quantization the ideal framework for describing hadron structures in terms of their constituent quark and gluon degrees of freedom. The LF wave functions not only encode the constituent spin but also capture the orbital angular momentum properties of hadrons. Furthermore, these wave functions depend on the LF boost-invariant mass, $M^{\p}_0$, defined in Eq.~\eqref{eq:kinec_mass}. The radial wave function is usually assumed to be of the Gaussian-type, \textit{i.e.},
	\begin{align}\label{eq:RWFs}
		\psi_s(x,\kb^{\p}) =&4\frac{\pi^{\frac{3}{4}}}{\beta^{\frac{3}{2}}} \sqrt{ \frac{\partial k_z^{\p}}{\partial x}}\exp\left[ -\frac{k_z^{\p 2}+\kb^{\p2}}{2\beta^2}\right]\,,
	\end{align}
	and
	\begin{align}\label{eq:RWFp}
		\psi_{p}(x,\kb^{\p\p}) =&\frac{\sqrt{2}}{\beta}\psi_s(x,\kb^{\p}) \,,
	\end{align}
	for $s$-wave and $p$-wave mesons, respectively \cite{Cheng:2003sm}. The relative momentum in the $z$-direction, $k_z^{\p}$, leads to 
	\begin{align}
		\label{QM3}
		\frac{\partial k_z^{\p}}{\partial x}
		= \frac{M_0^{\p}}{4 x (1-x)}[ 1-
		(\frac{m^{\p2}_{1} - m^2_{2}}{M^{\p2}_0})^2].
	\end{align}
	In both Eqs.~\eqref{eq:RWFs} and \eqref{eq:RWFp}, the shape parameter, $\beta$, plays a crucial role. It governs the momentum distribution of the constituents within the bound state and essentially determines the confinement scale ($\Lambda_{QCD}$), which reflects the characteristic “size” of the bound state \cite{Jaus:1999zv, Jaus:1996np}. The parameters $\beta$ are usually determined from experimental data \cite{Verma:2011yw}. It may be noted that $\Lambda_{QCD}$ signifies the energy scale where confinement and hadronization processes become dominant. In this regime, non-perturbative effects govern the strong interaction dynamics, leading to the breakdown of perturbative QCD \cite{Sun:2021gmi, Brambilla:2004jw}. Notably, the specific form of the LF wave functions ensures adherence to the covariant requirement, as demonstrated in \cite{Cheng:1997au}. The LF wave functions given by Eqs.~\eqref{eq:RWFs} and \eqref{eq:RWFp} are manifestly Lorentz invariant\footnote{Note that the commonly utilized BSW-type wave function \cite{Cheng:1997au} cannot preserve the Lorentz covariance, and violates the covariance requirement, therefore, can not be used in the LF QM.} as it is expressed in terms of the momentum fraction variables with plus component \cite{Cheng:1996if}. 
	
	The underlying link between the hadronic phenomena in QCD at large and small distances is the hadronic wave function, which determines the probability amplitudes and distributions of the quark and gluons that enter the short-distance sub-processes \cite{Brodsky:1981jv, Brodsky:1981fz, Brodsky:1980vj}. It may be remarked that phenomenological LF wave functions are employed to characterize the hadronic structure because they incorporate known properties of QCD, such as the expected fall-off at high relative transverse momentum and endpoint behavior, into a tractable functional form \cite{Brodsky:1997de}. These wave functions are crucial for calculating the hadronic matrix elements, which inherently encode the low-energy manifestations, \textit{i.e.}, non-perturbative contributions. Within the CLF framework, the structure of a $q\bar{q}$ meson bound state is modeled by a covariant vertex function rather than a conventional wave function. It is important to note that, in the CLF approach, the total four-momentum is conserved at each vertex where the quarks (antiquarks) are off-shell \cite{Cheng:2003sm, Jaus:2002sv}. Additionally, Jaus \cite{Jaus:1999zv} provided the correspondence between the manifestly covariant BS approach and LF QM to eliminate the spurious $\omega$-dependence. The inconsistency problem is resolved by utilizing the manifestly covariant BS approach as a guide for the calculation, \textit{i.e.}, substituting the multipole-type vertex function in the BS model with the Gaussian radial wave function. The detailed formalism for the CLF QM is described in Refs.~\cite{Jaus:1999zv, Jaus:2002sv, Cheng:2003sm, Choi:2013mda, Chang:2018zjq, Chang:2019mmh, Chang:2020wvs}. We will proceed to calculate the form factors for $P \to A$ transitions in the next subsection. 
	\subsection{Form factors}
	\label{sec:ff}
	The amplitudes for the $B_c\to A$ transition form factors can be obtained from the Feynman rules for the meson quark–antiquark vertices as shown in Figure~\ref{fig:feynman}. 
	These Feynman rules for vertices ($i\Gamma^\p_M$) of ground state $s$-wave mesons and low-lying $p$-wave mesons are summarized in Table~I of Ref.~\cite{Cheng:2003sm}. In general, the current matrix element in the CLF approach is given by \cite{Chang:2019mmh},
	\begin{align}\label{eq:amp}
		{\cal B} \equiv \la  M^{\p\p}(p^{\p\p}) | \bar{q}^{\p\p}_1 (k_1^{\p\p})\Gamma q^{\p}_1(k_1^{\p}) |M^{\p}(p^{\p}) \ra \,,
	\end{align} 
	where $\Gamma=\r^\u\,$ and $\r^\u\r_5$. The matrix element in Eq.~\eqref{eq:amp} is equivalent to the most general form of the hadronic matrix elements of $B_c\to A$ transition induced by vector and axial-vector currents given by \cite{Cheng:2003sm},
	\begin{eqnarray}
		\label{eq:AV_ISGW}
		\la A(p^{\p\p},\ve^*)|\bar q_1^{\p\p} \gamma^\mu q_1^{\p}|B_c(p^\p)\ra&=&i\left\{l(q^2)\ve^{*\mu}+ \ve^*\cdot P[P^\mu c_+(q^2)+q^\mu c_-(q^2)]\right\},\\
		\label{eq:AA_ISGW}
		\la A(p^{\p\p},\ve^*)|\bar q_1^{\p\p} \gamma^\mu\gamma_5 q_1^{\p}|B_c(p^\p)\ra&=&-q(q^2)\e^{\mu\nu\rho\sigma}\ve^{*}_{\nu}P_{\rho} q_{\sigma},
	\end{eqnarray}
	where the four-momentum transfer, $q^\mu=p^{\p}-p^{\p\p}$, $P=p^\p+p^{\p\p}$, and $\ve^*$ represents the polarization of axial-vector meson. Further, in terms of conventional Bauer-Stech-Wirbel (BSW) type form factors, the above matrix elements can be expressed as \cite{Cheng:2003sm}
	\begin{eqnarray}
		\label{eq:AV_BSW}
		\la A(p^{\p\p},\ve^*)|\bar q_1^{\p\p} \gamma^\mu q_1^{\p}|B_c(p^\p)\ra&=& -\Bigg\{2M_{A}
		V_0(\q2)\frac{\ve^*\cdot q}{\q2}\ q^\mu
		+(M_{B_c}-M_{A})V_1(\q2)\left(\ve^{*\mu}-\frac{\ve^*\cdot
			q}{\q2}\ q^\mu\right)\cr\cr
		&&-V_2(\q2)\frac{\ve^*\cdot q}{M_{B_c}-M_{A}}\left[p^{\p\mu}+
		p^{\p\p\mu}-\frac{M_{B_c}^2-M_{A}^2}{\q2}\ q^\mu\right]\Bigg\},\\
		\label{eq:AA_BSW}
		\la A(p^{\p\p},\ve^*)|\bar q_1^{\p\p} \gamma^\mu\gamma_5 q_1^{\p}|B_c(p^\p)\ra&=&\frac{2iA(\q2)}{M_{B_c}-M_{A}} \e^{\mu\nu\rho\sigma}\ve^*_\nu
		p^\p_{\rho} p^{\p\p}_{\sigma},
	\end{eqnarray}
	where $M_{B_c}(M_{A})$ represents the mass of initial $B_c$ (final axial-vector $A$) meson and the form factors satisfy the relations, \begin{equation}\label{V1V2V3}
		\begin{gathered}
			V_3(0)=V_0(0),\\
			V_3(\q2)=\frac{M_{B_c}-M_{A}}{2M_{A}}V_1(\q2)
			-\frac{M_{B_c}+M_{A}}{2M_{A}}V_2(\q2).
		\end{gathered}
	\end{equation} 
	These BSW-type form factors are related to $l(\q2), ~c_{+}(\q2), ~c_{-}(\q2)$, and $q(\q2)$ CLF QM form factors by the following relations:
	\begin{equation}\label{ff_CLF}
		\begin{gathered}
			A^{P A}(\q2)=-(M_{B_c}-M_{A}) q(\q2), \quad V_{1}^{P A}(\q2)=-\frac{l(\q2)}{M_{B_c}-M_{A}}, \\
			V_{2}^{P A}(\q2)=(M_{B_c}-M_{A}) c_{+}(\q2), \quad
			V_{3}^{P A}(\q2)-V_{0}^{P A}(\q2)=\frac{\q2}{2 M_{A}} c_{-}(\q2).
		\end{gathered}
	\end{equation}
	
	In LF QM, efficiently calculates non-perturbative quantities, like, form factors and decay constants, using the plus component of weak currents. The CLF QM extends this approach, achieving full Lorentz covariance by incorporating zero-mode effects in one-body currents, particularly when $q^+ = q^0 + q^3 = 0$. These zero-mode contributions are crucial for restoring Lorentz covariance in LF framework matrix elements~\cite{Jaus:2002sv}. Customarily, for $M^{\p}(p^{\p})\rightarrow M^{\p\p}(p^{\p\p})$ transition, it is convenient to use Drell–Yan–West frame\footnote{Drell and Yan \cite{Drell:1969km} demonstrated that choosing a frame where $q^+ =0$ simplifies calculations of form factors, structure functions, and other current matrix elements. In this frame, vacuum pair creation/annihilation is effectively suppressed, allowing for direct computation of matrix elements as overlaps of wave functions.}, $q^+ = 0$, which implies that the form factors are known only for space-like momentum transfer, $\q2=-\mathbf{q}_\bot^2 \leqslant 0$, and for the time-like region ($\q2=-\mathbf{q}_\bot^2 \geqslant 0$) one needs an additional $\q2$ extrapolation. Furthermore, we consider a Lorentz frame in which $\mathbf{p}_{\bot}'=0$ and $\mathbf{p}_{\bot}''=-\mathbf{q}_{\bot}$ leads to $\mathbf{k}_{\bot}''=\mathbf{k}_{\bot}'-\bar{x}\mathbf{q}_{\bot}$. Evaluating the hadronic matrix elements of the aforementioned $\Gamma$ operators described in Eqs.~\eqref{eq:AV_ISGW}-\eqref{eq:AA_BSW} is necessary for calculating the exclusive weak decays of the $B_c$ meson. Following the approach of \cite{Jaus:1999zv, Chang:2019mmh, Chang:2020wvs}, the form factors in Eqs.~\eqref{eq:AV_ISGW} and \eqref{eq:AA_ISGW} can be extracted from one-loop approximation as a momentum integral given by, 
	\begin{eqnarray}
		\label{eq:Bclf1}
		{\cal B}=N_c \int \frac{\d^4 k_1^{\p}}{(2\pi)^4} \frac{H_{M^{\p}}H_{M^{\p\p}}}{N_1^{\p}\,N_1^{\p\p}\,N_2}iS_{\cal B}\,,
	\end{eqnarray} 
	where $N_c$ denotes the number of colors. $H_{M', M''}$ are the bound-state vertex functions, those depend upon quark's momenta $\hat k^{\p 2}_1$ and $\hat k^2_2$, \textit{i.e.}, $H_{M^{(\p,\p\p)}} \equiv H_{M^{(\p,\p\p)}}(\hat k^{(\p,\p\p) 2}_1,\hat k^2_2)$. The denominators $N_{1}^{(\p,\p\p)}=k_{1}^{(\p,\p\p)2}-m_1^{(\p,\p\p)2}+i\e$ and $N_{2}=k_{2}^{2}-m_2^{2}+i\e$ (with infinitesimally small $\e>0$) represent the free fermionic propagators for $q_1^{(\p,\p\p)}$ and $\bar{q}_2$ (since the quarks are off-shell in CLF QM) in momentum space, corresponding to decaying and spectator quarks, respectively. The trace term $S_{\cal B}$ associated with the fermion loop in Eq.~\eqref{eq:Bclf1} is expressed as \cite{Chang:2019mmh}
	\begin{eqnarray}
		S_{\cal B}={\rm Tr}\left[\Gamma\, (\not\!k'_1+m'_1)\,(i\Gamma_{M'})\,(-\!\not\!k_2+m_2)\,(i\r^0{\Gamma}_{M''}^{\dag}\r^0) (\not\!k_1''+m_1'')\right]\,,
	\end{eqnarray}
	where the vertex operators $\Gamma_{M'}$ and ${\Gamma}_{M''}$ are given by~\cite{Cheng:2003sm}
	\begin{align}\label{eq:voCLF}
		i\Gamma_P&=-i\r_5\,,  & i\Gamma_V&=i\left[\gamma^\mu-\frac{ (k_1-k_2)^\mu}{D_{ V,{\rm con}}}\right]\,,\nonumber\\
		i\Gamma_{A_B}&=i\frac{(k_1-k_2)^\mu}{D_{A_B,{\rm con}}} \r_5\,, & i \Gamma_{ A_A}&=i\left[ \gamma^\mu+\frac{ (k_1-k_2)^\mu}{D_{A_A,{\rm con}}} \right]\r_5 \,,\\ 
		D_{V,\rm con}&=M+m_1+m_2\,, & D_{A_B,\rm con}&=2\,, \quad D_{A_A,\rm con}=\frac{{M}^2}{m_1-m_2}\,\non
	\end{align} 
	for $P$, $V$, $A_B$, and $A_A$ mesons, respectively. We employ the LF decomposition of the four-momentum vector to evaluate the four-momentum integrals in Eq.~\eqref{eq:Bclf1}, where $\d^4 k_1'=\frac{1}{2} \d k_1'^- \d k_1^{\p +} \d^2 \mathbf{k}_{\bot}'$. This transition from covariant Feynman momentum loop integrals to a LF approach is facilitated by the contour integration method, which eliminates the internal $k_1'^-= k_0'-k_3'$ momentum component in $\cal B$ (Eq.~\eqref{eq:Bclf1}) \cite{Chang:1968bh}. A fundamental assumption in deriving the LF expression from covariant perturbation theory is that the composite meson state is independent of the $k_1'^- $ component. It has been established that when the vertex function $H_{M^\p}$ is assumed to have no poles in the upper complex $k_1'^-$ plane, the covariant evaluation of meson properties and LF calculations yield equivalent results at the one-loop level \cite{Jaus:1999zv}. The contour integration method necessitates vertex functions free of singularities, with quark propagator singularities being the sole contributors within the integration contour. A class of covariant meson vertex functions, characterized by asymmetry in the constituent quark-antiquark pair variables, satisfies this criterion. The contour integration over the upper $k_1'^-$ plane under the condition $q^+ = 0$ ensures that the integrals are evaluated at the poles corresponding to the spectator quark's momentum. This effectively places the spectator on its mass shell, i.e., $k_2^2=\hat{k}_2^2=m_2^2$. The other momenta can be determined through momentum conservation: $k_{1}^{\p}=p^{\p}-k_2$ and $k_{1}^{\p\p}=p^{\p\p}-k_2$. It is important to note that the invariant mass $M^{\p}_0$ in Eq.~\eqref{eq:kinec_mass} of the composite meson state, represented by the $q_1^{\p}\bar{q}_2$ bound-state, generally differs from the mass $M$ of the meson satisfying $p^{\p 2}= M^{\p 2}$. This discrepancy arises because the meson, quark, and antiquark cannot all be simultaneously on-shell \cite{Hwang:2010hw}. The presence of the spectator propagator, $N_2$, in the denominator of Eq.~\eqref{eq:Bclf1} necessitates particular caution when performing the $k_1'^-$ contour integration. Thereafter, a conversion of the covariant BS formulation to LF is achieved by implementing the following replacements, as given in Refs.~\cite{Jaus:1999zv, Cheng:2003sm},
	\begin{eqnarray}
		\hat{N_2}&=&\hat{k}_{2}^{2}-m_2^{2}=0, \non\\
		N_1^{\p} \to \hat{N}_1^{\p}&=&\hat{k}_{1}^{\p2}-m_1^{\p2}=x \left( M^{\p2}-M_0^{\p2}\right),\non\\
		N_1^{\p\p} \to \hat{N_1^{\p\p}}&=& \hat{k}_{1}^{\p\p2}-m_1^{\p\p2}=x \left( M^{\p\p2}-M_0^{\p\p2}\right),
	\end{eqnarray}
	\begin{eqnarray}\label{eq:type-I}
		\chi_M = H_M/N_1\to h_M/\hat{N}_1\,,\qquad  D_{M,{\rm  con}} \to D_{M,{\rm  LF}}\,, 
	\end{eqnarray}
	where $h_M$ (or $\chi_M$) represents the LF covariant vertex function. Eq.~\eqref{eq:type-I} portrays the relationship between the manifestly covariant and SLF formalism. Jaus \cite{Jaus:1999zv} demonstrated that achieving the above mentioned equivalences necessitates the inclusion of two-body currents. These currents ensure compatibility between the total current operator and the chosen LF vertex function. Moreover, LF vertex functions denoted by $h_M$ that depend solely on $\hat{N_1^{\p}}$ and $\hat{N_1^{\p\p}}$ guarantee form factor expressions devoid of spurious $\omega$ contributions. This property stems from the covariant formalism's requirement for LF vertex functions to be functions exclusively of $\hat{N_1^{\p}}$ and $\hat{N_1^{\p\p}}$. Note that LF vertex functions $h_M$ of the bound state are asymmetric with respect to the constituent quark variables \cite{Jaus:1999zv}. This condition essentially dictates the specific structure of vertex functions employed within a covariant analysis. Furthermore, the $D$ factor appearing in the denominator of the vertex operator accounts for the model dependency of a meson \cite{Choi:2013mda, Chang:2018zjq}. For example, in the case of vector meson, $D_{M,{\rm con}}=M + m_1 + m_2$ and $D_{M,{\rm LF}}=M_0 + m_1 + m_2$. The transformation given by Eq.~\eqref{eq:type-I} is referred to as the type-I correspondence scheme between the covariant BS and LF approaches \cite{Jaus:1999zv, Choi:2013mda, Cheng:2003sm}. As previously mentioned, for the pseudoscalar transition form factor $P \to P$, Eq.~\eqref{eq:type-I} substantiates the correspondence between the CLF and SLF expressions through certain zero-mode independent quantities \cite{Jaus:1999zv, Cheng:2003sm}. However, this correspondence is confined solely to the $D$ factor \cite{Choi:2013mda}. It may be noted that our theoretical results and analyses are presented explicitly in CLF formalism for the plus component of the weak current. The definitions of $h_M$ and $D$ factors, for \textit{P}, $A_A$, and $A_B$ mesons are given by,
	\begin{eqnarray} \label{eq:vertex}
		h_{{P}}^{\p}/\hat{N}_1^{\p} & = &\frac{1}{\sqrt{2N_c}}\sqrt{\frac{\bar{x}}{x}}
		\frac{\psi_s}{\hat{M}^\p_0},
		\non\\
		h_{{A_A}}^{\p\p}/\hat{N}_1^{\p\p} & = &\frac{1}{\sqrt{2N_c}}\sqrt{\frac{\bar{x}}{x}}
		\frac{\hat{M}^{\p\p
				2}_0}{2\sqrt{2}M^{\p\p}_0}\frac{\psi_p}{\hat{M}^{\p\p}_0},
		\non\\
		h_{{A_B}}^{\p\p}/\hat{N}_1^{\p\p} & = &\frac{1}{\sqrt{2N_c}}\sqrt{\frac{\bar{x}}{x}}
		\frac{\psi_p}{\hat{M}^{\p\p}_0},
		\non\\
		D_{^{A_A}}^{\p\p}&=&\frac{\hat{M}^{\p\p2}_0}{m^{\p\p}_1-m_2},\quad
		D_{^{A_B}}^{\p\p}=2,
	\end{eqnarray}
	with $\hat{M}_0^{\p2} \equiv M_0^{\p2}-(m_1^{\p}-m_2)^2$. Eq.~\eqref{eq:vertex} then explicitly describes the connection between these wave functions and the corresponding vertex operators. It should be noted that Jaus~\cite{Jaus:1999zv} has described the relationship between CLF vertex operators and LF wave functions. The dependence on the invariant mass squared in Eq.~\eqref{eq:vertex} signifies that the form factors expressed as LF wave function overlap integrals are analytic functions of $\q2$. 
	
	Aforementioned, hadronic matrix elements are typically expressed in terms of constituent quark-antiquark bound-state wave functions. The Eq.~\eqref{eq:Bclf1} can be explained from factorization within QCD \cite{Collins:1989gx, Brodsky:1981fz, Buras:1979yt, Lepage:1980fj}. In QCD, the amplitude for a hadronic process can be separated into two distinct components: a) the long-distance hadronic bound state dynamics (non-perturbative contributions) associated with the hadron wave function; b) a hard-scattering amplitude for the transition of constituents $q_1^{\p}\to q_1^{\p\p}$ ($\bar{q}_2$ as a spectator), which only involves the short-distance gluon contribution\footnote{Due to the inherent confinement of quarks within hadrons, a universal definition of their on-shell mass is precluded. However, within the perturbative framework, by ignoring confinement effects, an ``on-shell'' mass can be approximated as the pole of the renormalized propagator (represented by $N_{1}^{\p(\p\p)}$ and $N_{2}$ in Eq.~\eqref{eq:Bclf1}) to a specific perturbative order \cite{Neubert:1993mb}.} to the form factor, at least to leading order in QCD \cite{Brodsky:1981fz, Brodsky:1981jv}. The LF formalism for being efficient in describing relativistic bound states provides a suitable framework for such analysis. The diagrammatic approach in light-front perturbation theory \cite{Lepage:1980fj, Michael:1981xx} permits the exact calculation of the one-loop Feynman diagram (Figure~\ref{fig:feynman}) as shown in~\cite{Jaus:1989au}. This is attributed to the correspondence between the covariant Feynman formalism and time-ordered perturbation theory in the infinite momentum frame \cite{Schmidt:1973ja}. The well-established perturbative methods, that rely on the notion of free particles, are only strictly applicable in a limited regime: the far-off-shell, short-distance domain \cite{Brodsky:1981fz}. However, the perturbative QCD breaks down near hadron production thresholds \cite{Mueller:1981sg, Brodsky:1981jv} due to the appearance of numerous coherent and incoherent transitions and non-perturbative effects \cite{Brodsky:1981jv}. All the non-perturbative bound state physics is isolated in the process-independent transition amplitudes expressed as the product of initial and final state LF covariant vertex functions \cite{Brodsky:1981fz}. Therefore, the hadron form factors can be written in a factorized form as a product of LF vertex functions that are expressed in terms of wave functions\footnote{Alternatively, Brodsky \textit{et al.} \cite{Brodsky:1997de} has demonstrated that hadron form factors can be expressed in a factorized form as the convolution of quark distribution amplitudes and a hard-scattering amplitude, where each hadron in the process is represented by its respective distribution amplitude.} (see Eq.~\eqref{eq:vertex}), one for each hadron involved in the amplitude, with a hard-scattering amplitude corresponding to the free fermionic propagator \cite{Lepage:1980fj, Lepage:1979zb, Lepage:1979za}. 
	
	\subsubsection*{\textbf{Effects of zero-mode contributions:}}
	After the $k_1'^-$ integration, one can then employ well-established vertex functions (Eq.~\eqref{eq:vertex}) within the conventional light-front approach, details of which are discussed in Ref.~\cite{Jaus:1999zv}. We proceed further to evaluate Eq.~\eqref{eq:Bclf1}, which after some simplifications (see Ref.~\cite{Chang:2019mmh}), can be written into a more convenient LF form,
	\begin{eqnarray}
		\label{eq:Bclf2}
		\hat{{\cal B}}=N_c \int \frac{\d x \d^2 \mathbf{k}_{\bot}^{\p}}{2(2\pi)^3}\frac{h_{M^{\p}}h_{M^{\p\p}}}{\bar{x} \hat{N}_1^{\p}\,\hat{N}_1^{\p\p}\,}\hat{S}_{\cal B}\,,
	\end{eqnarray}
	where, $\hat{N}_1^{\p}$ and $\hat{N}_1^{\p\p}$ describes the propagation of the two off-shell quarks with the same definition as in Eq.~\eqref{eq:Bclf1}. A peculiar property of the integrals $\hat{{\cal B}}$ (Eq.~\eqref{eq:Bclf2}) is their dependence on the LF, defined by the light-like four vector $\omega^\mu=(0, 2, 0_{\bot})$. It is essential to acknowledge that the transition from Eq.~\eqref{eq:Bclf1} to Eq.~\eqref{eq:Bclf2}, facilitated by the integration over $k_1^{\p -}=0$, implicitly neglects the contributions associated with zero modes. Chang et al.~\cite{Chang:1972xt} identified that despite contour integration analysis, the contribution of the zero-mode specifically arises from the $k_1^{\p +}=0$ region as all poles are in the lower complex $k_1^{\p-}$ plane at
	\begin{eqnarray}\label{eq:dispersion}
		k_1^{\p-}= \frac{m^{\p 2}_{1\bot}-i\e}{k_1^{\p+}}.
	\end{eqnarray}
	where $m^{\p 2}_{1\bot}= m^{\p 2}_{1}+k^{\p 2}_{1\bot}$ (for more details see \cite{Jaus:1999zv}). In Ref.~\cite{Chang:1972xt}, it is argued that these pole contributions vanish entirely for non-zero $k_1^{\p+}$. However, when $k_1^{\p+}=0$, the pole in Eq.~\eqref{eq:dispersion} resides at infinity. This pole cannot be eliminated by closing the contour in either the upper or lower half-plane. In this instance, the integration yields a term proportional to $\delta(k_1^{\p+})$. The zero-mode contributions corresponding to $\delta(k_1^{\p+})$ have been understood as the residues of virtual pair creation processes \cite{deMelo:1998an}. In order to isolate the $\omega$-dependent terms, Jaus \cite{Jaus:1999zv} proposed to decompose the terms related to $\hat{k_1^{\p}}$ in the integrand corresponding to $P=p^\p+p^{\p\p}$, $q=p^\p-p^{\p\p}$, and $\omega$. Hence, three coefficient functions are introduced: $A^{(i)}_{j}$, $B^{(i)}_{j}$, and $C^{(i)}_{j}$. While $A^{(i)}_{j}$ represents the $\omega$-independent components, $B^{(i)}_{j}$, and $C^{(i)}_{j}$ includes the $\omega$-dependent terms, which contribute to spurious components in physical observables (for their explicit expressions see Appendix~\ref{cov_ml}). These spurious contributions can be canceled after correctly performing the integration, namely, by the inclusion of the zero-mode contribution \cite{Chang:1972xt, Jaus:1999zv}, so that the result is guaranteed to be covariant. It has been established that the $\omega$-dependence defined by $\omega^\mu=(0,2,\mathbf{0}_\bot)$, mainly arising through $C^{(i)}_{j}$ functions, can be eliminated by incorporating appropriate zero-mode contributions. Furthermore, the light-like four-vector $\omega$-dependence is an inherent characteristic of the LF matrix element $\hat{\cal B}$, which exhibits physical and $\omega$-dependent (unphysical) components \cite{Jaus:1999zv}. These unphysical contributions, if non-zero, could violate the covariance of the matrix element. In the CLF QM, the majority of the $\omega$-dependent contributions are nullified by the zero-mode contributions. However, some residual $\omega$-dependencies linked to $B^{(i)}_{j}$ functions persist and are independent of zero-mode (due to the condition $x\hat{N}_2=0$).
	
	Recently, it was investigated in \cite{Cheng:2003sm, Choi:2013mda} that type-I correspondence, Eq.~\eqref{eq:type-I}, suffers from the problems of self-consistency and covariance. It has been argued that these issues are connected with zero-mode contributions in type-I correspondence. Furthermore, in addition to vector and axial-vector decay constants, the $P \to V$ meson transition form factors have self-consistency problems \cite{Choi:2013mda}. This inconsistency arises because of the contributions from $ B^{(i)}_{j}$ functions. Likewise, $P \to V$ transition form factors, these issues are expected to affect $P \to A$ transition form factors as well. However, as mentioned before, the type-II scheme has not yet been employed and tested for the resolution of the self-consistency problem in $P \to A$ transition form factors. 
	
	As proposed by Choi et al.~\cite{Choi:2013mda}, these issues can be resolved with type-II correspondence after replacing the meson mass ($M$) with the kinetic invariant mass ($M_0$). The effect of type-II correspondence is such that the contributions from $B^{(i)}_{j}$ functions should numerically vanish. In $P \to A$ transitions, the $l(q^2)$ and $c_{-}(q^2)$ form factors receive zero-mode contributions for the $\lambda=0$ polarization states through $B^{(2)}_{1}$ and $B^{(3)}_{3}$ functions, which are absent in the $\lambda=\pm$ polarization states. At this juncture, it is crucial to acknowledge the distinction between the LF quark models proposed by Choi and Ji \cite{Choi:2013mda} and Jaus \cite{Jaus:1999zv}. The authors in \cite{Choi:2013mda} and their subsequent works initiate their approach from the manifestly covariant Bethe-Salpeter wave function model. Crucially, they introduce a novel matching condition, termed ``type-II'' correspondence, during the transformation from the BS model to the LF QM framework utilizing a Gaussian wave function. It is of utmost importance to recognize that the type-II correspondence relies on the fact that, although the results obtained from the BS model allow the non-zero binding energy ($M^2-M_0^2$), the LF results are derived under the constraint of zero binding energy ($M \to M_0$) \cite{Choi:2013mda, Choi:2021qza}. Therefore, Choi and Ji~\cite{Choi:2013mda} have proposed a generalized type-II correspondence scheme given as 
	\begin{equation}
		\label{eq:type-II}
		\chi_M = H_M/N_1\to h_M/\hat{N}_1\,,\qquad M \to M_0.
	\end{equation}
	The type-II correspondence scheme effectively resolves the issues concerning self-consistency and covariance, as demonstrated in Ref.~\cite{Chang:2019mmh}, for the vector meson decay constants and $P \to V$ form factors. It should be noted that the $M \to M_0$ replacement is applicable in each and every term containing $M$ in the integrand of the form factors inclusive of $D$ factors given in Eq.~\eqref{eq:vertex}. Further, the type-II correspondence, in Eq.~\eqref{eq:type-II}, is expected to eliminate the unnecessary spurious contribution and show self-consistency in $P \to A $ transition form factors. Thus, their numerical evaluation should be free from zero-mode contributions. 
	
	The type-I correspondence scheme, typically, suffers from these residual $\omega$-dependencies, as shown in Refs. \cite{Chang:2018zjq, Chang:2019mmh}, which in turn leads to a violation of the covariance of the matrix element. However, in type-II correspondence, these unphysical $\omega$-dependent contributions vanish numerically restoring the covariance of the matrix element. Therefore, one can conclude that the self-consistency and covariance originate for the same reason \cite{Chang:2018zjq, Chang:2019mmh}. It may be noted that Jaus~\cite{Jaus:1999zv} presented a practical approach to deal with the $\omega$-dependencies, which involve several specific substitutions in the integrand $\hat{S}_{\cal B}$ of Eq.~\eqref{eq:Bclf2}. Furthermore, authors of \cite{Chang:2018zjq} have discussed the resolution of the covariance problem in type-II correspondence and highlighted the major substitutions signifying the presence of residual $\omega$-dependencies as shown in the Appendix~\ref{cov_ml}. Thus, Qin Chang~\textit{et al.}~\cite{Chang:2019mmh} have demonstrated that problems of self-consistency and covariance of matrix elements can be simultaneously resolved by employing type-II correspondence. A similar treatment followed in $P\to A$ transition matrix element, as explained in Appendix~\ref{cov_ml}, should resolve the covariance problem. It is then straightforward to extract the $P\to A$ transition form factor relations according to their Lorentz structure decomposition of the matrix elements as described by Eqs.~\eqref{eq:amp} and \eqref{eq:Bclf1}, which will further be modified due to type-II correspondence scheme as given in Eq.~\eqref{eq:Bclf2}. Note that $P\to A$ form factor relations can be derived by expanding the $P\to V$ framework through specific modifications described in \cite{Jaus:1999zv, Cheng:2003sm}. Following which, the matrix element ${\cal B}$ applicable to transitions from $P \to A_A,\,A_B$, where the axial-vector states $A_A$ is ${}^{3} P_{1}$ and $ A_B$ is ${}^{1} P_{1}$ (as described in Appendix~\ref{axial_mixing}), are expressed as:	       	
	\begin{align}
		{\cal B}^{P\,A_A}_\mu&=-i^3\frac{N_c}{(2\pi)^4}\int d^4 k^\p_1
		\frac{H^\p_P (-i H^\pp_{A_A})}{N_1^\p N_1^\pp N_2} S^{P\,A_A}_{\mu\nu}\,\ve^{*\nu},
		\\
		{\cal B}^{P\,A_B}_\mu&=-i^3\frac{N_c}{(2\pi)^4}\int d^4 k^\p_1
		\frac{H^\p_P (-i H^\pp_{A_B})}{N_1^\p N_1^\pp N_2} S^{P\,A_B}_{\mu\nu}\,\ve^{*\nu},
		\label{eq:BPA}
	\end{align}
	with
	\begin{align}\label{SP_AA}
		S^{P\,A_A}_{\mu\nu}
		&= (S^{P\,A_A}_{V}-S^{P\,A_A}_A)_{\mu\nu}
		\non\\
		&={\rm Tr}\left[\left(\gamma_\nu-\frac{1}{D^\pp_{A_A}}(k_1^\pp-k_2)_\nu\right)
		\gamma_5
		(\not \!k^\pp_1+m_1^\pp)
		(\gamma_\mu-\gamma_\mu\gamma_5)
		(\not \!k^\p_1+m_1^\p)\gamma_5(-\not
		\!k_2+m_2)\right]
		\non\\
		&={\rm Tr}\left[\left(\gamma_\nu-\frac{1}{D^\pp_{A_A}}(k_1^\pp-k_2)_\nu\right)
		(\not \!k^\pp_1-m_1^\pp)
		(\gamma_\mu\gamma_5-\gamma_\mu)
		(\not \!k^\p_1+m_1^\p)\gamma_5(-\not
		\!k_2+m_2)\right],
	\end{align}
	\begin{align}\label{SP_AB}
		S^{P\,A_B}_{\mu\nu} &=(S^{P\,A_B}_{V}-S^{P\,A_B}_A)_{\mu\nu}
		\non\\
		&={\rm Tr}\left[\left(-\frac{1}{D^\pp_{A_B}}(k_1^\pp-k_2)_\nu\right)
		\gamma_5
		(\not \!k^\pp_1+m_1^\pp)
		(\gamma_\mu-\gamma_\mu\gamma_5)
		(\not \!k^\p_1+m_1^\p)\gamma_5(-\not
		\!k_2+m_2)\right]
		\non\\
		&={\rm Tr}\left[\left(-\frac{1}{D^\pp_{A_B}}(k_1^\pp-k_2)_\nu\right)
		(\not \!k^\pp_1-m_1^\pp)
		(\gamma_\mu\gamma_5-\gamma_\mu)
		(\not \!k^\p_1+m_1^\p)\gamma_5(-\not
		\!k_2+m_2)\right],
	\end{align}
	where the slash notation refers to the trace term associated with the fermion loop \cite{Cheng:2003sm}. Eqs.~\eqref{eq:Bclf1}-\eqref{SP_AB} provide the theoretical formulas for the $B_c \to A$ transition form factors at the one-loop level. Furthermore, we used the following transformations to obtain the analytic expressions for $P\rightarrow A$ transition form factors from that of $P \rightarrow V$ in \cite{Chang:2019mmh}\footnote{Note that we use the notation of \cite{Chang:2019mmh}, \textit{i.e.}, $[\mathcal{F}(\q2)]_{full}$ to keep the consistency with analytical relations of the form factors in type-II correspondence that should be converted to $P \to A_A/A_B$ form factors.}: 
	\begin{equation}\label{eq:V_A_trans}
		\begin{aligned}
			& F^{A_A, A_B}(\q2)=[\mathcal{F}(\q2)]_{full} \text { with }\left(m_1^{\p\p} \rightarrow-m_1^{\p\p},~ \chi_V^{\p\p}\to \chi_{A_A, A_B}^{\p\p}, D^{\p\p}_V\to D^{\p\p}_{{{ }^{A_A,A_B}}}\right). 
		\end{aligned}
	\end{equation}
	Thus, we obtain the following relation for $P\rightarrow A$ transition form factors in type-II correspondence of the CLF QM, 
	\begin{align}\label{eq:Fq2}
		F^{A_A, A_B}(\q2)=N_c \int \frac{\d x \,  \d^2{\bf k}_{\bot}^{\p}}{2(2\pi)^3}\frac{\chi^{\p}_{P}\chi_{A_A, A_B}^{\p\p}}{\bar{x}}\,\widetilde{F}(x,{\bf k}_{\bot}^{\p},\q2)\,,
	\end{align} 
	where the function $\widetilde{F}(x,{\bf k}_{\bot}^{\p},\q2)=\widetilde{q}(x,{\bf k}_{\bot}^{\p},\q2), ~\widetilde{l}(x,{\bf k}_{\bot}^{\p},\q2)$, and $\widetilde{c}_{\pm}(x,{\bf k}_{\bot}^{\p},\q2)$ for $\q2 < 0$. We, therefore, express the form factors for $P \to A_A$ transitions as
	\be \label{eq:qfull}
	\widetilde{q}(x,{\bf k}_{\bot}^{\p},\q2)=&-2\left\{\bar{x} m_1^{\p} +xm_2+(m_1^{\p} + m_1^{\p\p} )\frac{{\bf k}_{\bot}^{\p}\cdot {\bf q}_\bot}{\q2}+\frac{2}{D^{\p\p}_{{{ }^{A_A}}}}\left[{{\bf k}_{\bot}^{\p}}^2 +\frac{({\bf k}_{\bot}^{\p} \cdot {\bf q}_{\bot})^2}{\q2} \right] \right\}\,, \en 
	\be \label{eq:cpfull}
	\widetilde{c}_{+}(x,{\bf k}_{\bot}^{\p},\q2)
	=&2 \bigg\{  (-m_1^{\p\p}-2xm_1^{\p}+m_1^{\p}+2xm_2)\, \frac{ {\bf k}_{\bot}^{\p}\cdot {\bf q}_{\bot }}{{\bf q}_{\bot }^2} + ( x-\bar{x})(\bar{x}m_1^{\p}+x m_2)   \nonumber\\
	& +\frac{2}{D^{\p\p}_{{{ }^{A_A}}}}   \frac{{\bf k}_{\bot}^{\p\p}\cdot {\bf q}_{\bot}}{\bar{x} {\bf q}_{\bot}^2}  \left[ {\bf k}_{\bot}^{\p}\cdot{\bf k}_{\bot}^{\p\p}+  (x m_2 + \bar{x}m_1^{\p\p})(x m_2+ \bar{x}m_1^{\p}) \right] \bigg\} \,, \en 
	\begin{align}
		\label{eq:lfull}
		\widetilde{l}(x,{\bf k}_{\bot}^{\p},\q2)=&-2\Bigg\{
		-(m_1^{\p}-m_1^{\p\p})^2 (m_1^{\p}-m_2) +(xm_2-\bar{x}m_1^{\p}) {M^{\p}}^2+(xm_2+\bar{x}m_1^{\p}) {M^{\p\p}}^2\nonumber\\ 
		& -x(m_2-m_1^{\p} )({M_0^{\p}}^2+{M_0^{\p\p}}^2)-2xm_1^{\p\p}{M_0^{\p}}^2  -4 \left(m_1^{\p}-m_2\right) \left({{\bf k}_{\bot}^{\p}}^2 +\frac{({\bf k}_{\bot}^{\p} \cdot {\bf q}_{\bot})^2}{\q2}\right)\nonumber\\ 
		&- m_2 \q2 -(m_1^{\p}-m_1^{\p\p})(\q2+q\cdot P)\frac{{\bf k}_{\bot}^{\p}\cdot {\bf q}_{\bot}}{\q2} {+4 (m_1^{\p}-m_2)  B_1^{(2)}}\non
	\end{align}
	\begin{align}
		&~~~~~~~~~~~~~+\frac{2}{D^{\p\p}_{{{ }^{A_A}}}} \bigg[ \left({{\bf k}_{\bot}^{\p}}^2 +\frac{({\bf k}_{\bot}^{\p} \cdot {\bf q}_{\bot})^2}{\q2}\right) \bigg((x-\bar{x}){M^{\p}}^2+{M^{\p\p}}^2-2(m_1^{\p}+m_1^{\p\p}) (m_1^{\p}-m_2) \nonumber\\ 
		&~~~~~~~~~~~~~+2x {M_0^{\p}}^2-\q2 -2 (\q2+q\cdot P)\frac{{\bf k}_{\bot}^{\p}\cdot {\bf q}_{\bot}}{\q2}\bigg) \nonumber\\ 
		&~~~~~~~~~~~~~- \left({M^{\p}}^2+{M^{\p\p}}^2-\q2 + 2(m_1^{\p}-m_2)(-m_1^{\p\p}+m_2)\right) B_1^{(2)} +2  B_3^{(3)}\bigg]\Bigg\}\,, 
	\end{align} 
	\begin{align}
		\label{eq:cmfull}
		\widetilde{c}_{-}(x,{\bf k}_{\bot}^{\p},\q2)=&-2\Bigg\{   (3-2x)(\bar{x}m_1^{\p}+xm_2) -  \left[(6x-7)m_1^{\p}+(4-6x)m_2-m_1^{\p\p}\right] \frac{{\bf k}_{\bot}^{\p} \cdot {\bf q}_{\bot}}{\q2}\nonumber\\ 
		& +4(m_1^{\p}-m_2)\left[2\left(\frac{{\bf k}_{\bot}^{\p} \cdot {\bf q}_{\bot}}{\q2}\right)^2+\frac{{{\bf k}_{\bot}^{\p}}^2}{\q2}\right] {- 4 \frac{(m_1^{\p}-m_2)  }{\q2}B_1^{(2)}}\nonumber\\ 
		&+\frac{1}{D^{\p\p}_{{{ }^{A_A}}}} \bigg[-2\left({M^{\p}}^2+{M^{\p\p}}^2-\q2 + 2(m_1^{\p}-m_2)(-m_1^{\p\p}+m_2)\right) (A_3^{(2)}+A_4^{(2)}-A_2^{(1)})\non
	\end{align}
	\begin{align}
		&~~~~~~~~~~~~~+ \left( 2{M^{\p}}^2-\q2-\hat{N}_1^{\p}+\hat{N}_1^{\p\p}- 2(m_1^{\p}-m_2)^2+(m_1^{\p} -m_1^{\p\p})^2 \right) \left( A_1^{(1)}+ A_2^{(1)} -1\right)\nonumber\\ 
		&~~~~~~~~~~~~~+2Z_2 \left( 2A_4^{(2)}-3A_2^{(1)} +1\right) + 2 \frac{q\cdot P}{\q2} \left( 4 A_2^{(1)}A_1^{(2)} - 3A_1^{(2)} \right) \nonumber\\ 
		&~~~~~~~~~~~~~~
		{+\frac{2}{\q2}  \left(\left({M^{\p}}^2+{M^{\p\p}}^2-\q2 + 2(m_1^{\p}-m_2)(-m_1^{\p\p}+m_2)\right) B_1^{(2)} -2  B_3^{(3)}\right)}\bigg] \Bigg\}\, ,
	\end{align}
	which corresponds to the $\lambda=0$ polarization state. It should be noted that the expressions for $\widetilde{l}$ and $\widetilde{c}_{-}$ as shown in Eqs.~\eqref{eq:lfull} and \eqref{eq:cmfull} will be affected by the presence of zero-mode contributions due to $B^{(i)}_{j}$ functions. To obtain the results for the case of $\lambda=\pm$ polarization state, one can derive them from these formulas by excluding the terms that are related to the $B^{(i)}_{j}$ functions. As mentioned before, the form factors $\widetilde{q}(x,{\bf k}_{\bot}^{\p},\q2)$ and $\widetilde{c}_{+}(x,{\bf k}_{\bot}^{\p},\q2)$ described in Eqs.~\eqref{eq:qfull} and \eqref{eq:cpfull} are independent of $M^{\p(\p\p)}$ and $B^{(i)}_{j}$, and thus, will be free from the zero-mode contributions in both type-I and type-II schemes. It should be noted that the form factors $l(\q2)$ and $c_{-}(\q2)$ as described in Ref.~\cite{Cheng:2003sm} are given for $\lambda=\pm$ polarization only; however, the terms associated with the contribution of $B^{(i)}_{j}$ functions should appear in $\lambda=0$ polarization in type-I. Therefore, it is expected that self-consistency issues will arise in these form factors due to the presence of $B^{(i)}_{j}$ in the type-I scheme, making them numerically different for $\lambda=0$ and $\lambda=\pm$ polarization states \cite{Chang:2019mmh}. As proposed by \cite{Choi:2013mda}, this problem is resolved by an additional transformation $M\to M_{0}$ such that the contributions from $B^{(i)}_{j}$ functions vanish numerically but exist formally in the expressions. This means that the form factors corresponding to $\lambda=0$ and $\lambda=\pm$ polarization states should be numerically equal to confirm the self-consistency in type-II correspondence. Thus, using the formulae discussed until Eq.~\eqref{eq:Fq2}, we obtain the form factors for $B_c\to A_A$ and $B_c\to A_B$ transitions. It should be noted that only the terms $1/D^{\p\p}_{{{ }^{A_B}}}$ are retained for the $B_c\to A_B$ form factors, as shown in Eq. \eqref{SP_AB}. One must exercise caution when considering the replacement of $m_1^\pp\to-m_1^\pp$ and ensure that it is not applied indiscriminately to $m_1^\pp$ in both $D^\pp$ and $\chi^\pp$ \cite{Cheng:2003sm}. Finally, using Eqs.~\eqref{eq:qfull}-\eqref{eq:cmfull} that represent general expressions for the form factors $~\widetilde{l}(q^2),~\widetilde{q}(q^2),~\widetilde{c}_{+}(q^2)$ and $\widetilde{c}_{-}(q^2)$ for the $P\to A$ transitions, we investigate the self-consistency by evaluating these form factors for $\lambda=0$ and $\lambda=\pm$ polarization states. A thorough numerical analysis of these form factors will be presented in the forthcoming sections. 
	
	We remark that the LF formalism, combined with time-ordered perturbation theory, provides a streamlined computational framework for higher-order QCD corrections compared to equal-time methods. Accurate form factor determination, particularly at time-like momentum transfers, requires consideration of the interplay between VMD and the constituent quark model, which in turn necessitates beyond leading-order approximations and incorporating higher-order QCD corrections \cite{Jaus:1999zv, Brodsky:1997de, Brodsky:2004tq, Zhang:1994ti, Wilson:1994fk}. The CLF approach primarily focuses on establishing a consistent framework for form factor calculations and ensuring Lorentz covariance, integrating higher-order QCD corrections necessitates a comprehensive understanding of the interplay between zero-mode contributions, vertex function selection, and renormalization schemes \cite{Jaus:1999zv, Jaus:1996np}. It should also be noted that higher-order contributions in the strong coupling constant, higher-twist effects, and nonleading anomalous dimensions are expected to modify leading-twist QCD predictions. However, empirical evidence suggests that these scale-breaking effects are relatively small \cite{Brodsky:2004tq}. Nevertheless, higher-order QCD corrections within the CLF formalism remain challenging \cite{Zhang:1994ti}. Non-perturbative effects necessitate extending renormalization beyond perturbative methods and incorporating higher Fock states. The singular nature and susceptibility to infrared divergences inherent to the light-front gauge necessitate rigorous regularization and renormalization procedures. \cite{Brodsky:2018vyy}.
	\section{Momentum dependence}
	\label{sec:ffq2} 
	The study of semileptonic $B_c$ decays involving axial-vector meson transitions presents certain challenges, particularly regarding the $\q2$ behavior of form factors. Therefore, a comprehensive analysis of these form factors is necessary to improve the understanding of the underlying decay processes. It is straightforward to evaluate the transition matrix elements, \textit{e.g.}, Eqs.~\eqref{eq:AV_BSW}-\eqref{eq:AA_BSW}, knowing both the state vectors and the current operator. In order to numerically evaluate the transition form factors, it is crucial to understand the momentum transfer in the available $\q2$ region in the CLF QM. Conventionally, a conducive way is to consider the meson transition in the Drell-Yan-West frame with $q^+=0$, which restricts the evaluation of the form factors to space-like momentum transfer, where $\q2=-\mathbf{q}_\bot^2 \leqslant 0$ \cite{Jaus:1989au, Jaus:1996np, Jaus:1999zv, Jaus:2002sv}. It is well-established that weak transition form factors at space-like momentum transfer are most simply evaluated from matrix elements of the plus component of the current in the preferred Lorentz frame \cite{Jaus:2002sv}. Despite this, essentially time-like ($\q2=-\mathbf{q}_\bot^2 \geqslant 0$) form factors alone are relevant for physical decay processes \cite{Jaus:1999zv, Cheng:2003sm}. Several studies \cite{Jaus:1996np, Cheng:2003sm, Choi:2021mni, Chang:2019mmh, Wang:2008xt} propose reformulating the form factors as explicit functions of $\q2$, analytically extending them from space-like ($\q2 \leqslant 0$) to time-like regions ($\q2 \geqslant 0$). These explicit forms are based on the assumption that the form factors are continuously differentiable functions of $\q2$, and the knowledge of the form factors in the vicinity of $\q2=0$ is essential \cite{Jaus:1989au}. Therefore, understanding the behavior of the transition form factors, which are expressed as wave function overlap between the initial and final state mesons near $\q2 = 0$, is significant. Furthermore, it has been argued that form factors obtained directly in the time-like region (with $q^+>0$) are equivalent to those acquired through analytic continuation from the space-like region \cite{Bakker:2003up}. However, the transition form factors calculated in the $\q2 \geqslant 0$ (in the frame of $q_{\bot}=0$) have some theoretical uncertainties because of the nonvalence configuration. The nonvalence contributions from the \textit{Z}-graph represent the quark-antiquark pair creation from the vacuum. These contributions are hard to calculate in the CLF QM formalism because of the unknown non-wave function vertex in the nonvalence diagram; however, a little progress has been made in this direction so far \cite{Choi:2021mni, Bakker:2000rd, Bakker:2000pk}. Furthermore, a recent study by Heger \textit{et al.} \cite{Heger:2021gxt}, made attempts to estimate \textit{Z}-graph contributions by a vector meson dominance (VMD) like decay mechanism for time-like momentum transfers. The form factors are parameterized as meromorphic functions of $\q2$, which are analytically continued from $\q2 < 0$ to $\q2 > 0$. It has been argued that such an analytic continuation provides a reasonable account of the form factors at time-like momentum transfers. Interestingly, the consideration of frames with purely transverse momentum transfer, \textit{i.e.}, $q^+=0$, resolves this issue as the nonvalence contributions are expected to be diminish. In addition, as discussed before, these transition form factors will be affected by zero-mode contribution, which is resolved by the type-II self-consistent CLF approach.
	
	As already mentioned, the theoretical expressions discussed in the previous section, in the $q^+ = 0$ frame, are sufficient to calculate the form factors only in the space-like region, for $\q2 \leqslant 0$. A description of the form factors as explicit functions of squared momentum dependence ($\q2$ parameterization) is required to extrapolate the space-like transition form factors to the time-like region \cite{Jaus:1996np}. The two descriptions of form factors in space-like and time-like regions are complimentary to each other and thus provide insight into the complete decay dynamics of the transition process in the full $\q2$ range. Correspondingly, several $\q2$ dependence formulations have been proposed and used in literature \cite{Melikhov:2000yu, Cheng:2003sm, Melikhov:2001zv, Becirevic:1999kt, Bourrely:2008za, HFLAV:2022esi} to parameterize and reproduce the transition form factors in space-like region and then extrapolate to physical form factors for $\q2 \geqslant 0$. 
	
	In general, the parameterization used to describe the physical form factors, appearing in hadronic matrix elements, receive contributions from momentum transfer ($\q2$) and are expressed as a function of the same. The most conventional dependence of the form factor on $\q2$ is typically expressed as $F(\q2)=F(0)/(1-\frac{\q2}{M_{p}^2})$, where a pole structure arises from VMD \cite{Wirbel:1985ji}, and is commonly referred to as the BSW-type monopole approximation. Here, the pole mass, $M_{p}$, is the mass of the lowest-lying meson with the vector quantum numbers that will yield the most significant contribution to the form factor for the $q_1^{\p} \rightarrow q_1^{\p\p}$ transition. However, it has been observed that fit to the experimental data does not reproduce the expected vector meson masses \cite{ParticleDataGroup:2022pth, HFLAV:2022esi, Hill:2006ub}. Further, it has been argued that the monopole form alone cannot explain the $\q2$ behavior and higher resonance contributions to the form factor are expected \cite{Burdman:1992hd}. Furthermore, it was argued that the validity of the nearest pole dominance assumption depends on the specific system under consideration. In some cases, multiple resonances may contribute significantly to the form factor, and neglecting the contributions of higher-mass poles may lead to inaccuracies in the description of the physical process \cite{Richman:1995wm, Burdman:1996kr}. We wish to remark that it is difficult to anticipate the non-perturbative physics that governs the $\q2$ dependence; therefore, the most accurate way to express $\q2$ dependence is typically not evident. The models make reasonable assumptions only for certain ranges of $\q2$. Therefore, the assumptions should only be seen as reasonable approximations in the models. The most generic assumption to parameterize $\q2$ involves the use of a simple pole and the summation of effective poles \cite{HFLAV:2022esi}, which involves multiple undetermined expansion parameters to be determined from the experiment.
	
	It may be noted that compared to $P \to P$ transitions, which are simpler from a first-principles point of view, the study of the transitions to vector or orbitally excited states is more challenging because they involve multiple invariant form factors, such as in Eqs.~\eqref{eq:AV_BSW} and \eqref{eq:AA_BSW}. In addition, the form factors’ analytic structure is complicated due to the unstable nature of the excited mesons and may be influenced by sub-threshold contributions \cite{Hill:2005ju, Hill:2006ub, Melikhov:2001zv}. Furthermore, the inclusion of higher-order poles becomes especially relevant when the final meson is in a radially or orbitally excited state. For instance, the momentum dependence of $A(\q2)$ is governed by the mass of the nearest resonance with $J^P=1^+$ pole, while $V_0(\q2)$ and $V_i(\q2)$ (for $i=1,2$) include contributions from $J^P=0^+$ and $J^P=1^-$ pole, respectively \cite{Bauer:1988fx, Gubernari:2022hrq}.
	\begin{table}[!ht]
		\caption{Pole masses for the transition form factor parameterization.}
		\label{tab1}
		\setlength{\tabcolsep}{5pt}
		\begin{tabular}{|c|c|c|c|c|c|} \hline
			\multirow{2}{*}{$J^P$} & \multirow{2}{*}{Form factor} &
			\multicolumn{4}{c|}{\text{Pole masses (GeV)}}  \\
			\cline{3-6}	
			& & $b\bar{u}$ & $b\bar{c} $ & $c\bar{d}$ & $c\bar{s}$ \\				
			\hline	
			$1^{+}$ & $A(\q2)$ & 5.71 & 6.87 & 2.42 & 2.46 \\	
			\hline
			$0^{+}$ & $V_0(\q2)$ & 5.70 & 6.84 & 2.34 & 2.32 \\	
			\hline
			$1^{-}$ & $V_1(\q2), V_2(\q2)$ & 5.32 & 6.47 & 2.01 & 2.11 \\			
			\hline						
		\end{tabular}
	\end{table}
	In the present work, the available $\q2$ range for the bottom-conserving $B_c \to B_{(s)1}$ transitions is $0 \leqslant \q2 \lesssim  0.30$ GeV${}^2$; however, for bottom-changing $B_c \to D_{1}(B_c \to\chi_{c1})$ transitions, the $\q2$ range is significantly larger than that of the bottom-conserving modes, \textit{i.e.}, $0 \leqslant \q2 \lesssim  15$ GeV${}^2$ ($0 \leqslant \q2 \lesssim  8$ GeV${}^2$). Therefore, to establish a coherent theoretical framework for $B_c$ decays to axial-vector mesons, it is crucial to accurately ascertain the $\q2$ dependence of the decay amplitudes across the full range of kinematics. The squared momentum dependence of form factors throughout the space-like region can be reproduced reliably and extrapolated to the time-like region by using the VMD-like three-parameter form \cite{Melikhov:2000yu}:
	\begin{align}
		\label{eq:q2_ff}
		{F}(\q2)=\frac{{F}(0)}{\left(1-\q2/M_{p}^2\right)\left(1-a(\q2/M_{p}^2)+b(q^4/M_{p}^4)\right)}.
	\end{align}
	First, we numerically evaluate the $P \rightarrow A$ form factors (Eq.~\eqref{eq:q2_ff}) at five $\q2$ values within the $\q2$ range of  $-10\,{\rm GeV}^2\leq \q2\leq 0$ for charm decays and $-20\,{\rm GeV}^2\leq \q2\leq 0$ for bottom decays to extract the parameters $a$, $b$, and $F(0)$. We took five $\q2$ points as $-0.01, -0.1, -1.0, -5.0, ~\text{and } -10.0 ~ \textrm{GeV}^2 $ for $c$-quark decays, and $\q2 =  -0.01, -0.1, -5.0, -10.0,~\text{and } -20.0 ~ \textrm{GeV}^2 $ for $b$-quark decays. The corresponding values of the form factors obtained are tabulated in Table~\ref{ff_space} of Appendix~\ref{FF_space-like}. Subsequently, we performed a five-point fit using Eq.~\eqref{eq:q2_ff} to determine these parameter values for the specified ranges. We used pole masses given in Table~\ref{tab1} for the parameterization. Additionally, we fit the form factor $F(\q2)$ at $\q2=\q2_{max}$ for various transitions to obtain $F(\q2_{max})$. Thus, the $B_c$ to axial-vector meson transition form factors are first obtained in the space-like region $(\q2 \leqslant 0)$ and are then extrapolated to the time-like region using Eq.~\eqref{eq:q2_ff}.  It should be remarked that the terms involving the slope parameters \textit{a} and \textit{b} signify effective poles ($M_p$), which represent the deviation from the single resonance contribution in the $q_1^{\p} \rightarrow q_1^{\p\p}$ transition and capture the effects of higher-order resonances~\cite{HFLAV:2022esi}. Typically, the parameterization Eq.~\eqref{eq:q2_ff} is identified as a four-parameter fit, where $F(0)$, \textit{a}, \textit{b}, and $M_p$ are expected to be determined from the available experimental data. However, in the absence of experimental information, to ensure the reliability of our calculations and the proper selection of quark-model parameters, we set the pole mass $M_p$ to the mass of the nearest pole according to the values given in Table~\ref{tab1} and describe it as a three-parameter fit \cite{Melikhov:2000yu, Melikhov:2001zv}. The phenomenological accuracy and reliability of $\q2$ dependence, given by Eq.~\eqref{eq:q2_ff}, have been extensively discussed in Refs. \cite{Melikhov:2000yu, Melikhov:2001zv}. The use of above said parameterization is not surprising in bottom-changing decays due to a larger $\q2$ range, where the contributions from bottom and bottom-charmed resonances could be significant. Thus, in bottom-changing transitions with a substantial momentum transfer with $\q2_{max}\simeq15$ GeV${}^2$, the inclusion of additional higher-order contributions is necessary to accurately predict the dynamics of the physical decay processes. Therefore, the form factors spanning a high $q^2$ range cannot be described by considering only a few initial physical poles ~\cite{Becirevic:2014kaa}. Furthermore, the above-stated picture depicts the confining interaction between $b$ and $\bar{u}/ \bar{c}$ in $b\Bar{u}/ b\Bar{c}$ current resulting in $B/ B_c $ resonances. Consequently, the poles associated with the form factors are situated at $\q2 = M^2_{p}$ (as listed in Table \ref{tab1}), specifically at unphysical values of the time-like momentum transfer that are away from the $\q2_{max}$. Furthermore, the parameterization described in Eq.~\eqref{eq:q2_ff} can be applied to $B_c \to B_{1}/B_{s1}$ transitions, considering the fact that the production threshold for mesons, such as $D/D_s$ resonances being lightest, from the $c \to d/s$ current occurs at $\q2$ values where the poles are located far outside the physical region of $\q2 \lesssim  0.30$ GeV${}^2$. Consequently, the influence of the poles on the form factor can be effectively incorporated into the polynomial\footnote{ Note that the parameterization proposed in Eq.~\eqref{eq:q2_ff} has been used in literature \cite{Cheng:2003sm, Melikhov:2000yu} as a solution to the abnormal behavior of the form factors involving transitions with heavy spectator quarks. The abnormal behavior in the form factors corresponds to exceedingly large values of the form factors at $\q2=0$ and/or slope parameters, as reported in \cite{Cheng:2017pcq, Chang:2019obq}.}. This allows for the exploration of the entire range of physical momentum transfer, which is expected to significantly enhance the accuracy.
	
	We wish to remark that various other theoretical works have used the following $\q2$ structure \begin{align}
		\label{eq:q2_t1}
		{F}(\q2)=\frac{F(0)}{1-a(\q2/M_{B_c}^2)+b(q^4/M_{B_c}^4)},
	\end{align}
	with the parent pole mass \cite{Cheng:2003sm}. The parameterization given by Eq.~\eqref{eq:q2_t1} can be regarded as an approximation of Eq.~\eqref{eq:q2_ff} with modified slope parameters. The parameterization in Eq.~\eqref{eq:q2_t1} is also expected to be valid for the region that lies farther away from the physical decay region \cite{Melikhov:2001zv}. We will discuss the numerical results of the form factors and their $\q2$ behavior in later section.
	\section{Decay rate and helicity amplitudes}
	\label{sec:drha}
	The semileptonic decay amplitude, $\mathcal{M}\left(B_c \rightarrow A \ell \nu_\ell\right)$, can be expressed as \cite{Richman:1995wm}
	\begin{equation}
		\mathcal{M}\left(B_c \rightarrow A \ell \nu_\ell\right)=\frac{G_F}{\sqrt{2}}\left|V_{q_1^{\p} q_1^{\p\p}}\right| \mathrm{L}^\mu \mathrm{H}_\mu,
	\end{equation}
	where $V_{q_1^{\p} q_1^{\p\p}}$ is the CKM matrix element for $q_1^{\p} \rightarrow q_1^{\p\p}$ transition, namely, $c \rightarrow d/s$ and $b \rightarrow u/c$; $G_F$ is the Fermi coupling constant, $\ell$ denotes the lepton flavor, and $\mathrm{L}^\mu$ is the lepton current defined by  
	\begin{equation}
		\mathrm{L}^\mu=\bar{u}_{\ell}\gamma^\mu\left(\mathbb{1}-\gamma_5\right) v_\nu,
	\end{equation}
	where $u_{\ell}$, $v_{\nu}$ are Dirac spinors. $\mathrm{H}_\mu$ is the hadronic current given in terms of the matrix elements between $B_c$ and axial-vector meson states as 
	\begin{equation}
		\mathrm{H}_\mu=\la A\left|\gamma^\mu\left(\mathbb{1}-\gamma_5\right)\right| B_c\ra.
	\end{equation}
	The hadronic current, $\mathrm{H}_\mu$, can further be expressed in terms of the form factors as shown in Eqs.~\eqref{eq:AV_BSW} and \eqref{eq:AA_BSW}. The branching ratio of semileptonic $B_c$ decays involving axial-vector mesons can be written as\footnote{We define our expressions for $\ell^{+}$.}
	\begin{equation}
		\label{eq:tdr}
		\mathcal{B}(B_c\to
		A\ell\nu_\ell)=\frac{\tau_{B_c}}{\hbar}\int_{m_{\ell}^2}^{\q2_{max}}\left(\frac{d\Gamma(B_c\to
			A\ell\nu_\ell)}{d\q2}\right)\,d\q2,
	\end{equation}
	where $\tau_{B_c}$ denote $B_c$ meson lifetime. The differential decay rate for $B_c\to A\ell\nu_\ell$ is \cite{Hernandez:2006gt, Ivanov:2005fd}
	\begin{equation}
		\label{eq:ddr}
		\frac{d\Gamma(B_c\to
			A\ell\nu_\ell)}{d\q2}=\frac{G_F^2}{(2\pi)^3}
		|V_{q_1^{\p} q_1^{\p\p}}|^2\frac{\lambda^{1/2}\q2}{24M_{B_c}^3}\left(1-\frac{m_\ell^2}{\q2}\right)^2{\cal H}_{\rm Total},
	\end{equation}
	where $m_\ell$ is the lepton mass. The three-momentum distribution of the hadron in the $B_c$ rest frame is given by the \text{Källén function} $\lambda\equiv\lambda(M_{B_c}^2, M_A^2,\q2)=(M_{B_c}^2+M_A^2-\q2)^2-4M_{B_c}^2M_A^2$. The total helicity amplitude ${\cal H}_{\rm Total}$ is expressed in terms of helicity components ${\cal H}_i (i=U, L, P, S, SL$) as, 
	\begin{equation}
		\label{eq:htot}
		{\cal H}_{\rm Total}= ({\cal H}_U+{\cal
			H}_L)\left(1+\frac{m_\ell^2}{2\q2}\right) +\frac{3m_\ell^2}{2\q2}{\cal H}_S.
	\end{equation}
	Further, the unpolarized–transverse ($U$), longitudinal ($L$), parity–odd ($P$), scalar ($S$) and scalar– longitudinal interference ($SL$) helicity components can be expressed in terms of four helicity amplitudes $H_i$ ($i=+,-,0,t$) as \cite{Ivanov:2006ni}
	\begin{equation}
		\label{eq:hh}
		{\cal H}_U=|H_+|^2+|H_-|^2, \quad {\cal H}_L=|H_0|^2, \quad {\cal H}_P=|H_+|^2-|H_-|^2, \quad
		{\cal H}_S=|H_t|^2, \quad {\cal H}_{SL}=Re(H_0H_t^\dag).
	\end{equation}
	The helicity amplitudes are related to hadronic transition form factors as given below:
	\begin{eqnarray}
		\label{eq:haa}
		H_\pm(\q2)&=& -(M_{B_c}-M_A)V_1(\q2)\pm \frac{\lambda^{1/2}}{M_{B_c}-M_A}A(\q2),\cr
		H_0(\q2)&=&\frac1{2M_{A}\sqrt{\q2}}\left[-(M_{B_c}-M_A)
		(M_{B_c}^2-M_{A}^2-\q2)V_1(\q2)+\frac{\lambda}{M_{B_c}-M_A}V_2(\q2)\right], \cr
		H_t&=& -\frac{\lambda^{1/2}}{\sqrt{\q2}}V_0(\q2).
	\end{eqnarray}
	It should be noted that $\q2 = \q2_{max}$ indicates zero-recoil of the final meson in the rest frame of the $B_c$ meson, while $\q2 = 0$ corresponds to the maximum recoil of the final meson. In addition to the branching ratio, we also study multiple (experimentally important) physical observables for a comprehensive understanding of semileptonic $B_c$ decays as given below:
	\begin{itemize}
		\item[{1.}]{Forward-backward asymmetry $(A_{FB})$:} It is defined as the difference between the number of particles produced in the forward and backward directions. The expression for $A_{FB}(\q2)$ is given as \cite{Hernandez:2006gt, Ivanov:2005fd, Becirevic:2016hea} 
		\begin{equation}
			\label{eq:afb}
			A_{FB}(\q2)=\frac34\frac{{\cal
					H}_P+2\frac{m_\ell^2}{\q2}{\cal H}_{SL}}{{\cal H}_{\rm Total}}.
		\end{equation}
		\item[{2.}]{Lepton-side convexity parameter $(C_{F}^\ell)$:} The convexity parameter gives the angular decay distribution over $\cos\theta$ dependence and is defined as \cite{Ivanov:2019nqd, Zhang:2020dla}
		\begin{equation}
			\label{eq:clf}
			C^\ell_F(\q2)=\frac34\left(1-\frac{m_\ell^2}{\q2}\right)\frac{{\cal H}_U-2{\cal H}_L}{{\cal H}_{\rm Total}}. 
		\end{equation}
		\item[{3.}]{Polarization asymmetry $ (P_{L}^\ell) $:} The longitudinal lepton-polarization asymmetry corresponding to the charge of the final state lepton is given by
		\cite{Ivanov:2019nqd, Zhang:2020dla} 
		\begin{equation}
			\label{eq:ple}
			P_L^\ell(\q2)=\frac{({\cal H}_U+{\cal
					H}_L)\left(1-\frac{m_\ell^2}{2\q2}\right) -\frac{3m_\ell^2}{2\q2}{\cal H}_S}{{\cal H}_{\rm Total}}.
		\end{equation}
		\item[{4.}] {Longitudinal (transverse) polarization fraction $(F_{L(T)})$:} $F_{L}$ for the final axial-vector meson can be expressed as 
		\begin{equation}
			\label{eq:fl}
			F_L(\q2)=\frac{{\cal
					H}_L\left(1+\frac{m_\ell^2}{2\q2}\right) +\frac{3m_\ell^2}{2\q2}{\cal H}_S}{{\cal H}_{\rm Total}},
		\end{equation}
		such that, $F_T(\q2)=1-F_L(\q2)$ \cite{Ivanov:2019nqd}.
		
		\item[{5.}]{Asymmetry parameter $(\alpha^{*})$:} The transverse (longitudinal) composition of the final state mesons in the decay process are governed by asymmetry parameter \cite{Ivanov:2005fd}, \textit{i.e.},
		\begin{equation}
			\label{eq:alpha}
			\alpha^{*}(\q2)=\frac{{\cal H}_U+\bar{\cal H}_U-2({\cal H}_L+\bar{\cal H}_L+3\bar{\cal H}_S)}{{\cal H}_U+\bar{\cal H}_U+2({\cal H}_L+\bar{\cal H}_L+3\bar{\cal H}_S)},
		\end{equation}
		where $\bar{\cal H}_i=\frac{m^2_\ell}{2\q2}{\cal H}_i$. 
		
		\item[{6.}]{Lepton flavor universality ratio ($R_A$):} In the SM, the semileptonic decay modes involving different lepton flavors are expected to be approximately equal due to the lepton flavor universality, which treats all leptons equally. The LFU ratio, $R_A$, is defined by \cite{Wang:2022yyn},
		\begin{equation}
			\label{eq:lfu}
			R_A=\frac{\int_{m_{\ell^{\p}}^2}^{\q2_{max}}\left(\frac{d\Gamma(B_c\to
					A\ell^{\p}\nu_{\ell^{\p}})}{d\q2}\right)\,d\q2}{\int_{m_{\ell}^2}^{\q2_{max}}\left(\frac{d\Gamma(B_c\to
					A\ell\nu_\ell)}{d\q2}\right)\,d\q2},
		\end{equation}
		where $\ell^{\p}$ and $\ell$ represents heavy and light leptons, respectively. This ratio can be used as a test for the SM predictions.
	\end{itemize}
	We want to emphasize that $A_{FB}(\q2)$, $ C^{\ell}_F(\q2)$, $ P^{\ell}_L(\q2)$, $ F^{\ell}_L(\q2)$, and $\alpha^{*}(\q2)$ observables discussed above are independent. We present the expectation values of all the aforementioned physical observables by individually integrating the numerator and denominator over $\q2$ inclusive of kinematical factor $\lambda^{1/2}\q2(1-m_\ell^2/\q2)^2$.
	\section{Numerical results and discussion}
	\label{sec:NR&D}
	In this study, we investigated the $B_c \to A_A(A_B)$ transition form factors in the self-consistent CLF approach, as described in Sec.~\ref{methodology}. Consequently, we predicted the branching ratios of semileptonic weak decays of $B_c$ involving $A$ meson in the final state for bottom-conserving and bottom-changing decay modes. As described in Sec.~\ref{methodology} and Sec.~\ref{sec:ffq2}, we calculated the form factors using self-consistent (type-II) CLF QM in the $\q2<0$ region (in the frame $q^+=0$) and extrapolated them to the time-like region (with $\q2>0$) using the parameterization influenced by VMD as defined in Eq.~\eqref{eq:q2_ff}. In addition, it is crucial to know the values of the form factors in the vicinity of $\q2=0$. We studied the form factors over their entire $\q2$ range and performed a comprehensive analysis of their momentum dependence. The self-consistency proposed in the type-II CLF approach \cite{Chang:2019mmh} confirms that the full and valence contribution results are equivalent for $P\to V$ transitions because they are not affected by zero-mode contributions arising from $B^{(i)}_{j}$ functions. The issue of self-consistency arises not only in vector mesons but also in axial-vector mesons \cite{Chang:2018zjq, Chang:2019mmh}. Therefore, we conducted a thorough analysis of the self-consistency problem by examining the $B_c\to A_A(A_B)$ form factors, considering both bottom-conserving and bottom-changing transitions. We calculated the $B_c \to A_{A}(A_{B})$ transition form factors ($A(\q2)$, $V_{0}(\q2)$, $V_{1}(\q2)$, and $V_{2}(\q2)$) using the numerical inputs, \textit{i.e.}, constituent quark masses and $\beta$ parameters, derived from Gaussian-type LF wave functions \cite{Verma:2011yw, Chang:2019mmh} as listed below. We use 
	\begin{equation}
		\begin{gathered}
			\label{eq:mq_input}
			m_u=m_d=0.27\pm 0.04 ,\ m_s=0.48\pm 0.05 ,\\
			m_c=1.6\pm 0.2 , \text{ and } m_b=4.8\pm 0.2 \\
		\end{gathered}
	\end{equation}
	as quark masses (in $\mathrm{GeV}$), and the $\beta$'s (in $\mathrm{GeV}$) are taken from\footnote{Note that Ref.~\cite{Verma:2011yw} distinctly list $\beta$ parameters for orbitally excited \textit{p-} wave mesons. }~\cite{Verma:2011yw},
	\begin{equation}
		\begin{gathered}
			\label{eq:beta_input}
			\beta_{B_{c}}= 0.921\pm 0.092,\ \beta_{D_1}=0.389\pm 0.039, \\
			\beta_{\chi_{c1}}=0.420\pm 0.042,\ \beta_{B_1}=0.500\pm 0.050, \\
			\text{ and } \beta_{B_{s 1}}=0.550\pm 0.055,
		\end{gathered}
	\end{equation}
	except for pseudoscalar $B_c$ meson \cite{Verma:2011yw, Shi:2016gqt, Chang:2019mmh}. In practice, as outlined in Sec.~\ref{methodology}, the central value of the shape parameters $\beta$ for initial and final state mesons are typically determined by fitting to experimental meson decay constant data corresponding to the input quark masses \cite{Cheng:2003sm}. For pseudoscalar mesons, ample experimental data is available. Usually, the decay constants are determined from leptonic and radiative-leptonic decays of hadrons \cite{Wang:2015bka, Chen:2015csa, Chang:1999gn, Chang:1997re}. However, such experimental information does not exist for the $B_c$ meson. On the other hand, for the $B_c$ meson, there exists a wide range of theoretical estimates for the decay constants in the literature, spanning from $f_{B_c} \sim (371~-~ 489)$ MeV across various theoretical models \cite{Zhang:2023ypl, Chang:2018zjq, Narison:2019tym, Narison:2020guz, Shi:2016gqt, Colquhoun:2015oha}. It should be noted that earlier LQCD calculations predicted a larger $B_c$ meson decay constant, with $f_{B_c} = (489 \pm 4 \pm 3)$ MeV \cite{Chiu:2007km}, which has more recently improved to $f_{B_c}=(434 \pm 15)$ MeV \cite{Colquhoun:2015oha}. In addition, recent QCD spectral sum rules have suggested a smaller value of $f_{B_c} = (371\pm17)$ MeV \cite{Narison:2019tym, Narison:2020guz}. Given this wide domain of predictions, we have adopted $\beta_{B_c} = (0.9207 \pm 0.0921)$ GeV, where the central value\footnote{It is worth mentioning that we confirm that the $\beta$ value, i.e., $\beta_{B_c} = (0.921 \pm 0.0921)$~GeV correspond to $f_{B_c}= (420\pm 50)$ MeV with $10\%$ uncertainty, which in turn yields the extreme values of $f_{B_c}$ found in literature.} is in good agreement with the recent LQCD estimate for the decay constant ($f_{B_c} = (434 \pm 15)$ MeV). In addition, $\beta_{B_c}$ reasonably close to the latest results obtained in the self-consistent CLF QM approach \cite{Chang:2019mmh}. Furthermore, our work employs a wider range of theoretical uncertainties in $\beta$ parameters as compared to some recent studies \cite{Zhang:2023ypl, Chang:2018zjq}. Notably, the numerical values of the different transition form factors depend on the overlap of wave functions between the initial and final mesons. Additionally, the wave function is influenced by the quark masses and $\beta$ parameters, apart from ${\bf k}_{\bot}^{\p}$. As a result, the overlap factor of the meson wave functions is expected to be flavor-dependent due to the different quark masses and $\beta$ parameters of the initial and final states. Thus, variation in the valence quark masses of the involved mesons (and their charge conjugates) described by the quark model \cite{ParticleDataGroup:2022pth}, \textit{i.e.}, \[B_{c}^{+}= c\bar{b}, ~ B_{s1}^{0(\p)}= s\bar{b}, ~B_1^{0(\p)}= d\bar{b}, ~\chi_{c1}^{(\p)}= c\bar{c}, ~D_1^{0(\p)}= c\bar{u}, \] and the $\beta$ parameters are allowed to account for the effect of flavor dependence on form factors and branching ratios. The CKM matrix elements were sourced from Particle Data Group (PDG)~\cite{ParticleDataGroup:2022pth}. The axial-vector meson masses and their mixing angles used in the numerical calculation are given in Appendix~\ref{axial_mixing} and the lepton masses used are $m_e = 0.511$ MeV, $m_\mu = 105.66$ MeV, and $m_\tau = 1776.86$ MeV \cite{ParticleDataGroup:2022pth}. The uncertainties in the masses of mesons (leptons) have been disregarded owing to their significantly smaller magnitude compared with the uncertainties in both quark masses and shape parameters. An important aspect of our investigation is the evaluation of transition form factors using the self-consistent CLF QM and employing the $\q2$ dependence formulation Eq.~\eqref{eq:q2_ff}, which can uniformly describe the form factors for both bottom-conserving and bottom-changing transitions. The results for the type-II approach are presented in the respective columns of Tables \ref{tab:ff_bb} and \ref{tab:ff_bc}, which are comprehensively elucidated in the subsequent sections. To facilitate comparison, we computed the type-I form factors\footnote{Note that in the type-I scheme, we used the $B_c$ pole mass to comply with Eq.~\eqref{eq:q2_t1} as it is. In addition, we used Eq.~\eqref{eq:q2_ff} with pole masses defined in Table~\ref{tab1} for the type-I scheme (referred to as type-I$^*$) to estimate the exact magnitude of self-consistency effects, as shown in Tables \ref{tab:ff_bb} and \ref{tab:ff_bc}.} using the $\q2$ dependence formulation given by Eq.~\eqref{eq:q2_t1}. This allows us to contrast the results obtained from the type-I and type-II approaches within the CLF QM. A comprehensive comparison between these two types is presented in Tables \ref{tab:ff_bb} and \ref{tab:ff_bc}, providing valuable insights into the discrepancies associated with $\q2$ formulation and zero-mode contributions. Furthermore, it should be emphasized that the form factors listed in Tables \ref{tab:ff_bb} and \ref{tab:ff_bc} will also be affected by the mixing of $A_A$ and $A_B$ axial-vector states, which will be thoroughly examined in the forthcoming analysis of semileptonic decays. After evaluating the form factors, we extend our analysis to calculate the semileptonic branching ratios, where we have employed all the helicity amplitudes. In addition, we computed the expectation values of significant physical observables, namely $A_{FB}$, $C_{F}^\ell$, $P_{L}^\ell$, $F_{L}$, and $\alpha^{*}$. These observables provide a reasonable understanding of the decay processes and have the potential for experimental verification in future observations.
	\subsection{ Analysis of transition form factors}
	\label{ff}
	We studied the form factors of $B_c \rightarrow A_{A}(A_{B})$ in bottom-conserving CKM-enhanced $(\Delta b = 0, \Delta C =-1, \Delta S = -1)$ and suppressed $(\Delta b = 0, \Delta C =-1, \Delta S = 0)$ modes, as well as bottom-changing CKM-enhanced $(\Delta b = -1, \Delta C =-1, \Delta S = 0)$ and suppressed $(\Delta b = -1, \Delta C =0, \Delta S = 0)$ modes using the self-consistent type-II CLF approach. By incorporating the type-II correspondence, we computed the form factors numerically. As described in Sec.~\ref{sec:ffq2}, we calculate the space-like Type-II bottom-conserving and bottom-changing transition form factors at various $\q2_{\bot}$ values as listed in Table~\ref{ff_space} of Appendix~\ref{FF_space-like} and extrapolate to the physical $\q2 \geqslant 0$ region by fitting Eqs.~\eqref{eq:q2_ff}. Furthermore, we established the validation of the type-II scheme for $P \to A$ form factors through the self-consistency of $B_c \to A$ transition form factors in our approach. We found that the numerical values of $l(\q2)(V_1(\q2))$ and $c_{-}(\q2)(V_{0}(\q2))$ form factors calculated using Eqs.~\eqref{eq:lfull} and \eqref{eq:cmfull} for the $\lambda=0$ and $\lambda=\pm$ polarization states in type-II CLF QM are equal, which confirms their self-consistency. Moreover, as described in Appendix~\ref{cov_ml}, the covariance issue is resolved in a straightforward manner for $P \to A$ transition matrix element in type-II correspondence. Finally, the results by utilizing Eqs.~\eqref{eq:q2_ff} and~\eqref{eq:q2_t1} (for both type-II and type-I, respectively) are presented in Tables \ref{tab:ff_bb} and \ref{tab:ff_bc}. We list our observations below:
	
	\subsubsection{\textbf{Bottom-conserving transition form factors}}
	
	\begin{itemize}	
		\item [i.] The bottom-conserving weak decays of $B_c$ are governed by $c$-quark decays, for which the available phase space is much smaller than that of the $b$-quark decays. Therefore, the observed $\q2$ range for bottom-conserving $B_c\rightarrow B_{(s)1}$ transitions is limited to a narrow interval of $0 \leqslant \q2 \leqslant (M_{B_c}-M_{B_{(s)1}})^2\simeq 0.30$ GeV${}^2$. Consequently, these form factors are anticipated to demonstrate minimal variations within the permissible $\q2$ range. Furthermore, Table \ref{tab:ff_bb} provides the numerical values of the form factors $A(0)$, $V_0(0)$, $V_{1}(0)$, and $V_{2}(0)$ for $B_c \to A_A(A_B)$ transitions without considering any mixing between the $A_A$ and $A_B$ states. However, these transition form factors will exhibit mixing when they appear in the decay processes.
		
		\item [ii.] We plotted all the bottom-conserving $B_c \to B_{(s)1}$ transition form factors to observe their variation with respect to $\q2$ described by Eq.~\eqref{eq:q2_ff}, as shown in Figures \ref{fig:B1} and \ref{fig:Bs1}. As expected, we observe that these form factors show negligible variation with respect to $\q2 $. In addition, to examine the effect of mixing on the magnitude and $\q2$ dependence of the transition form factors involving $A_A$ and $A_B$ states, we have also plotted the mixed transition form factors $B_c \to A^{(\p)}$. It is worth mentioning that the magnitude of the form factor $V_1(\q2)$ is larger for both mixed and unmixed transitions compared to other form factors.
		
		\item [iii.] In general, one of the main ingredients in the LF approach is the relativistic hadron wave functions, which generalize distribution amplitudes through transverse momenta. It should be noted that the wave functions, mainly, depend on internal degrees of freedom, such as transverse momentum distributions and constituent quark masses. In addition to this, meson-to-meson transitions have contributions originating from vertex functions and current operators represented as the full overlap integrand in CLF QM (Eq.~\eqref{eq:Fq2}). Hence, this overlap integrand contains all the information of $B_c \to A_A{(A_B)}$ transitions. Since both initial and final state mesons in $B_c \to B_{(s)1}$ transitions are heavy, maximum overlap should occur at the zero-recoil point ($\q2=\q2_{max}$). The magnitude of the $B_c \to A_A{(A_B)}$ form factors is significantly influenced by the overlap between the initial and final state meson wave functions. Hence, it becomes imperative to first analyze the meson wave function overlap\footnote{The normalization of our Gaussian-type radial wave function of meson can be described as $\int _0^1 {dx}\int {d^2 \mathbf{k}^{\p}_\bot\over 2(2\pi)^3}\,|\psi(x,\kb^{\p})|^2=1$ \cite{Choi:2021mni, Choi:2009ai}.} excluding vertex functions and other factors. To assess the overlap between the initial and final wave functions (in Eqs.~\ref{eq:RWFs} and~\ref{eq:RWFp}, respectively), we numerically integrate out the momentum $\kb^{\p2}$ at $\q2 = 0$. The resulting overlap functions are visualized in Figure~\ref{fig:bb_WOL}, and the respective numerical overlap factors are also listed. It is obvious that the difference in the constituent quark masses is large for $\psi_{B_{(s)1}}(x)$ that peaks narrowly about $x \sim (\frac{1}{2}-\frac{m_2^2-m_1^{\p(\p\p)2}}{2 M^{\p(\p\p)2}}) \sim 0$, as per Eq.~\eqref{eq:int_kz}. However, the difference in the constituent quark masses is relatively smaller for $B_c$, therefore, the peak for $\psi_{B_c}(x)$ is located at $x \sim 0.25$ \textit{i.e.}, away from the origin~\cite{Hwang:2010hw, Cheung:1996qt, Choi:2009ai}. It should be noted that the asymmetry in the constituent quark masses determines the width of the corresponding peak. In addition, the overlap integrand that results in specific values for the transition form factors includes the combined effect of the vertex functions and other factors. Therefore, to facilitate a more comprehensive analysis, we additionally plot the full integrand defined by Eq.~\eqref{eq:Fq2} with respect to $x$ for all $B_c \to A_A$ transitions at $\q2 = 0$, which incorporates the mass factors from Eq.~\eqref{ff_CLF}, as illustrated in Figure~\ref{fig:Ap_Fx_vs_x}. As expected, the full integrand of transition form factors, such as $B_c \to A_A$, exhibits the peaked structure that corresponds to the spatial overlap region governed by the initial and final state wave functions. We observe that the overlap factors within the integrand significantly influence the overall magnitude of the resulting form factors. The plots (in Figure~\ref{fig:Ap_Fx_vs_x}) reveals that the integrands, denoted by $V_{i}(x)$ (where $i=0,1,2$), depict both positive and negative values as $x$ varies, except for $A(x)$. This suggests that the magnitudes of the transition form factors for $V_{i}(x)$ arise from the cancellation (destructive interference) between the positive and negative contributions enclosed by the area under the curves. In contrast, the form factor for $A(x)$ likely results from the constructive interference of the corresponding wave functions. To obtain the overall magnitude of the full integrand, these areas under the curves must be added with their respective signs. It is worth noting that among the different bottom-conserving ${B_c\rightarrow B_{(s)1A}}$ transition form factors, the amplitude of the $V_{1}(x)$ integrand is largest (see Figure~\ref{fig:V1x}), indicating the large numerical values for the $V_{1}$ transition form factors, as listed in Table \ref{tab:ff_bb}. Similar conclusions can be made for the remaining integrands of the transition form factors. Note that the behavior of the ${B_c\rightarrow B_{(s)1B}}$ integrands (see Figure~\ref{fig:An_Fx_vs_x}) can be explained in an analogous way and consequently, have been excluded from further analysis. Thus, the full integrand plots help to understand the actual behavior of form factors at $\q2 = 0$. Additionally, it is anticipated that the magnitude of the full integrand will rise as a function of $\q2$ to reach a maximum at $\q2_{max}$ in the physical region. However, due to the small range of available $\q2$, the overlap integrand is anticipated to be roughly equivalent to its value at $\q2 = 0$ when evaluated at $\q2 = \q2_{max}$.
		
		\item [iv.] We wish to emphasize that due to the imposition of the $q^+ = 0$ frame, the form factors are obtained only for space-like momentum transfer $\q2 \leqslant 0$, as explained in Sec.~\ref{sec:ffq2}. However, we need to know the form factors in time-like region ($\q2 \geqslant 0$) to understand the physical decay process\footnote{For time-like region ($\q2 > 0$), the impact of the Z-graph contributions on form factors becomes more significant~\cite{Heger:2021gxt}. These Z-graph contributions can be approximated by a VMD-like decay process, where the emitted quark-antiquark pair annihilates into a W boson. This process is influenced by intermediate vector mesons. The Z-graph contributions becomes more important than the valence contributions when the intermediate resonance lies closer to the physical region. Unlike scattering, where we can choose a reference frame to minimize the Z-graph's effect, decay processes don't offer this flexibility. The total system's invariant mass is always fixed by the decaying particle's rest mass \cite{Heger:2021gxt}.}. Moreover, evaluation of the form factors in the physical region ($0 \leqslant \q2 \leqslant \q2_{max}$) is dominated by higher resonant structures \cite{Jaus:1996np}, which describe the long-distance effects, are nontrivial for being non-perturbative and therefore, lead to lack of proper description for the hadronic form factors. Nevertheless, the effective change in the $B_c \to B_{(s)1}$ form factors in the region $0 \leqslant \q2 \leqslant(M_{B_c}-M_{B_{(s)1}})^2$ can be explained through the confining interaction between $c$ and $\bar{d}$ (or between $c$ and $\bar{s}$) to produce $D/D_s$ meson resonances that fluctuate into $W$-boson. Therefore, we parameterize the form factors in the space-like region in terms of explicit functions of $\q2$, \textit{i.e.}, Eq.~\eqref{eq:q2_ff}, and extrapolate them in the $\q2 > 0$ region by properly locating a series of meson poles (throughout the kinematic range). We stress that although the two approaches provide independent descriptions of form factors in space-like and time-like regions, they complement each other \cite{Jaus:1996np}. It is worth noting that, as expected, the behavior and numerical values of form factors for bottom-conserving transitions show a strong correlation with the space-like region at low $\q2$. This is evident from a comparison of Tables \ref{ff_space} and \ref{tab:ff_bb} where values for $\q2_{\perp} \approx 0$ closely approximate $F(0)$, with a few exceptions. We also observe a decreasing trend in form factor values is observed as $\q2_{\perp}$ increases toward the endpoint ($\q2_{\perp} = 10.0$ GeV$^2$). Furthermore, We observe lower numerical values for the $A^{B_c \to B_{(s)1B}}(0)$ transition form factors, which can be attributed to the decreasing behavior of the form factor as the form factor passes from space-like region to physical region. Thus, a thorough understanding of the form factors' behavior near $\q2 = 0$ (as exemplified by the full integrand plots in Figure~\ref{fig:Ap_Fx_vs_x}) is critical to elucidate their overall $\q2$ dependence, particularly for $\q2 > 0$. At $\q2 = 0$, the full overlap integrand between initial and final states can be understood by considering the significant mass difference between the $c$ and $b$-quarks. In $B_c \to B_{(s)1}$ transitions, where $M_0^\p \sim m_b$, the energy released to the final state is very small ($E_1^\p \sim m_c$) as compared to the mass of the $b$-quark. Consequently, the $b$-quark remains largely unaltered during the transition. For $\q2>0$, we use resonances $D_{(s)}^{**}$ as pole masses with definite spin-parity quantum numbers (see Table~\ref{tab1}), for analyzing $\q2$ behavior throughout the available range. Since the $\q2$ range is small, we took the nearest pole contributions from the meson production threshold in $\q2$, which is expected to be important in the vicinity of $\q2_{max}$. However, it is worth remarking that the pole at $M_{D_{(s)}^{**}}^2$ lies far from the maximum $\q2$, in fact, $\q2_{max}$ is only $\sim 3\% ~\text{to}~ 7 \%$ of the $M_{D_{(s)}^{**}}^2$. Therefore, we expect the fit for Eq.~\eqref{eq:q2_ff} to be insensitive to the value of the squared resonance mass at the nearest poles, where $q^4/M_{p}^4\approx \mathcal{O}(10^{-4})$. Thus, the $\q2$ behavior can be reliably estimated by a simple VMD-type pole. Nevertheless, for accurate determination of the numerical values of the form factors, the three-parameter fit described by Eq.~\eqref{eq:q2_ff} is necessary\footnote{Semileptonic decays exhibit a distinctive characteristic, where resonances not only manifest directly within the kinematic region of meson decay but also extend beyond the available $\q2$ region \cite{Melikhov:1996du}.}. Hence, the form factors do not receive substantial contributions from the far away poles and, in fact, these contributions do not rise near the $\q2_{max}$ due to the remarkably small $\q2$ available. Therefore, the form factors appear flat in this region, as shown in Figures~\ref{fig:B1} and \ref{fig:Bs1}.  
		
		\item [v.] Furthermore, we notice that the numerical values of all the bottom-conserving transition form factors are positive, except for $V_2^{B_c B_{(s)1B}}(0)$. Interestingly, the magnitude of the numerical values of the slope parameter \textit{a} is less than unity and positive for most of the form factors, with a few exceptions. The sign and magnitude of the slope parameter\footnote{Note that the mixing among the slope parameters corresponding to $A_A$-type and $A_B$-type axial-vector mesons have been taken into account.} signifies how sharply the form factor varies with respect to allowed $\q2$. On the other hand, the slope parameter \textit{b} takes positive values for all form factors, with $b > a$ consistently. Moreover, $b$ slope parameters are greater than unity for many of (roughly half) the transitions, which indicates that the form factors do not receive significant contributions from higher-order resonances (note that $q^4/M_{p}^4\approx \mathcal{O}(10^{-4})$). Therefore, higher-order resonances will have a negligible impact on these form factors. Thus, the form factors for bottom-conserving transitions at $\q2=0$ differ marginally compared to those observed at their maximum $\q2$, \textit{i.e.}, $\q2_{max}$, therefore, leading to a flat behavior in the available $\q2$. Furthermore, the slope parameters for $A^{B_c B_{(s)1B}}(\q2)$ are exceptionally large because the form factor takes negligible values very close to zero, due to their decreasing trend towards the $\q2_{max}$. One may suspect the influence of the spectator quark mass on $B_c\rightarrow B_{(s)1}$ transition form factors ($F(\q2)$), however, lattice QCD (LQCD) calculations have recently shown that the bottom-conserving form factors do not depend upon the spectator quark mass \cite{Cooper:2020wnj}. In addition, we evaluate these form factors by considering uncertainties in the quark masses and $\beta$ parameters given by~\eqref{eq:mq_input} and~\eqref{eq:beta_input}, respectively. Form factors evaluated at $\q2 = 0$ show relatively low sensitivity to variations in quark masses and $\beta$ parameters, whereas their corresponding slope parameters, $a$ and $b$, are notably influenced by the chosen $\beta$ range. Overall, uncertainties in form factors are generally modest with few exceptions. Interestingly, unlike quark masses, the $\beta$ parameter uncertainties significantly impact wave functions; however, their influence on form factors is understood from the full integrand (Eq.~\eqref{eq:Fq2}). This integral is primarily determined by initial and final state vertex functions (Eq.~\eqref{eq:vertex}) and the function $\widetilde{F}(x,{\bf k}_{\bot}^{\p},\q2)$. Consequently, despite imposing a standard $10\%$ uncertainty on the input parameter $\beta$, the induced uncertainty in the $V_1^{B_c B_{(s)1A}}$ form factor is significantly small. This reduction can be attributed to the combined effects of vertex functions, $\widetilde{F}(x,{\bf k}_{\bot}^{\p},\q2)$, and $\q2$-fitting, which effectively attenuate the $\beta$ uncertainty. 
		\item [vi.] We also investigate the form factors in the type-I scheme \cite{Cheng:2003sm} for the most commonly used $\q2$ dependence given by Eq.~\eqref{eq:q2_t1}, to simultaneously emphasize the effect of type-II correspondence and the sensitivity of the form factors with respect to the choice of $\q2$ dependence formulation\footnote{We notice that a simple monopole or dipole approximation shows abnormal behavior for more than one form factors in the bottom-conserving mode. However, the use of Eq.~\eqref{eq:q2_ff} resolves all such issues. We also observed that the $z$-series parameterization \cite{ HFLAV:2022esi} in $P \to V$ form factor analysis shows $\q2$ behavior similar to Eq.~\eqref{eq:q2_ff}, as shown in Ref.~\cite{S:2024adt}.}. The numerical results for type-I CLF QM are presented in columns $6-9$ of Table \ref{tab:ff_bb}. We found that the numerical values of all the form factors for the type-I scheme are less than one, except for $V_1(0)$. The form factors $V_1(0)$ have large numerical values for all transitions in both type-I and type-II schemes using Eq.~\eqref{eq:q2_t1} and Eq.~\eqref{eq:q2_ff}, respectively. Although the numerical values of $V_1(0)$ between type-I and type-II differ roughly by $(20 - 25)\%$, the slope parameters are substantially different, with the former taking large values for both $a$ and $b$. Specifically, the slope parameter $b$ takes significantly large values ranging roughly from $50 - 300$ for $V_1(\q2)$ in type-I. Similar observations can be made for all the remaining form factors, where the slope parameters $a$ and $b$ are typically large for the type-I scheme. In addition, $V_0(0)$ form factors show a significant change in the magnitude of their numerical values (along with sign flip for $V_0^{B_c B_{1A}}(0)$) ranging from $(25-80)\%$ as compared to the type-II scheme. Also, we notice that the form factor $A^{B_c B_{(s)1B}}(\q2)$, which is not affected by the self-consistency problem, is approaching zero numerically in both schemes.
		
		\item [vii.] We wish to point out that the numerical values of the form factors in type-I correspondence suffer from a two-fold problem when compared to type-II correspondence. First, it has large values of slope parameters for most of the form factors; second, the form factors $V_1(0)$ and $V_0(0)$ vary significantly in magnitude and one of the form factor flips sign. The variation in magnitude of form factors ranges roughly from $\sim (20-80)\%$. One may expect that consideration of appropriate pole masses may resolve the issues within the type-I scheme; however, the problem persists. Thus, for a one-to-one comparison between type-I and type-II schemes, we use Eq.~\eqref{eq:q2_ff} in the type-I scheme with transition pole masses (which we refer to as type-I*). We observed that the inclusion of transition pole masses leads to relatively small values of the slope parameters, yet large as compared to the type-II scheme. Also, the numerical values of the $V_1(0)$ and $V_0(0)$ transition form factors are increased, but the difference as compared to type-II numerical values remains large in magnitude, \textit{i.e}, ranging from $\sim (15-70)\%$ and stays negative for type-I*. Most recently, Li\textit{ et al.} \cite{Li:2023wgq} have obtained these transition form factors using different $\q2$ parameterization, which suffer from a similar issue where the form factors at $ \q2=0$ are small while the corresponding parameters take exceedingly large numerical values\footnote{Note that the exceedingly large values of the $b$ parameters indicate that the form factors in $B_c \to B_{(s)1}$ follow a simple VMD pole like behavior irrespective of the different $\q2$ parameterizations \cite{Melikhov:2000yu, Cheng:2003sm, Melikhov:2001zv, Becirevic:1999kt, Bourrely:2008za, HFLAV:2022esi}, because of the smaller available $\q2$. However, the exclusion of zero-mode contributions in the $B_c \to B_{(s)1}$ transitions could be erroneous.}. Thus, we conclude that it is very difficult to accurately predict the behavior of the form factors within type-I correspondence. The same can be observed from the numerical results of CLF QM type-I analysis \cite{Shi:2016gqt}. Furthermore, we found that such dubious behavior can be attributed to zero-mode contributions in the type-I scheme, which affects the form factors $V_0(\q2)$ and $V_1(\q2)$. We also observe that the self-consistency effects are large for bottom-conserving transitions. Therefore, we should emphasize that efficient resolution for the above-stated problems can be reached by employing type-II correspondence, as the zero-mode contributions appearing through $B^{(i)}_{j}$ functions numerically vanish when $M \to M_0$ \cite{Choi:2013mda}.
		
		\item [viii.] Dynamically, the contributions to the branching ratios of semileptonic decays mainly come from the terms containing form factors $V_1(\q2)$, $V_2(\q2)$, and $A(\q2)$, while the impact of the term with form factor $V_0(\q2)$ is negligible. It is important to note that in vector meson emitting decays of the {$B_c$} meson, the contribution of the $A_2(\q2)$ form factor is usually disregarded because of its small magnitude resulting from division by $(M_{B_c}+M_V)$ in the amplitude. However, such contributions cannot be overlooked in $B_c \to A$ semileptonic decays, where it is divided by $(M_{B_c}-M_A)$ as shown in Eq.~\eqref{eq:haa}. In general, substantial changes in numerical values for $V_1(\q2)$ form factors cause corresponding changes in the amplitudes, and therefore, large branching ratios involving these transition form factors. Therefore, for the type-I scheme, the branching ratio of $B_c \to A(A^\p)$ semileptonic decays are expected to be overestimated (underestimated) for $B_c \to B_{1}\ell \nu_{\ell}$ and $B_c \to B_{s1}\ell \nu_{\ell}$ decays. We want to emphasize that numerically the magnitude of $V_1(\q2)$ form factors for $B_c \to B_{(s)1}$ transition is predominantly large and thus overshadows significant variation in other form factors. As a result, the effect of subtle variation in $V_0(\q2)$ form factor, which has a negligible impact on semileptonic decays, cannot be seen efficiently. However, $V_0(\q2)$ form factor can be substantially relevant in nonleptonic decays. Therefore, type-II correspondence is expected to affect the nonleptonic decays even more significantly.
	\end{itemize} 
	
	\subsubsection{\textbf{Bottom-changing transition form factors}}
	
	\begin{itemize}
		\item[i.] For bottom-changing transitions, particularly in $B_c\rightarrow D_1$ form factors, the $\q2$ range is expected to be significantly broader with $0 \leqslant \q2 \lesssim 15$ GeV${}^2$. In contrast to bottom-conserving transitions, this extensive range provides an opportunity to analyze how the dependence on $\q2$ affects the form factors. For the reasons stated in Sec. \ref{sec:ffq2}, we adopt $\q2$ parameterization provided in Eq.~\eqref{eq:q2_ff} for which the resulting numerical values of the form factors and slope parameters are listed in columns $2-5$ of Table \ref{tab:ff_bc}. We have plotted mixed and unmixed transition form factors involving axial-vector charm mesons, as shown in Figure \ref{fig:D1}. Throughout the accessible range of $\q2$, we observe a consistent pattern in the behavior of all the form factors, with slight deviations for some of the mixed form factors. All unmixed form factors have positive numerical values except for $V_2(0)$ for $B_c \to D_{1B}$. It may be noted that $A^{B_c D_{1B}}(0)$ form factor acquires a small numerical value, indicating that its magnitude is relatively small and approaches zero at $\q2_{max}$. The bottom-changing type-II $A^{B_c \to D_{1B}}$ transition form factor exhibits a consistent decline across both space-like and time-like regions, culminating in a minimal value at $\q2_{max}$. This behavior is similar to that of bottom-conserving transitions, as shown in Table~\ref{ff_space} in Appendix~\ref{FF_space-like}. Furthermore, It is interesting to note that, unlike bottom-conserving form factors, the bottom-changing form factor demonstrates a uniform pattern throughout the space-like $\q2$ domain. Further, these form factors show significantly different $\q2$ behavior after mixing. Since the contribution to decay width corresponding to $V_0(\q2)$ form factor is negligible, the variation of $V_1(\q2)$ form factor with $\q2$ has significant implications and is a matter of special interest in semileptonic decay analysis.
		
		\item [ii.] Similar to the bottom-conserving transitions, to enhance our understanding of the $B_c \to D_{1}$ transition form factors, we plot the wave function overlap between initial $\psi_{B_c}(x)$ and final $\psi_{D_{1}}(x)$ wave functions at $\q2 = 0$ and their full integrands, as shown in Figures~\ref{fig:D1_WOL} and \ref{fig:Ap_Fx_vs_x}, respectively. Figure~\ref{fig:D1_WOL} shows that $\psi_{D_{1}}(x)$ peaks near $x \sim 1/4$ with a larger width (corresponding to the mass difference of constituent quarks), whereas the peak for $\psi_{B_c}(x)$ lies at $x \sim 3/4$. The numerical overlap factor for $B_c \to D_{1}$ transition at $\q2=0$ is lowest in magnitude as compared to those of $B_c \to B_{(s)1}$ and $B_c \to \chi_{c1}$ transitions. Furthermore, we note that the full integrands of the bottom-changing transition form factors reveal substantially small amplitudes in comparison to their bottom-conserving counterparts (Figure~\ref{fig:Ap_Fx_vs_x}). It is intriguing to note that the full overlap integrand amplitude for $B_c \to D_{1}$ transition is comparatively reduced for $V_1(x)$. Therefore, we expect smaller numerical values of these form factors at $\q2=0$ as compared to $B_c \to B_{(s)1}/\chi_{c1}$ transitions. In addition, it can be noted that the available $\q2$ for $B_c$ to $D_1$ channel is significantly larger than the bottom-conserving decays, \textit{i.e.}, $0 \leqslant \q2 \lesssim 15$ GeV${}^2$. Hence, the energy released to the final state is very large ($E_1^\p \sim m_b$) and the initial $c$ quark in the bound-state is almost unaffected which further explains the small overlap for $B_c \to D_1$ transitions. As mentioned before, the $B_c$ to $D_1$ transitions are governed by $b \to u$ current, which conventionally involves $B^{**}$ poles. Moreover, the $\q2_{max}$ is roughly $50\%$ of the $M_{B^{**}}^2$, which is not far away from the $\q2_{max}$ as compared to the scenario in $B_c \to B_{(s)1}$ transitions. Therefore, we expect reasonable contributions from the resonance poles, which are expected to rise with increasing $\q2$ and will be maximum at zero-recoil as shown in Figure~\ref{fig:D1}. Therefore, the form factors will have larger numerical values at $\q2_{max}$ as can be seen from Table~\ref{tab:ff_bc}. Furthermore, the form factors $V_{1}^{B_c D_{1}}(\q2)$ and $V_{2}^{B_c D_{1}}(\q2)$ receive contributions from the vector $(J^P=1^-)$ $B^{**}$ pole, while $A^{B_c D_{1}}(\q2)$ and $V_{0}^{B_c D_{1}}(\q2)$ form factors receive contributions from the axial-vector $(1^+)$ and scalar $(0^+)$ $B^{**}$ poles, respectively (see Table~\ref{tab1}). Since the scalar and axial-vector poles lie farther, the form factors $A^{B_c D_{1}}(\q2)$ and $V_{0}^{B_c D_{1}}(\q2)$ are expected to vary less sharply, which can be also confirmed from their numerical values at $\q2=0$ and $\q2=\q2_{max}$, as given in the Table~\ref{tab:ff_bc}. 
		
		\item[iii.] One of the most peculiar aspects of bottom-changing transition form factors is that they have larger values of $a$ and $b$ parameters due to the smaller magnitude of form factors as compared to bottom-conserving ones. Both $a$ and $b$ acquire positive values greater than one, with few exceptions, and $a < b$ for all type-II transitions. A similar trend can be observed for type-I and type-I* results using $\q2$ dependence given by Eq.~\eqref{eq:q2_t1} and Eq.~\eqref{eq:q2_ff}, respectively, as shown in columns $6-9$ of Table \ref{tab:ff_bc}. It is worth mentioning that even though the central values of form factors $A(0)$ and $V_{2}(0)$ are similar between type-I (or type-I*) and type-II schemes, their respective slope parameters differ significantly with larger magnitudes observed exclusively for parameter $b$. This shows that the form factors with $\q2$ dependence given by Eq.~\eqref{eq:q2_t1} are more sharply varying. Moreover, the utilization of $\q2$ dependence expressed by Eq.~\eqref{eq:q2_ff} (with transition pole masses) in the type-I scheme effectively addresses this issue, as observed in bottom-conserving transitions. It is worth noting that the type-I scheme has a negligible effect on the central values of all form factors at $\q2 = 0$ for both Eq.~\eqref{eq:q2_ff} and Eq.~\eqref{eq:q2_t1} (between type-I* and type-I). However, the impact of the type-II scheme on the numerical values of $V_{0}(0)$ and $V_{1}(0)$ form factors at $\q2=0$ is such that their values change roughly by $(15-95)\%$ along with significant variation in the slope parameters. Another characteristic feature of zero-mode contributions as can be observed in the variation of affected form factors throughout the large $\q2$ range, \textit{i.e.}, they show considerable change in numerical values at $\q2_{max}$ for transition pole masses in type-I*, as can also be seen in the results of Ref.~\cite{Li:2023wgq}. It should be emphasized that, likewise bottom-conserving transition form factors, we observe significant numerical variation in magnitudes of the form factors $V_0(0)$ and $V_1(0)$, and their slope parameters between the two schemes. Therefore, the effect of self-consistency cannot simply be determined from the numerical values of affected form factors at $\q2 = 0$. Furthermore, the form factor $A^{B_c D_{1B}}(0)$ is negligibly small and therefore possesses exceptionally large slope parameters for both schemes, as shown in Table \ref{tab:ff_bc}. As a result, the mixed form factors $A^{B_c D_{1}^{(\p)}}(\q2)$ approach negligible values at large $\q2$ as shown in Figure \ref{fig:D1}. It is interesting to note that after mixing the form factors corresponding to $B_c \to D_1$ transitions become negative except for $V_0(\q2)$ form factor.
		
		\item[iv.] In the case of charmonium states, it may be noted that the transition form factors $B_c \to \chi_{c1}^{(\p)}$ are free from mixing effects\footnote{As mentioned in the Appendix~\ref{axial_mixing}, we are denoting $h_c$ as $\chi_{c1}^{\p}$.}. The $\q2$ dependence of these transition form factors is shown in Figure \ref{fig:chi}. Also, we plot the wave function overlap of $\psi_{B_c}(x)$ and $\psi_{\chi_{c1}}(x)$ at $\q2 = 0$ and full integrands, as shown in Figures~\ref{fig:Chic1_WOL} and \ref{fig:Ap_Fx_vs_x}, respectively. It is interesting to note that the wave function overlap between the initial and final state mesons for $B_c \to \chi_{c1}^{(\p)}$ is larger than that of $B_c $ and $D_{1}$, which can be described in a similar way to $B_c \to B_{(s)1}$ meson transitions. In the $B_c \to \chi_{c1}^{(\p)}$ transition, a peak is expected near $x \sim 1/2$ for $\psi_{\chi_{c1}}(x)$ with a relatively narrow width for symmetric quark structure. Furthermore, the $\psi_{\chi_{c1}}(x)$ peak lies closer to $\psi_{B_c}(x)$ (\textit{i.e.}, situated $ x \sim 3/4$), leading to a larger overlap between them compared to the overlap of $\psi_{B_c}(x)$ and $\psi_{D_{1}}(x)$. Similarly, analysis of the full integrand plot (Figure \ref{fig:Ap_Fx_vs_x}) reveals that the form factor for $B_c \to \chi_{c1}$ displays an intermediate amplitude, with the exception of $A(x)$ and $V_2(x)$ which exhibit prominent peaks attributed to the inclusion of mass factors given by Eq.~\eqref{ff_CLF}. These peaks are even larger than those observed for $B_c \to B_{(s)1}$ transitions indicating the significance of the mass factors (Eq.~\eqref{ff_CLF}). Consequently, this leads to larger numerical values of form factors for $B_c \to \chi_{c1}$ transitions at $\q2 = 0$, when compared to $B_c \to D_{1}$ and $B_c \to B_{(s)1}$ transitions. However, one exception can be observed for the $B_c \to \chi_{c1}^{\p}$ transition. Furthermore, in the physical region, the $\q2$ range in $B_c \to \chi_{c1}$ transitions is roughly half of $B_c \to D_{1}$ transitions with $\q2_{max} \sim 8$ GeV${}^2$. Also, $\q2_{max} \sim 16\%$ of $M_{B_c^{**}}^2$, where $B_c^{**}$ is the lightest pole associated with $b\bar{c}$ current. This indicates that the $B_c^{**}$ poles lies far away from the maximum available $\q2$. Therefore, we do not expect any sharp variation in $B_c \to \chi_{c1}$ transition form factors in the entire $\q2$ region, as shown in Figure~\ref{fig:chi}. Although most of the form factors for $B_c \to \chi_{c1}^{(\p)}$ transitions are nearly flat, the form factors $A^{B_c \chi_{c1}}(\q2)$, $V_{0}^{B_c \chi_{c1}^\p}(\q2)$, and $V_{1}^{B_c \chi_{c1}^\p}(\q2)$, gradually increase with respect to $\q2$. Further, we also observe that the form factors $V_{1}^{B_c \chi_{c1}}(\q2)$ and $V_{2}^{B_c \chi_{c1}^\p}(\q2)$ decrease with increasing $\q2$. In addition, we observe that $B_c \to \chi_{c1}^{(\p)}$ form factors are least affected by the problem of self-consistency as compared to other transitions, as can be observed from Table~\ref{tab:ff_bc}.
		
		\item[v.] In general, a comparison of the numerical values of the form factors between type-I and type-II correspondence reveals that the effect of self-consistency and covariance leads to significant changes in the numerical values of $V_{0}(\q2)$ and $V_{1}(\q2)$. Such deviations between the two schemes are expected to be decisive for the study of weak semileptonic decays. We also observe that the effects of self-consistency on bottom-conserving transition form factors are larger than those of bottom-changing transition form factors. As a result, the quest of self-consistency in the decays is of utmost importance to make reliable predictions, given the strong dependence on $V_1(\q2)$ form factor.
	\end{itemize}
	We have employed the type-II correspondence to axial-vector emitting transitions for both bottom-changing and bottom-conserving decays. Finally, we ensure that on the application of type-II correspondence to axial-vector emitting transition form factors for both bottom-changing and bottom-conserving $B_c \to A$ transitions are self-consistent, \textit{i.e.}, zero-mode contributions vanish numerically. We now proceed to calculate the branching ratios of semileptonic $B_c \to A^{(\p)}\ell \nu_{\ell} $ decays involving $B_c \to A^{(\p)}$ transition form factors.
	\subsection{Semileptonic decays}
	\label{semi_decay}
	In this subsection, we predict the branching ratios of the semileptonic weak decays of $B_c$ meson to the bottom and charm axial-vector mesons in the final state, as shown in Table \ref{semilep_Br}. To evaluate these branching ratios, we utilize the decay rate formulae given by Eqs.~\eqref{eq:tdr} and \eqref{eq:ddr} and the numerical values of the form factors as given in Tables \ref{tab:ff_bb} and \ref{tab:ff_bc}. In addition, we calculate the expectation values of the observables, such as the forward-backward asymmetry $\la A_{\rm FB} \ra$, lepton-side convexity parameter $\la C_F^\ell \ra$, polarization parameters ($\la P_L^\ell \ra,~ \la F_L \ra$), and asymmetry parameter $\la\alpha^{*}\ra$, which are of significant experimental importance. Predictions corresponding to these observables are given in the respective columns of Table \ref{semilep_obs}.
	\subsubsection{\textbf{Bottom-conserving decay modes }}
	The bottom-conserving CKM-enhanced $(\Delta b = 0, \Delta C =-1, \Delta S = -1)$ and CKM-suppressed $(\Delta b = 0, \Delta C =-1, \Delta S = 0)$ semileptonic decay modes of $B_c$ mesons, which involve the bottom mesons in the final states, undergo kinematic suppression due to their heavier masses. These semileptonic decay processes provide an excellent opportunity to observe the effects of form factors on the branching ratios and, therefore, to test the theoretical models. As mentioned earlier, the branching ratios are affected by the numerical values of the form factor as well as their signs. In addition, kinematic and CKM factors, and the mixing of final state axial-vector mesons play a significant role in determining their magnitude. We analyzed $B_c \to B_{(s) 1}^{(\p)}\ell\nu_{\ell}$ semileptonic decays using the self-consistent CLF approach. We list our major findings as follows:
	\begin{itemize}
		\item[i.] Despite the kinematic suppression of bottom-conserving decays, our predictions indicate that the branching ratios for these decay channels are of the order of $\mathcal{O}(10^{-4})$ to $\mathcal{O}(10^{-6})$. Among them, the CKM-enhanced semileptonic decays $B_c^+ \to B_{s1}^{0\p} e^+\nu_e$ and $B_c^+ \to B_{s1}^{0\p} \mu^+\nu_\mu$ are most dominant with branching ratios $({5.89}^{+0.54 +1.06}_{-0.55 -1.41}) \t 10^{-4}$ and $({4.43}^{+0.39 +0.81}_{-0.41 -1.06}) \t 10^{-4}$, respectively. In these CKM-enhanced decays, the dominant CKM factor ($V_{cs}$) overcomes the kinematic suppression and leads to significant branching ratios similar to those observed in $b \to u/c$ transitions. However, the branching ratios of the CKM-suppressed $B_c \to B_{1}^{(\p)}\ell \nu_{\ell}$ decays are smaller due to the lower value of the associated CKM factor ($V_{cd}$). Among these, $B_c^+ \to B_{1}^{0\p} e^+\nu_e$ and $B_c^+ \to B_{1}^{0\p}\mu^+\nu_\mu$, have the larger branching ratios given by $({8.69}^{+1.17 +2.06}_{-0.79 -2.78}) \t 10^{-5}$ and $({7.20}^{+0.97 +1.72}_{-0.79 -2.32}) \t 10^{-5}$, respectively.
		
		\item[ii.] In addition, we observe that the branching ratios of $B_c$ decays to primed axial-vector mesons are roughly larger by an order of magnitude than those of decays involving unprimed axial-vector mesons. This difference between their branching ratios can be attributed mainly to the combined effect of axial-vector meson mixing and the numerical values of the form factors. Therefore, experimental determination of the mixing angle can provide valuable insights into these decay channels. We wish to point out that we have also estimated the uncertainties in the branching ratios corresponding to errors in form factors. The uncertainties in the branching ratios corresponding to form factors because of quark masses and $\beta$ parameters are dealt with independently. It is interesting to note that despite the nearly symmetric uncertainties in the form factors, the uncertainties in the branching ratios are highly asymmetric, particularly for bottom-conserving decays. This behavior can be attributed to the mixing angle. Although the impact of uncertainties on branching ratios is mostly incremental (with few exceptions), we noticed that variations in quark masses result in changes opposite to those caused by $\beta$ uncertainties. However, for the sake of uniformity in the presentation, we have given the respective mass uncertainties similar to those of $\beta$. As previously mentioned, we anticipated significant uncertainties due to the inclusion of a wide range of values for the $\beta$ parameter, in addition to variation in quark masses. Interestingly, the branching ratios show greater sensitivity to variations due to uncertainties corresponding to the $\beta$ parameter, for primed-type bottom-conserving decays. However, uncertainties in the quark masses cause larger variations in the branching ratios of unprimed-type bottom-conserving decays. The uncertainties in $\beta$ (quark mass) lead to a maximum change in branching ratios roughly by $\sim 50\%$ ($\sim 76\%$). Taking into account both types of uncertainties, we notice that the maximum increase (roughly by a factor of $2$) in the branching ratios is observed for the decays $B_c^{+}\to B_{s1}^0 \ell^{+}\nu_{\ell}$. On the other hand, uncertainties for $B_c \to B_{1}^{(\p)}\ell\nu_{\ell}$ are relatively smaller, with a maximum increase (approximately by $80\%$) observed for $B_c \to B_{1}e\nu_{e}$ and $B_c \to B_{1}\mu\nu_{\mu}$. It is worth noticing that in the $B_c \to B_{s1}^{\p}\ell\nu_{\ell}$ decays uncertainties range from $\sim (30-40)\%$. This further highlights the sensitivity of branching ratios to form factors, slope parameters, and mixing. Note that uncertainties in the slope parameters play a significant role in overall errors in the branching ratios. Such an expanded range of uncertainties will provide a reasonable domain for experimental searches.
		
		\item [iii.] In order to assess the impact of self-consistency effects, we ignored the mixing between the final state axial-vector mesons for bottom-conserving decays, as shown in Appendix~\ref{Br_unmix}. We found that, within type-I scheme, the branching ratios of these decays for $\q2$ formulations described by Eq.~\eqref{eq:q2_ff} and Eq.~\eqref{eq:q2_t1} (\textit{i.e.}, type-I* and type-I, respectively) are of the same order, however, reflect a change in magnitude by $\sim (15-35)\%$. In addition, when compared to type-II results without mixing, their branching ratios change by $\sim (25-30)\%$ for Eq.~\eqref{eq:q2_ff}. These variations in the magnitude of branching ratios can also be observed in the presence of mixing (except for $B_c^+ \to B_{1}^0\mu^+\nu_{\mu}$ decay), with a maximum change ranging up to $\sim (40 - 55)\%$ for some of the decays, as shown in Table \ref{semilep_Br}. This indicates that similar to form factors, the self-consistency affects the branching ratios of bottom-conserving decays. It may be noted that in the type-I scheme the choice of mixing angle $\theta_{B_{1}} = -55^{\circ}$ shows an extraordinarily large branching ratio $ \sim \mathcal{O}(10^{-3})$ for $B_c^+ \to B_{1}^0\mu^+\nu_{\mu}$ decay, which is resolved in type-II correspondence. This drastic change is due to the abnormal behavior observed in $V_0(0)$ form factor after mixing within the type-I scheme for both Eq.~\eqref{eq:q2_ff} and Eq.~\eqref{eq:q2_t1} ($\q2$ formulations). Recall that we already noticed a sign flip in addition to the change in the magnitude of the $V_0(0)$ form factor (without mixing) in type-I correspondence as compared to type-II, which signifies the effect of zero-mode contributions in type-I. Hence, it is crucial to highlight the significance of the considerable variations in form factors (branching ratios) between type-I and type-II correspondences, along with the inclusion of mixing effects, which cannot be ignored.
		
		\item [iv.] The effect of the lepton mass on semileptonic decay to axial-vector mesons is also noteworthy. It should be asserted that the difference in mass between the electron and muon has a significant impact on the branching ratios as well as other parameters of these transitions. The branching ratios in these modes display an opposite trend compared to the lepton mass, as expected. In the decays $B_c^+ \to B_{1}^{0(\p)}/B_{s1}^{0(\p)} \ell^+ \nu_{\ell}$ (as shown in Table \ref{semilep_Br}), the branching ratios decrease approximately by $(17 - 28)\%$ with increasing lepton mass. Furthermore, the effect of lepton mass is more pronounced in $B_c \to B_{s1}^{(\p)} \ell\nu_{\ell}$ decays. Consequently, this observation highlights the importance of considering the lepton mass difference when studying these decays and provides valuable insights into the contribution of lepton masses to the branching ratios. Similar remarks can also be made for other observables.
		
		\item[v.] We also calculated the forward-backward asymmetry, $A_{FB}$, for bottom-conserving decays, as shown in column $2$ of Table \ref{semilep_obs}. As mentioned before, we have followed \cite{Hernandez:2006gt, Ivanov:2005fd, Becirevic:2016hea} for $A_{FB}$ analyses in \textit{p}-wave emitting semileptonic decays of $B_c$. We observe that $A_{FB}(B_c\to B_{1}^{(\p)}\ell\nu_{\ell})$ are negative, except for $B_c\to B_{1}^{\p}\mu\nu_{\mu}$ decay that has positive $A_{FB}$ with small magnitude. The negative values of $A_{FB}(B_c\to B_{1}e\nu_{e})$ reflect the dominance of the parity-violating helicity structure-function, $\mathcal{H}_P$, with a larger contribution from $H_-$ amplitude. The $\mathcal{H}_{SL}$ contributions are negligible for electrons in the final state. Further, the effect of lepton mass can be observed for $A_{FB}(B_c\to B_{1}\mu\nu_{\mu})$, which changes to a smaller negative value due to increased contribution, \textit{i.e.}, $\sim 35\%$, from scalar-longitudinal interference ($\mathcal{H}_{SL}$). On the other hand, $A_{FB}(B_c\to B_{1}^{\p}e\nu_{e})$ acquires a predominant contribution from $\mathcal{H}_P$ with larger magnitude for $H_-$ amplitude. However, the contribution of $H_+$ amplitude increases substantially, leading to smaller values of $A_{FB}(B_c\to B_{1}^{\p} e\nu_{e})$. Similarly, $A_{FB}(B_c\to B_{s1}e\nu_{e})$ receives a dominant contribution from $H_+$ amplitude. This dominance leads to a positive but reduced magnitude of $\mathcal{H}_P$ (along with negligible $\mathcal{H}_{SL}$), resulting in an overall smaller positive numerical value. Furthermore, the effect of lepton mass in $B_c\to B_{s1}\mu\nu_{\mu}$ decay is such that the contribution from $\mathcal{H}_{SL}$ increases to $70\%$ raising the overall numerical value of $A_{FB}$ to $0.27$. Likewise ${B_c\to B_{1}^{\p}e\nu_{e}}$ decay, $A_{FB}({B_c\to B_{s1}^{\p}e\nu_{e}})$ receives contribution only from $\mathcal{H}_P$ with larger magnitude for $H_-$ amplitude as compared to $H_+$. In addition, for $B_c\to B_{s1}^{\p}\mu\nu_{\mu}$ decay $\mathcal{H}_{SL}$ contributions, \textit{i.e.}, $\sim 60\%$, are larger than $\mathcal{H}_{P}$, leading to $A_{FB} = 0.04$. The uncertainties in $A_{FB}$ are also given corresponding to the errors in the form factors. We observe that the uncertainties in the $A_{FB}$ involving $B_c\to B_{1}^{(\p)}\ell\nu_{\ell}$ and $B_c\to B_{s1}^{\p}\ell\nu_{\ell}$ decays are roughly symmetric, whereas, $B_c\to B_{s1}\ell\nu_{\ell}$ decays are asymmetric. In addition, the summed over errors cause sign flip in $A_{FB}(B_c\to B_{1}\ell\nu_{\ell})$ and $A_{FB}(B_c\to B_{s1}\ell\nu_{\ell})$, while $A_{FB}(B_c\to B_{s1}^{\p}\mu\nu_{\mu})$ numerically becomes zero with combined errors.
		
		\item[vi.] The bottom-conserving semileptonic decay modes have a negative lepton-side convexity parameter, $C_{F}^\ell$. For $B_c \to B_{s1} e\nu_{e}$ and $B_c \to B_{s1} \mu\nu_{\mu}$, the magnitude of $C_{F}^\ell$ is large compared to their primed partners. It is worth noting that as the mass of the lepton increases, the numerical magnitude of the observable decreases. Note that we have ignored the error due to quark masses and $\beta$ in $C_{F}^\ell$ and other observables.
		
		\item[vii.] The longitudinal polarization fraction of the final state meson, $F_{L}^\ell$, for the bottom-conserving semileptonic modes ranges from $48\%$ to $76\%$, except for $B_c \to B_{1}^\p \ell\nu_{\ell}$. It decreases with increasing lepton mass for both unprimed and primed axial-vector mesons, except for $B_c \to B_{1} \ell\nu_{\ell}$ where it shows a marginal increase.
		
		\item[viii.] The asymmetry parameter, $\alpha^{*}$, in bottom-conserving mode is consistently negative for all decays, indicating the dominance of the longitudinal character of the daughter meson. This dominance increases for decays involving $B_c \to B_{1}^\p$ transitions, however, reduces for $B_c \to B_{s1}^\p$. In fact, for $B_c \to B_{1} \ell\nu_{\ell}$ transition the transverse character is marginally greater than the longitudinal. 
		
		\item[ix.] We have plotted the $\q2$ variation of the branching ratios, $A_{FB}$, $C^\ell_F$, $P^\ell_L$, $F_L$, and $\alpha^*$, respectively, as shown in Figures \ref{fig:obs_B1}, \ref{fig:obs_B1p}, \ref{fig:obs_Bs1}, and \ref{fig:obs_Bs1p}. The plots show a comparative variation of the said observables with respect to lepton mass. Note that for the semileptonic decay processes, these observables depend on the final lepton mass, where $\q2_{min} = m^2_{\ell}$ (neglecting neutrino masses); therefore, they are crucial indicators for investigating the impact of lepton masses. The differential branching ratio plots show different peaks (for available $\q2$) pertaining to $m_{\ell}$, and show the same endpoints at $\q2_{max}$, as expected. The kinematic phase space is more seriously affected in $B_{s1}^{(\p)}$ decays as compared to $B_{1}^{(\p)}$ decays as can be observed from the peaks of differential branching, as shown in Figure~\ref{fig:obs_B1} and \ref{fig:obs_Bs1}. The same can be ascertained for rest the observables. We observe that some observables, such as $A_{FB}$ ($P^\ell_L$) exhibit a distinct rise near $\q2_{min}$, particularly when the decay process involves an electron, \textit{e.g.}, see top (middle) right panel of Figure~\ref{fig:obs_B1} and Figure~\ref{fig:obs_B1p}. For instance, in the case of $A_{FB}$, this exceptional behavior can be explained by the contribution from ${\cal H}_{SL}$ as shown in Eq.~\eqref{eq:haa}, where both $H_{0}$ and $H_{t}$ receive significant but finite contribution near $\q2 = m_{e}^2$. The high precision computation indicates that $A_{FB}$ approaches zero numerically at $\q2=0$. It should be noted that the minute squared mass of an electron ($ \mathcal{O}(10^{-7})~\text{GeV}^2$) lies very close to the point of maximum recoil (\textit{i.e.}, at $\q2 = 0$). Therefore, we see sharp rise in numerical value of $A_{FB}$ for $B_c\to B_{1}^{(\p)}e\nu_{e}$ due to large contributions from ${\cal H}_{SL}$. Notice that such a sharp increase in $A_{FB}$ is not present for heavy leptons for $\q2$ being away from the point of maximum recoil, although ${\cal H}_{SL}$ contribution tend to dominate at smaller $\q2$. As expected, the $A_{FB}$ approaches zero at $\q2_{max}$ due to the reason that $\lambda$ vanishes at zero recoil limit. On the other hand, the $P^\ell_L$ show distinctive rise toward a negative value through zero near $\q2=m_{e}^2$ (and finally acquire numerical value equal to $-1.0$ at $\q2=0$), owing to the predominant contribution from ${\cal H}_{S}$ near $\q2=m_{e}^2$. Thus, we notice that ${\cal H}_{S}$ acquire dominant contributions for smaller $\q2$ as a result $P^\ell_L$ tends to deviate from unit value and becomes negative. It is interesting to note that $C^\ell_F(\q2)$ shows a distinctive negative but finite peak near zero due to dominant contributions from ${\cal H}_{L}$, which decreases and becomes zero precisely at $\q2=0.1237~\text{GeV}^2$ and thereafter, $C^\ell_F(\q2)$ becomes positive for $\q2>0.124~\text{GeV}^2$ for $B_c\to B_{1}e\nu_{e}$ decays. However, $C^\ell_F(\q2)$ stays negative for the rest of the bottom-conserving decays throughout the available $\q2$ indicating predominance of ${\cal H}_{L}$, as shown in the left middle panel of Figure~\ref{fig:obs_B1} and Figure~\ref{fig:obs_B1p}. These strange behaviors can also be observed in other works \cite{Sun:2023iis, Colangelo:2021dnv}. In addition, we observe that the longitudinal polarization fraction ($F_L$) decreases with increasing $\q2$ for all the bottom-conserving decays. However, $B_c \to B_1 \ell \nu_{\ell}$ shows an increase in longitudinal character near $\q2 \geqslant 0.23~\text{GeV}^2$. A similar behavior can be confirmed from variation of asymmetry parameter with $\q2$, where rise in transverse character near $\q2_{max}$ can be observed for all bottom conserving decays, except for $B_c \to B_1 \ell \nu_{\ell}$, as shown in Figures~\ref{fig:obs_B1} and \ref{fig:obs_B1p}.
	\end{itemize}
	
	\subsubsection{\textbf{Bottom-changing decays}}
	In this subsection, we focus on bottom-changing CKM-enhanced $(\Delta b = -1, \Delta C =-1, \Delta S = 0)$ and CKM-suppressed $(\Delta b = -1, \Delta C =0, \Delta S = 0)$ decay modes, \textit{i.e.}, $B_c$ meson decaying to the charm meson final state. One of the most interesting features of bottom-changing decays is that, apart from decays involving $e^+ \nu_e$ and $\mu^+ \nu_{\mu}$ lepton pairs, they include $\tau^+ \nu_{\tau}$ pair in the final state. Kinematically, these semileptonic decays are more favorable than bottom-conserving decays. A few of the bottom-changing semileptonic $B_c$ decays that are both kinematically and CKM-enhanced have the most dominant branching ratios. We have analyzed and listed our major findings on $B_c \to D_{1}/\chi_{c1}$ semileptonic transitions as follows:
	\begin{itemize}
		\item[i.] The order of branching ratios for these semileptonic decay channels ranges from $\mathcal{O}(10^{-3})$ to $\mathcal{O}(10^{-6})$, as shown in Table \ref{semilep_Br}. The predominant decays have the following branching ratios: Br$(B_c^+ \to \chi_{c1}^{\p} e^+\nu_{e}) = ({1.05}^{+0.01 +0.27}_{-0.05 -0.29}) \t 10^{-3}$ and Br$(B_c^+ \to \chi_{c1}^{\p} \mu^+\nu_{\mu})= ({1.04}^{+0.01 +0.27}_{-0.05 -0.29}) \t 10^{-3}$. As stated before, for the CKM-enhanced mode ($b \to c$), the kinematic factors contribute significantly to increase the branching ratios of $B_c^+\to\chi_{c1}^{(\p)}\ell^+\nu_{\ell}$ decays. Furthermore, the CKM-suppressed $B_c^+ \to D_{1}^{(\p)}\ell^+\nu_{\ell}$ decays, involving $b \to u$ transition, have branching ratios of the order $\mathcal{O}(10^{-5})$ to $\mathcal{O}(10^{-6})$, which are comparable to most of the bottom-conserving decays.
		
		\item[ii.] As observed in bottom-conserving decays, the branching ratios of transitions involving primed axial-vectors in the final state are roughly higher by an order of magnitude than those of decays with unprimed axial-vector mesons, except for $B_c^+\to\chi_{c1}^{(\p)}\tau^+\nu_{\tau}$ decays. The branching ratio of $B_c^+\to\chi_{c1}^{\p}\tau^+\nu_{\tau}$ decay is enhanced by $\sim 35\%$ as compared to $B_c^+\to\chi_{c1}\tau^+\nu_{\tau}$ decay. As observed before, these branching ratios decrease with increasing lepton mass, $B_c^+\to\chi_{c1}^{(\p)}\ell^+\nu_{\ell}$ decays are most influenced by the lepton mass effect. It may be noted that the decays involving the $\tau$ lepton have the lowest branching ratios among all the decays because of the significantly larger mass of the $\tau$ lepton. 
		
		\item[iii.] Likewise, for bottom-conserving decays, the branching ratios of bottom-changing decays have uncertainties as large as $\sim 155\%$ inclusive of both types of uncertainties. Moreover, uncertainties in these decays caused by variations in quark masses are less asymmetric compared with those of bottom-conserving decays. In contrast to bottom-conserving decays, where uncertainties mostly cause an increase in branching ratios, uncertainties in bottom-changing decays lead to noticeable changes through gradual increases or decreases in the branching ratios of these decays, except for $B_c^+ \to D_1\ell^+\nu_{\ell}$. Moreover, the uncertainties are less than $\sim 40\%$ and are nearly symmetric, for $B_c^+ \to \chi_{c1}^{(\p)}\ell^+\nu_{\ell}$ decays. 
		
		\item[iv.] As pointed out earlier, the self-consistency effects are expected to be larger in semileptonic decays involving $B_c\to D_{1}^{(\p)}$ transitions, wherein we observe that the branching ratios of unmixed states are differently affected by zero-mode contributions. In comparison to the type-II scheme, the branching ratios of unmixed $B_c^+\to D_{1}^{(\p)} \ell^+\nu_{\ell}$ decays experience changes ranging from $\sim ~23\%$ to $92\%$ for $\q2$ dependence using Eq.~\eqref{eq:q2_ff} and $57\%$ to $98\%$ with Eq.~\eqref{eq:q2_t1} in type-I CLF QM, where the $98\%$ variation is observed in the branching ratios for $B_c^+\to D_{1} \ell^+\nu_{\ell}$ decays. In addition, we find that these effects are equivalently prominent in the presence of mixing, where branching ratios of $B_c^+\to D_{1}^{(\p)} \ell^+\nu_{\ell}$ decays increase roughly by the same order. Therefore, such effects for $D_{1}^{(\p)}$ emitting decays of $B_c$ can be significant and cannot be ignored. Nevertheless, the self-consistency has a minimal effect on the branching ratios of semileptonic decays of $B_c$ to $\chi_{c1}^{(\p)}$ states, with variation of $(2 - 40)\%$ across both type-I* and type-I results, when compared to type-II results.
		
		\item[v.] The numerical values of $A_{FB}$ parameters for bottom-changing transitions are listed in column $2$ of Table~\ref{semilep_obs}. The $A_{FB}$ for bottom-changing decays are consistently negative in numerical values because of the predominant contribution from $\mathcal{H}_P$ with larger magnitude for $H_-$ helicity amplitude. The exception is evident in the decays that involve the pairing of a $\tau$ lepton and a primed axial-vector meson in the final state. In all bottom-changing $B_c\to D_{1}e\nu_{e}/\chi_{c1}e\nu_{e}$ decays, $\mathcal{H}_{SL}$ contributions are negligible due to the large $\q2$ value for electrons in the final state. Further, as the mass of the lepton increases, the $A_{FB}({B_c\to D_{1}\tau \nu_{\tau}})$ receives approximately $54\%$ contribution from $\mathcal{H}_{SL}$, however, this contribution increases (from being negligible) to $\sim 10\%$ for $B_c\to \chi_{c1}\tau \nu_{\tau}$. In addition, we observe a sign flip in $\mathcal{H}_{SL}$ with $\tau$ lepton mass for decays including $D_1$. The positive values of $A_{FB}$ for ${B_c\to D_{1}^{\p}\tau\nu_{\tau}}/\chi_{c1}^{\p}\tau\nu_{\tau}$ decays signify the dominance of $\mathcal{H}_{SL}$, with negligible contributions from $\mathcal{H}_P$. On comparison with \cite{Hernandez:2006gt, Ivanov:2005fd}, we found that our results of $A_{FB}$ for $B_c \to \chi_{c1}^{(\p)} \ell\nu_{\ell}$ are consistent with \cite{Hernandez:2006gt}, except for $B_c \to \chi_{c1}^{\p} e\nu_{e}$ decay, which is negligibly small in their case. However, our results exhibit opposite signs for unprimed types as compared to those of Ref.~\cite{Ivanov:2005fd}. Furthermore, the uncertainties corresponding to $\beta$ values show exceptional variation in $A_{FB}(B_c\to D_{1}\ell\nu_{\ell})$ making them more negative with large numerical values. Such exceptional variation in $A_{FB}$ highlights the importance of interference between $\mathcal{H}_P$ and $\mathcal{H}_{SL}$ contributions and mixing in the form factors. However, uncertainties in masses behave in a symmetrical manner with negligibly small variations. For $A_{FB}({B_c\to \chi_{c1}^{\p}\mu\nu_{\mu}})$, we observe a possible sign flip corresponding to large $\beta$ error.  
		
		\item[vi.] In contrast to bottom-conserving decays, the numerical values of $C_{F}^\ell$ in bottom-changing semileptonic decays acquire both positive and negative signs, \textit{i.e.}, primed-type decays have negative values with larger magnitudes compared to the unprimed-type. The lepton mass has a substantial effect on these numerical values, as can be seen for the decays involving $\tau$ lepton. Furthermore, the longitudinal polarization fraction ($F_{L}^\ell$) for unprimed axial-vector mesons lies between $18\%$ and $30\%$, on the other hand, the primed states fractions vary from $62\%$ to $89\%$.
		
		\item[vii.] Interestingly, the asymmetry parameter $\alpha^{*}$ for all bottom-changing transitions involving primed axial-vector mesons in the final state are dominant in longitudinal character, whereas unprimed states favor transverse asymmetry, as shown in Table \ref{semilep_obs}. The unprimed states for $B_c \to D_{1}\ell\nu_{\ell}$ decays are marginally transverse in nature, and this behavior reduces further for decays involving $\tau$ leptons. In addition to the lepton mass, the mixing effect plays a significant role in these decays. However, the asymmetry in $B_c \to \chi_{c1} \ell\nu_{\ell}$ decays is transverse with a negligible effect of the lepton mass.
		
		\item[viii.] It is noteworthy to mention that most of the observables for bottom-conserving as well as bottom-changing semileptonic decays show a dominant longitudinal behavior, with few exceptions, especially for heavier leptons. The plots for all the observables, including branching ratios, showing variation with $\q2$ and lepton mass, are shown in Figures \ref{fig:obs_D1}, \ref{fig:obs_D1p}, \ref{fig:obs_chic1}, and \ref{fig:obs_chic1p}. The plots for differential branching ratios involving lighter leptons ($e$ and $\mu$) in the case of bottom-changing decays do not differ much owing to large $\q2$ range available for these decays. However, for the decays involving $\tau$, which is very heavy compared to other two leptons, we observe substantially small peaks due to the smaller phase space available, specifically for decays to $\chi_{c1}^{(\p)}$ states. Therefore, for the sake of clarity, we have given the plots only for $B_c \to \chi_{c1}^{(\p)} \tau \nu_{\tau}$ separately in Figures \ref{fig:obs_chic1} and \ref{fig:obs_chic1p}. It may be noted that for similar reasons stated in the bottom-conserving decay discussion, we observe sharp numerical changes in $A_{FB}$ and $P^\ell_L$ near $\q2=m_{\ell}^2$ (see top and middle right panel of Figures \ref{fig:obs_D1}, \ref{fig:obs_D1p}, \ref{fig:obs_chic1}, and \ref{fig:obs_chic1p} for $A_{FB}$ and $P^\ell_L$, respectively). Furthermore, it is worth pointing out that the $C^\ell_F$ show similar distinct negative peaks for bottom-changing decays at low $\q2$. However, turns positive for the decays involving unprimed final states (except for $\tau$ lepton) before falling to zero at endpoint, as shown in the left middle (bottom) panel of Figure \ref{fig:obs_D1} (\ref{fig:obs_chic1}) for $B_c \to D_{1} (B_c \to \chi_{c1})$ decays involving lighter leptons. In addition, for decays involving $D_1$ meson in the final state, the polarization fraction ($F_L$) is highly longitudinal for smaller $\q2$ and becomes transverse near $\q2 \sim 2~\text{GeV}^2$. Thereafter, $F_L$ shows a small rise in longitudinal behavior for $B_c \to D_1$ decays with lighter leptons. Interestingly, for tauonic decay, the longitudinal polarization fraction does not vary substantially, as shown in Figure \ref{fig:obs_D1}. Roughly, similar behavior can also be seen for $B_c \to \chi_{c1} \ell \nu_{\ell}$ decays. On the other hand, longitudinal polarization fraction falls linearly (non-linearly) for leptonic decays with $D_1^\p (\chi_{c1}^\p)$ meson in the final state, as shown in Figure \ref{fig:obs_D1p} (\ref{fig:obs_chic1p}). Furthermore, the variation of asymmetry parameter with $\q2$ reconfirms the above stated observations for $F_L$, see Figures \ref{fig:obs_D1}, \ref{fig:obs_D1p}, \ref{fig:obs_chic1}, and \ref{fig:obs_chic1p}.
		
		\item[ix.] Recently, some experiments involving \textit{B} meson decay have shown deviations in ${R}_{D^*}$ from the SM expectation. While the measured numerical value is ${R}_{D^*}^{exp}=0.281 \pm 0.018 \pm 0.024$ \cite{BaBar:2012obs, BaBar:2013mob, Belle:2015qfa, Belle:2016ure, Abdesselam:2016xqt, LHCb:2015gmp, LHCb:2023zxo}, the average theoretical value is ${R}_{D^*}^{SM}=0.254 \pm 0.005$ \cite{HFLAV:2022esi}. Thus, the study of semileptonic $B_c$ decays involving the heavy to heavy $B_c\to A$ transition may also suffer from similar anomalies. Motivated by these intriguing measurements, we obtain the LFU ratios for axial-vector meson emitting decays of $B_c$ meson by considering the masses of the leptons:
		\[ R_{D_{1}} = \frac{\mathcal{B}(B_c^{+} \rightarrow D_{1}^{0} \tau^{+} \nu_\tau)}{\mathcal{B}(B_c^{+} \rightarrow D_{1}^{0} e^{+} \nu_e)}= 0.50^{+0.23 +0.40}_{-0.33 -0.80}; ~R_{D_{1}^{\p}} = \frac{\mathcal{B}(B_c^{+} \rightarrow D_{1}^{0\p} \tau^{+} \nu_\tau)}{\mathcal{B}(B_c^{+} \rightarrow D_{1}^{0\p} e^{+} \nu_e)} =  0.45^{+0.27 +0.45}_{-0.31 -0.65};\]
		\[R_{\chi_{c1}}=\frac{\mathcal{B}(B_c^{+} \rightarrow \chi_{c1} \tau^{+} \nu_\tau)}{\mathcal{B}(B_c^{+} \rightarrow \chi_{c1} e^{+} \nu_e)}= 0.11^{+0.01 +0.01}_{-0.01 -0.01};~
		R_{\chi_{c1}^{\p}}=\frac{\mathcal{B}(B_c^{+} \rightarrow \chi_{c1}^{\p} \tau^{+} \nu_\tau)}{\mathcal{B}(B_c^{+} \rightarrow \chi_{c1}^{\p} e^{+} \nu_e)}= 0.08^{+0.01 +0.04}_{-0.00 -0.03};\]
		
		Similarly, $c\to d/s$ transitions yield,
		\[R_{B_{1}}=\frac{\mathcal{B}(B_c^{+} \rightarrow B_{1}^{0} \mu^{+} \nu_\mu)}{\mathcal{B}(B_c^{+} \rightarrow B_{1}^{0} e^{+} \nu_e)}= 0.83^{+0.26 + 0.02}_{-0.60 -0.34};~
		R_{B_{1}^{\p}}=\frac{\mathcal{B}(B_c^{+} \rightarrow B_{1}^{0\p} \mu^{+} \nu_\mu)}{\mathcal{B}(B_c^{+} \rightarrow B_{1}^{0\p} e^{+} \nu_e)} = 0.83^{+0.12 +0.40}_{-0.16 -0.28};\]	
		\[R_{B_{s1}} = \frac{\mathcal{B}(B_c^{+} \rightarrow B_{s1}^{0} \mu^{+} \nu_\mu)}{\mathcal{B}(B_c^{+} \rightarrow B_{s1}^{0} e^{+} \nu_e)} = 0.70^{+0.40 +0.40}_{-0.80 -0.50};~
		R_{B_{s1}^{\p}} = \frac{\mathcal{B}(B_c^{+} \rightarrow B_{s1}^{0\p} \mu^{+} \nu_\mu)}{\mathcal{B}(B_c^{+} \rightarrow B_{s1}^{0\p} e^{+} \nu_e)} = 0.75^{+0.10 +0.25}_{-0.10 -0.19}.\]
	\end{itemize}
	
	Finally, we compare our predictions for branching ratios with those of existing theoretical works from various models. It is noteworthy that no analyses have been reported so far in the literature (to the best of our knowledge) for physical observables $\la A_{\rm FB} \ra, ~\la C_F^\ell \ra, ~\la P_L^\ell \ra,~ \la F_L \ra$, and $\la\alpha^{*}\ra$ of semileptonic weak decays involving axial-vector meson emitting decays of $B_c$, except for orbitally excited charmonia. Therefore, such analyses highlight the importance of the current work. As mentioned before, there has been limited research on bottom-conserving semileptonic decays to \textit{p}-wave mesons until recently. The results for bottom-conserving and bottom-changing decays from other models, namely, \text{RQM} \cite{Ebert:2010zu}, \text{pQCD} \cite{Rui:2018kqr}, \text{BS} \cite{Wang:2011jt}, \text{NRQM} \cite{Hernandez:2006gt}, \text{RQM} \cite{Ivanov:2005fd}, \text{RCQM} \cite{Ivanov:2006ni}, \text{QCDSR} \cite{Azizi:2009ny}, \text{CLF QM} \cite{Wang:2009mi}, \text{modified Godfrey-Isgur model (MGI)-CLF QM} \cite{Li:2023wgq}, and \text{CLF QM} \cite{Shi:2016gqt} are given in Tables \ref{other_Br} and \ref{other_Br_chi}, respectively. We observe that our predictions for $B_c \to B_{(s)1}\ell \nu_{\ell}$ compare well with those of RQM~\cite{Ebert:2010zu}. However, our predictions for $B_c \to B_{(s)1}^{\p}\ell \nu_{\ell}$ are larger than those of RQM~\cite{Ebert:2010zu}. When compared to the type-I CLF QM~\cite{Shi:2016gqt}, we observe that our branching ratios for unprimed axial-vectors are smaller by an order of magnitude (except for $B_c^+ \to B_{s1}^0 e^+ \nu_{e}$), while those for primed types are comparatively larger with same order. In addition, a similar trend in branching ratios can be observed in the absence of mixing, however, the magnitudes vary by a factor of $\sim 2 ~\text{to}~ 5$ compared to their results. In general, our predictions are larger as compared to Li \textit{et al.}~\cite{Li:2023wgq} (MGI with type-I CLF QM) for both unmixed and mixed cases. It should be noted that besides the type-I correspondence, differences exist between their approaches and ours regarding the formulation of $\q2$ dependence as well as other factors such as quark masses and flavor dependence of $\beta$ parameters. For bottom-changing $B_c \to D_{1}^{\p}e \nu_{e}/D_{1}^{\p}\mu \nu_{\mu}$ decays, our results are larger than RQM~\cite{Ebert:2010zu}; however, compare well within the uncertainties for the rest of the decays. On the other hand, our predictions show an opposite trend with respect to the order of magnitude as compared to \cite{Li:2023wgq} for $B_c \to D_{1}^{(\p)}\ell \nu_{\ell}$. Furthermore, when compared to the estimates of QCDSR \cite{Khosravi:2015tea} for $B_c \to D_{1}^{0} (2420)/D_{1}^{0} (2430)\ell \nu_{\ell}$, our predictions for $B_c \to D_{1}^{(\p)}\ell\nu_{\ell}$ decays are smaller by more than an order of magnitude. Additionally, when considering decays to charmonium states, the branching ratios of $B_c \rightarrow \chi_{c1}^{(\p)} \ell \nu_\ell$ decays from various models (see Table \ref{other_Br_chi}), including ours, demonstrate a reasonably consistent pattern among results obtained for $e \text{~and~} \mu$ leptons. However, all these models present a scattered range of branching ratios for the case of $\tau$ lepton in the final state, which could be of immense experimental significance for near-future experiments. In addition, we compared our predictions for LFU ratios with available results from other models. We observe that the CLF QM predictions of $R_{\chi_{c1}} = 0.11$ and $R_{\chi_{c1}^{\p}} = 0.07$ \cite{Wang:2009mi} and relativistic constituent quark model predictions of $R_{\chi_{c1}}$ = $0.12$ and $R_{\chi_{c1}^{\p}}$ = $0.08$ \cite{Ivanov:2005fd} agree well with our results. 
	\section{Summary}
	\label{sec:summary}
	In the present work, we have done an in-depth analysis of the self-consistency issues in $B_c\to A^{(\p)}$ transition form factors and semileptonic $B_c\to A^{(\p)}\ell\nu_{\ell}$ decays employing type-II correspondence in CLF QM. The self-consistent type-II correspondence incorporates the resolution of existing inconsistencies associated with zero-mode contributions and violation of covariance of matrix element in type-I CLF QM. These issues have been studied only for bottom-changing transition form factors involving vector mesons \cite{Chang:2019mmh}. However, problems related to self-consistency and covariance of matrix elements were also expected to be present in axial-vector meson-emitting transitions. The quantitative extent of these issues has not been analyzed in both bottom-conserving as well as bottom-changing transition form factors (and semileptonic decays) involving axial-vector mesons. Therefore, as a first attempt, we have comprehensively examined these problems in the current work for both bottom-conserving and bottom-changing transition form factors. In order to examine the impact of self-consistency via type-II correspondence, we have done a thorough analysis of semileptonic weak decays involving $B_c\to A^{(\p)}$ transitions. Aforementioned, in the type-I scheme \cite{Jaus:1999zv, Cheng:2003sm}, it is important to note that the form factors $A(\q2)(q(\q2))$ and $V_2(\q2)(c_{+}(\q2))$ are self-consistent, \textit{i.e.}, they are free from zero-mode contributions. However, the challenges regarding self-consistency and strict covariance still exist in $V_{0}(\q2)(c_{-}(\q2))$ and $V_1(\q2)(l(\q2))$ form factors in the CLF QM. Furthermore, these form factors are inconsistent due to non-vanishing $\omega$-dependent spurious contributions associated with $B^{(i)}_{j}$ functions and the non-zero residual $\omega$-dependent components violate the covariance of matrix elements in the traditional type-I correspondence \cite{Chang:2019mmh}. Nevertheless, the type-II correspondence suggests that both of these issues share a common source, and it resolves them simultaneously. We have also done a comparative study of form factors and branching ratios between type-I and type-II CLF QM to gauge the impact of inconsistencies. In addition to the prediction of the branching ratios of $B_c\to A^{(\p)}\ell\nu_{\ell}$ decays in type-II CLF QM, we have calculated the expectation values of physical observables such as $\la A_{\rm FB} \ra, ~\la C_F^\ell \ra, ~\la P_L^\ell \ra,~ \la F_L \ra$, and $\la\alpha^{*}\ra$. Further, we have plotted the $\q2$ variation of form factors, branching ratios, and the rest of the observables for different generations of leptons. We list our important findings as follows:
	\begin{itemize}
		\item We observed that the form factors $V_{0}(\q2)(c_{-}(\q2))$ and $V_1(\q2)(l(\q2))$ in type-I CLF QM, acquire zero-mode contributions through $B^{(i)}_{j}$-functions, \textit{i.e.}, $B_1^{(2)}$ and $B_3^{(3)}$. Consequently, the form factors corresponding to longitudinal and transverse states yield different numerical values leading to inconsistency in the type-I scheme. As expected, the application of type-II correspondence resolves such issues. We confirm that although zero-mode contributions associated with $B^{(i)}_{j}$-functions exist formally in form factor relations, they vanish numerically in type-II correspondence. Therefore, the form factors exhibit self-consistency and obey the covariance of matrix elements in type-II correspondence.
		\item Numerically speaking, the effect of zero-mode contributions is overwhelming in both bottom-conserving and bottom-changing $V_{0}(\q2)$ and $V_1(\q2)$ transition form factors. It should be emphasized that $V_1^{B_c B_{(s)1}}(\q2)$ form factors, due to their large magnitude, have profound ramifications on the branching ratios of bottom-conserving semileptonic $B_c\to A^{(\p)}\ell\nu_{\ell}$ decays. The contributions of $V_1^{B_c B_{(s)1}}(\q2)$ form factors to these decays become even more important given the fact that contributions from $V_0^{B_c B_{(s)1}}(\q2)$ form factors are negligible. On the other hand in bottom-changing decays, although the numerical magnitude of $V_1^{B_c D_{1}}(\q2)$ form factors are considerably smaller as compared to their bottom-conserving counterparts, their effect on the decays are substantially larger.  
		\item We observe that there exists a two-fold problem in type-I form factors, irrespective of $\q2$ dependence formulation, when compared to type-II correspondence. The form factors at $\q2=0$ show significant variation in magnitude, including sign flip, and the slope parameters consistently acquire large numerical values. In addition, the choice of pole masses plays a significant role in lowering the slope parameters to some extent, however, the form factors still show large variation due to zero-mode contributions. Similar observations can be made from the numerical values of the form factors in other analyses in the type-I scheme \cite{Li:2023wgq, Shi:2016gqt}. Therefore, we conclude that while comparing type-I and type-II schemes, the effects of type-II correspondence can not simply be adjudged from the numerical values of the form factors at $\q2=0$ alone, the variation of the slope parameters also play a crucial role in the overall assessment of self-consistency issues. All the above-stated issues can be resolved by employing type-II CLF QM.
		\item In a comparative study of type-I and type-II correspondence, we also observe a sequence in type-I inconsistencies in different flavor transition form factors. The form factors for $B_c \to D_{1}$ transitions are more seriously affected by zero-mode contributions as compared to $B_c \to B_{(s)1}$ transitions. On the other hand, the effect is least in the case of $B_c \to \chi_{c1}$ transitions. A similar trend, as expected, has been observed in corresponding branching ratios. 
		\item The analyses of semileptonic $B_c\to A^{(\p)}\ell\nu_{\ell}$ decays pose more serious challenges in addition to inconsistency, \textit{i.e.}, corresponding to $\q2$ dependence of the form factors, and the mixing of final states. The branching ratios of all the decays involving unprimed axial-vector meson states are small as compared to decays including their primed partners. The numerical effects of self-consistency are profound in bottom-changing decays as compared to bottom-conserving decays. Interestingly, the $B_c \to \chi_{c1}^{(\p)} \ell\nu_{\ell}$ decays are affected minimally from self-consistency issues. Further, the branching ratios of self-consistent type-II CLF QM differ up to $\sim 98\% ~(\sim 30\% )$ as compared to type-I CLF QM for bottom-changing (bottom-conserving) modes. This effect can even be observed in the absence of mixing. It is worth noting that the branching ratios of some of the $B_c \to B_{(s)1}^{(\p)}\ell\nu_{\ell}$ decays are comparable to those of $B_c \to D_{1}^{(\p)}\ell\nu_{\ell}/\chi_{c1}^{(\p)}\ell\nu_{\ell}$ decays, which could be of experimental significance.
		\item Finally, we have analyzed multiple physical observables, like, $\la A_{\rm FB} \ra$, $\la C_F^\ell \ra$, $\la P_L^\ell \ra$, $\la F_L \ra$, and $\la\alpha^{*}\ra$. In addition, we also estimated LFU within the SM through the lepton mass effects in semileptonic decays. It should be emphasized that the mass of $e$ and $\mu$ has a considerable effect on the bottom-conserving decays. In contrast, for bottom-changing decays, the effect is prominent only when $\tau$ lepton is involved. Subsequently, we obtained LFU ratios that would be useful to highlight anomalies in comparison to future experiments.
	\end{itemize}
	The numerical results obtained from CLF QM and the phenomenological analysis offer crucial insights into the underlying physics and can guide future experimental investigations to validate and refine our understanding of these semileptonic decay processes.
	\section*{Acknowledgment}
	We express our sincere thanks to Gautam Bhattacharyya and J. P. B. C. de Melo for useful discussions. The authors gratefully acknowledge the computational resources provided by the High-Performance Computing Centre (HPCC) at SRM Institute of Science and Technology (SRMIST). RD acknowledges the financial support by the Department of Science and Technology (SERB:CRG/2018/002796), New Delhi.
	\appendix
	\section{${}^{3} P_{1}$ and ${}^{1} P_{1}$ mixing}
	\label{axial_mixing}
	Based on \textit{C}-parity, there exist two types of spectroscopic axial-vector mesons, \textit{i.e.}, ${}^{3} P_{1} ~(J^{PC} =1^{++})$ and ${}^{1} P_{1}~(J^{PC} =1^{+-})$, and they behave well with respect to the quark model $q_1\bar{q_2}$ assignments. We denote ${}^{3}P_{1}$ and ${}^{1}P_{1}$ as $A_A$ and $A_B$, respectively. Due to the absence of a quantum number that prohibits the mixing of strange and charmed states, they are expressed as a mixture of $A_A$ and $A_B$. However, due to their opposite \textit{C}-parity, the non-strange and uncharmed mesons $A_A$ and $A_B$ cannot undergo mixing. The following non-strange and uncharmed mesons have so far been observed experimentally \cite{ParticleDataGroup:2022pth}, \textit{i.e.},
	\begin{enumerate}
		\item[i.] $A_A$ multiplet: Isovector $a_{1} (1.230)$ and four isoscalars $f_{1} (1.282)$, $f_{1} (1.426)$, $f{^{\p}}_{1} (1.518)$, and $\chi_{c1} (3.511)$, where the numerical values given in the parentheses represent the masses (in GeV) of the individual mesons.
		\item[ii.] $A_B$ multiplet: Isovector $b_{1} (1.230)$ and three isoscalars $h_{1} (1.166)$, $h{^{\p}}_{1} (1.416)$, and $h_{c1} (3.525)$. \textit{C}-parity of $h^{{^{\p}}}_{1} (1.416)$, and spin and parity of the $h_{c1} (3.525)$\footnote{ We denote $h_{c} (3.525)$ as $\chi_{c1}^{\p} (3.525)$.} remains to be confirmed. 
	\end{enumerate}
	
	In the current work, we are focusing on charm and bottom mesons, therefore we consider the mixing of these states as follows\footnote{The form factors and slope parameters will exhibit mixing as per Eq.~\eqref{eq:A1}, which has been duly considered in our numerical calculations.}:
	\begin{equation} \label{eq:A1} 
		\begin{array}{l} {A = A_A   \sin \theta     +  A_{B}   \cos \theta ,} \\ 
			{A ^\p =  A_A  \cos \theta  -  A_{B}  \sin \theta, } 
		\end{array} 
	\end{equation} 
	where $\theta$ is the mixing angle between $A_A$ and $A_B$, while $A^{(\p)}$ denotes physical states.
	The charmed physical states $D_{1} (2.422)$ and ${D}^\p_{1} (2.427)$ mesons are then given as, 
	\begin{equation} \label{eq:D1} 
		\begin{array}{l} {D_{1} (2.422)  =  D_{1A}   \sin \theta _{D_{1} }     +  D_{1B}   \cos \theta _{D_{1} } ,} \\ 
			{{D}^\p_{1} (2.427)  =  D_{1A}   \cos \theta _{D_{1} } -  D_{1B}   \sin \theta _{D_{1} } ,} \end{array} 
	\end{equation} 
	where the mixing angle $\theta_{D_{1}} =-58^{\circ }$ \cite{Godfrey:2015dva, Chen:2016spr, Wang:2022tuy}. Note that the measured mass for ${D}^\p_{1}$ state is $2.412$~$\mathrm{GeV}$ \cite{LHCb:2019juy}, which has been used for calculation in our work. Similarly, for
	\textit{b}-flavored mesons we use the following:
	\begin{equation} \label{eq:B1} 
		\begin{array}{l} 
			{B_{1} (5.710) = B_{1A}\sin \theta_{B_{1}} + B_{1B} \cos\theta_{B_{1}},} \\ {B^{\p}_{1} (5.726)  =  B_{1A} \cos \theta_{B_{1}}     -B_{1B} \sin\theta_{B_{1}};} \end{array} 
	\end{equation}  
	and
	\begin{equation} \label{eq:Bs1} 
		\begin{array}{l} 
			{B_{s1} (5.820) = B_{s1A}\sin\theta_{B_{s1}} + B_{s1B}\cos\theta_{B_{s1}},} \\ {B^{\p}_{s1} (5.830) = B_{s1A}\cos\theta_{B_{s1}} - B_{s1B}\sin\theta_{B_{s1}}.} 
		\end{array} 
	\end{equation} 
	Based on previous studies that have reported various mixing angles for the bottom states, we adopt $\theta_{B_{1}}= -55^{\circ}$ and $\theta_{B_{s1}}= -55^{\circ}$ in our analysis \cite{Chen:2016spr, li:2021hss, Li:2018eqc}.
	\section{Zero-mode contributions and covariance of matrix element}
	\label{cov_ml}
	Aforementioned, Jaus~\cite{Jaus:1999zv} proposed a more pragmatic method to deal with the $\omega$-dependence inherent in the LF formalism that is manifest in the decomposition of the four-momenta product. This method accounts for the missing zero-mode contributions and, consequently, reinstates the covariance of matrix elements. These contributions are directly proportional to the $\omega^\mu$ and are associated with the $B^{(i)}_{j}$ functions as described in Sec.~\ref{methodology}. In addition, the $B^{(i)}_{j}$ functions play a special role since, on the one hand, it is combined with $\omega$, on the other hand, there is no zero-mode contribution associated with $B^{(i)}_{j}$. Such remnant contributions violate the covariance of the LF matrix element in the type-I scheme. Therefore, the self-consistency and covariance problems are of the same origin. Extending this method~\cite{Jaus:1999zv, Jaus:2002sv}, Qin Chang \textit{et al.}~\cite{Chang:2019mmh} presented the mechanism to neutralize these residual $\omega$-dependence and showed that both these problems can be simultaneously resolved by employing type-II correspondence in vector mesons. The proposed method ~\cite{Jaus:1999zv, Jaus:2002sv, Chang:2019obq} requires some proper replacements of the four-momenta decomposition for spin-$1$ system in the $\hat{S}_{\cal B}$, defined by Eq.~\eqref{eq:Bclf2}, under integration as described below:
	\begin{equation}
		\label{eq:repFF1}
		\hat{k}_1'^{\mu} \to P^\u A_1^{(1)}+q^\u A_2^{(1)} \,,
	\end{equation}
	\begin{equation}
		\hat{k}_1'^{\mu}\hat{k}_1'^{\nu} \to g^{\u\v}A_1^{(2)}+P^\u P^\v A_2^{(2)}+(P^\u q^\v+q^\u P^\v)A_3^{(2)}+q^\u q^\v A_4^{(2)}
		+\frac{P^\u\omega^\v+\omega^\u P^\v}{\omega\cdot P}B_1^{(2)}\,,
	\end{equation}
	\begin{align}
		\hat k_1'^{\mu}\hat k_1'^{\nu}\hat k_1'^{\alpha}\to&\left(g^{\mu \nu}P^\alpha+g^{\mu \alpha}P^\nu+g^{\nu\alpha}P^\mu\right)A_1^{(3)}+\left(g^{\mu \nu}q^\alpha+g^{\mu \alpha}q^\nu+g^{\nu\alpha}q^\mu\right)A_2^{(3)}\nonumber\\
		&+P^\mu P^\nu P^\alpha A_3^{(3)}+\left(P^\mu P^\nu q^\alpha+P^\mu q^\nu P^\alpha+q^\mu P^\nu P^\alpha\right)A_4^{(3)}\nonumber\\
		&+\left(q^\mu q^\nu P^\alpha+q^\mu P^\nu q^\alpha+P^\mu q^\nu q^\alpha\right)A_5^{(3)}+q_\mu q_\nu q_\alpha A_6^{(3)}\nonumber\\
		&+\frac{1}{\omega \cdot P}\left(P^\mu P^\nu\omega^\alpha+P^\mu \omega^\nu P^\alpha+\omega^\mu P^\nu P^\alpha\right)B_1^{(3)}\nonumber\\
		&+\frac{1}{\omega\cdot P}\left[\left(P^\mu q^\nu+q^\mu P^\nu\right)\omega^\alpha+\left(P^\mu q^\alpha+q^\mu P^\alpha\right)\omega^\nu+\left(P^\alpha q^\nu+q^\alpha P^\nu\right)\omega^\mu\right]B_2^{(3)}\,,
	\end{align}
	\begin{align}
		k_1'^{\mu}\hat{N}_2\to& q^\u\left(A_2^{(1)}Z_2+\frac{q\cdot P}{q^2}A_1^{(2)} \right) \,,\\
		\hat{k}_1'^{\mu}\hat{k}_1'^{\nu}\hat{N}_2\to &g^{\u\v}A_1^{(2)}Z_2+q^\u q^\v\left( A_4^{(2)}Z_2+2\frac{q\cdot P}{q^2}A_2^{(1)}A_1^{(2)}\right)+\frac{P^\u\omega^\v+\omega^\u P^\v}{\omega\cdot P}B_3^{(3)}\,,
	\end{align}
	\begin{align}
		\hat k_1'^{\mu}\hat k_1'^{\nu}\hat k_1'^{\alpha}\hat N_2\to &\left(g^{\mu \nu}q^\alpha+g^{\mu \alpha}q^\nu+g^{\nu \alpha}q^\mu\right)\left(A_2^{(3)}Z_2+\frac{q\cdot P}{q^2}A_1^{(4)}\right)+q^\mu q^\nu q^\alpha \left(A_6^{(3)}Z_2+3\frac{q\cdot P}{q^2}A_4^{(4)}\right)\nonumber\\
		&+\frac{1}{\omega\cdot P}\left[\left(P^\mu q^\nu+q^\mu P^\nu\right)\omega^\alpha+\left(P^\mu q^\alpha+q^\mu P^\alpha\right)\omega^\nu+\left(P^\alpha q^\nu+q^\alpha P^\nu\right)\omega^\mu\right]B_5^{(4)},
		\label{eq:repFF}
	\end{align}
	where $A^{(i)}_{j}$, $B^{(i)}_{j}$, and $Z_2$ are dependent on $x, ~\kb^{\p 2}, ~\kb^\p \cdot \qb$, and $\q2$ and their definitions are given by \cite{Jaus:1999zv} 
	\begin{align}\label{AB_fun}
		A_1^{(1)}&=  \frac{x}{2}\,,\qquad
		A_2^{(1)}=\frac{x}{2} -\frac{\kb' \cdot \qb }{q^2}\,;\\
		A_1^{(2)}&=-\kb'^2 -\frac{(\kb' \cdot \qb)^2}{q^2}\,,\qquad
		A_2^{(2)}=(A_1^{(1)})^2\,,\qquad
		A_3^{(2)}=A_1^{(1)}A_2^{(1)}\,,\qquad
		A_4^{(2)}=(A_2^{(1)})^2\,;\\
		A_1^{(3)}&=A_1^{(1)}A_1^{(2)}\,,\qquad
		A_2^{(3)}=A_2^{(1)}A_1^{(2)}\,,\qquad
		A_3^{(3)}=A_1^{(1)}A_2^{(2)}\,,\qquad
		A_4^{(3)}=A_2^{(1)}A_2^{(2)}\,,\nonumber\\
		A_5^{(3)}&=A_1^{(1)}A_4^{(2)}\,,\qquad
		A_6^{(3)}=A_2^{(1)}A_4^{(2)}-\frac{2}{q^2}A_2^{(1)}A_1^{(2)}\,;\\
		A_3^{(4)}&=A_1^{(1)}A_2^{(3)}\,;\\
		B_1^{(2)}&=\frac{x}{2}Z_2-A_1^{(2)}\,;\\
		B_2^{(3)}&=\frac{x}{2}Z_2A_2^{(1)}+A_1^{(1)}A_1^{(2)} \frac{P\cdot q}{q^2}-A_2^{(1)}A_1^{(2)},
		~~B_3^{(3)}=B_1^{(2)}Z_2+\left(P^2-\frac{(q\cdot P)^2}{q^2}\right)A_1^{(1)}A_1^{(2)};\\
		B_1^{(4)}&=\frac{x}{2}Z_2A_1^{(2)}-A_1^{(4)}\,,\qquad
		B_5^{(4)}=B_2^{(3)}Z_2+\frac{q\cdot P}{q^2}B_1^{(4)}+\left[P^2-\frac{\left(q\cdot P\right)^2}{q^2}\right]A_3^{(4)} \,;\\
		Z_2&=\hat{N}_1'+m_1'^2-m_2^2+(\bar{x}-x)M'^2+(q^2+q\cdot P)\frac{\kb' \cdot \qb}{q^2}\,.
	\end{align}
	Thus, the above-mentioned specific entities, Eqs.~\eqref{eq:repFF1}-\eqref{eq:repFF}, utilized by \cite{Chang:2019mmh, Chang:2019obq, Cheng:2003sm} to analyze the matrix element ${\cal B}$. This approach can be uniformly applied to the spin-$1$ system \cite{Jaus:2002sv}. Following their approach, it is easy to observe that Eq.~(56) in Ref.~\cite{Chang:2019mmh}, for spin-$1$ mesons, is applicable to $P \to A$ transitions (using appropriate transformations proposed in Eq.~\eqref{eq:V_A_trans}, except for $D^\pp$ and $\chi^\pp$ \cite{Cheng:2003sm}, and the above-stated relations) and thus leads to similar conclusions, as in \cite{Jaus:2002sv}, for $P \to A$ matrix elements. Therefore, the residual $\omega$-dependent components present in the traditional type-I correspondence scheme, which are non-zero and violate the covariance of the matrix element, numerically vanish in the type-II scheme for $P \to A$ transitions.
	\section{Space-like transition form factor}
	\label{FF_space-like}
	The Table~\ref{ff_space} given below represents the Type-II $B_c^{+}\to A_A/A_B$ transition form factor results (given by Eq.~\eqref{eq:q2_ff}) at space-like region before extrapolated into time-like region.
	\begin{table}[!ht]
		\centering
		\caption{Space-like Type-II form factors of $B_c\to A_A/A_B$ transitions.}
		\label{ff_space}
		\setlength{\tabcolsep}{0.5pt}
		\begin{tabular}{|c|c|c|c|c|c|c|c|c|c|c|c|}
			\hline
			\multicolumn{1}{|c|}{F} & $\q2_{\bot}= 0.01$ & $\q2_{\bot}=0.1$ & $\q2_{\bot}=1.0$ & $\q2_{\bot}=5.0$ & $\q2_{\bot}=10.0$ & \multicolumn{1}{c|}{F} & $\q2_{\bot}= 0.01$ & $\q2_{\bot}=0.1$ & $\q2_{\bot}=1.0$ & $\q2_{\bot}=5.0$ & $\q2_{\bot}=10.0$ \\
			\hline
			$A^{B_c B_{1A}}$        & 0.18 & 0.17 & 0.12 & 0.03 & 0.01  & $A^{B_c B_{1B}}$           & 0.04 & 0.04 & 0.03 & 0.01 & 0.000002 \\ 
			
			$V_{0}^{B_c B_{1A}}$    & 0.20 & 0.19 & 0.15 & 0.05 & 0.02  & $V_{0}^{B_c B_{1B}}$       & 0.55 & 0.53 & 0.38 & 0.10 & 0.02 \\
			
			$V_{1}^{B_c B_{1A}}$    & 5.85 & 5.91 & 5.96 & 3.30 & 1.27  & $V_{1}^{B_c B_{1B}}$       & 7.45 & 7.19 & 5.11 & 1.39 & 0.36 \\
			
			$V_{2}^{B_c B_{1A}}$    & 0.09 & 0.08 & 0.06 & 0.02 & 0.005 & $V_{2}^{B_c B_{1B}}$       & $-0.18$ & $-0.18$ & $-0.12$ & $-0.02$ & $-0.005$ \\
			\hline
			$A^{B_c B_{s1A}}$       & 0.16 & 0.16 & 0.12 & 0.04 & 0.01  & $A^{B_c B_{s1B}}$          & 0.04 & 0.03 & 0.02 & 0.01 & 0.000006 \\ 
			
			$V_{0}^{B_c B_{s1A}}$   & 0.16 & 0.15 & 0.13 & 0.05 & 0.02  & $V_{0}^{B_c B_{s1B}}$      & 0.57 & 0.55 & 0.39 & 0.11 & 0.03 \\
			
			$V_{1}^{B_c B_{s1A}}$   & 5.74 & 5.93 & 6.88 & 4.81 & 2.15  & $V_{1}^{B_c B_{s1B}}$      & 9.93 & 9.62 & 7.09 & 2.18 & 0.63 \\
			
			$V_{2}^{B_c B_{s1A}}$   & 0.06 & 0.06 & 0.05 & 0.02 & 0.01  & $V_{2}^{B_c B_{s1B}}$      & $-0.18$ & $-0.17$ & $-0.12$ & $-0.03$ & $-0.01$ \\
			\hline
			\multicolumn{1}{|c|}{F} & $\q2_{\bot}= 0.01$ & $\q2_{\bot}=0.1$ & $\q2_{\bot}=5.0$ & $\q2_{\bot}=10.0$ & $\q2_{\bot}=20.0$ & \multicolumn{1}{c|}{F} & $\q2_{\bot}= 0.01$ & $\q2_{\bot}=0.1$ & $\q2_{\bot}=5.0$ & $\q2_{\bot}=10.0$ & $\q2_{\bot}=20.0$ \\ 
			\hline
			$A^{B_c D_{1A}}$        & 0.12 & 0.12 & 0.07 & 0.05 & 0.02 & $A^{B_c D_{1B}}$           & 0.02 & 0.02 & 0.01 & 0.01 & 0.00002 \\ 
			
			$V_{0}^{B_c D_{1A}}$    & 0.08 & 0.08 & 0.05 & 0.04 & 0.02 & $V_{0}^{B_c D_{1B}}$       & 0.20 & 0.19 & 0.12 & 0.08 & 0.04 \\
			
			$V_{1}^{B_c D_{1A}}$    & 0.30 & 0.30 & 0.23 & 0.17 & 0.09 & $V_{1}^{B_c D_{1B}}$       & 0.11 & 0.11 & 0.07 & 0.04 & 0.02 \\
			
			$V_{2}^{B_c D_{1A}}$    & 0.09 & 0.09 & 0.06 & 0.04 & 0.02 & $V_{2}^{B_c D_{1B}}$       & $-0.06$ & $-0.06$ & $-0.03$ & $-0.02$ & $-0.01$ \\
			\hline
			$A^{B_c \chi_{c1}}$     & 0.21 & 0.21 & 0.15 & 0.10 & 0.05 & $A^{B_c \chi_{c1}^\p}$     & 0.04 & 0.04 & 0.02 & 0.02 & 0.01 \\ 
			
			$V_{0}^{B_c \chi_{c1}}$ & 0.07 & 0.07 & 0.05 & 0.04 & 0.03 & $V_{0}^{B_c \chi_{c1}^\p}$ & 0.35 & 0.34 & 0.23 & 0.16 & 0.08 \\
			
			$V_{1}^{B_c \chi_{c1}}$ & 0.47 & 0.47 & 0.48 & 0.45 & 0.34 & $V_{1}^{B_c \chi_{c1}^\p}$ & 0.30 & 0.30 & 0.22 & 0.16 & 0.09 \\
			
			$V_{2}^{B_c \chi_{c1}}$ & 0.08 & 0.08 & 0.07 & 0.05 & 0.03 & $V_{2}^{B_c \chi_{c1}^\p}$ & $-0.17$ & $-0.16$ & $-0.11$ & $-0.07$ & $-0.03$ \\
			\hline
		\end{tabular}
	\end{table}
	\newpage
	\section{Unmixed branching ratios }
	\label{Br_unmix}
	The Table~\ref{semilep_Br_unmix} given below represents the unmixed (for zero mixing angle in relations given by Eqs.~\eqref{eq:D1}-\eqref{eq:Bs1}) branching ratios of $B_c$ meson decaying to axial-vector mesons in the final states.
	\begin{table}[!ht]
		\centering
		\caption{Unmixed branching ratios of $B_c^{+}\to A^{(\p)}\ell^{+}\nu_{\ell}$ decays. For Type-II, Type-I, and Type-I*, the definitions provided in the caption of Table \ref{tab:ff_bb} apply.}
		\label{semilep_Br_unmix}
		\setlength{\tabcolsep}{5pt}
		\begin{tabular}{|l|l|l|}
			\hline
			\multicolumn{1}{|c|}{Decays} & \multicolumn{1}{c|}{Type-II} & \multicolumn{1}{c|}{Type-I [Type-I*]}     \\ \hline
			$B_c^{+}\to B_{1}^0 e^{+}\nu_{e}$            & $ 7.63\t 10^{-5}$ & $ 3.75\t 10^{-5}$ [$ 5.72\t 10^{-5}$] \\ \hline
			$B_c^{+}\to B_{1}^{0\p} e^{+}\nu_{e}$        & $ 2.75\t 10^{-5}$ & $ 1.61\t 10^{-5}$ [$ 1.95\t 10^{-5}$] \\ \hline
			$B_c^{+}\to B_{1}^0 \mu^{+}\nu_{\mu}$        & $ 6.33\t 10^{-5}$ & $ 2.98\t 10^{-5}$ [$ 4.67\t 10^{-5}$] \\ \hline
			$B_c^{+}\to B_{1}^{0\p} \mu^{+}\nu_{\mu}$    & $ 2.32\t 10^{-5}$ & $ 1.33\t 10^{-5}$ [$ 1.62\t 10^{-5}$] \\ \hline
			$B_c^{+}\to B_{s1}^0 e^{+}\nu_{e}$           & $ 5.63\t 10^{-4}$ & $ 3.37\t 10^{-4}$ [$ 4.29\t 10^{-4}$] \\ \hline
			$B_c^{+}\to B_{s1}^{0\p} e^{+}\nu_{e}$       & $ 1.30\t 10^{-4}$ & $ 8.11\t 10^{-5}$ [$ 9.25\t 10^{-5}$] \\ \hline
			$B_c^{+}\to B_{s1}^0 \mu^{+}\nu_{\mu}$       & $ 4.24\t 10^{-4}$ & $ 2.45\t 10^{-4}$ [$ 3.18\t 10^{-4}$] \\ \hline
			$B_c^{+}\to B_{s1}^{0\p} \mu^{+}\nu_{\mu}$   & $ 1.00\t 10^{-4}$ & $ 6.11\t 10^{-5}$ [$ 6.99\t 10^{-5}$] \\ \hline
			$B_c^{+}\to{D}_{1}^0 e^{+}\nu_{e}$           & $ 1.72\t 10^{-5}$ & $ 1.78\t 10^{-6}$ [$ 2.98\t 10^{-6}$] \\ \hline
			$B_c^{+}\to{D}_{1}^{0\p} e^{+}\nu_{e}$       & $ 2.77\t 10^{-5}$ & $ 1.19\t 10^{-5}$ [$ 2.13\t 10^{-5}$] \\ \hline
			$B_c^{+}\to{D}_{1}^0 \mu^{+}\nu_{\mu}$       & $ 1.72\t 10^{-5}$ & $ 1.79\t 10^{-6}$ [$ 3.00\t 10^{-6}$] \\ \hline
			$B_c^{+}\to{D}_{1}^{0\p} \mu^{+}\nu_{\mu}$   & $ 2.77\t 10^{-5}$ & $ 1.19\t 10^{-5}$ [$ 2.13\t 10^{-5}$] \\ \hline
			$B_c^{+}\to{D}_{1}^0 \tau^{+}\nu_{\tau}$     & $ 6.54\t 10^{-6}$ & $ 1.43\t 10^{-7}$ [$ 5.18\t 10^{-7}$] \\ \hline
			$B_c^{+}\to{D}_{1}^{0\p} \tau^{+}\nu_{\tau}$ & $ 1.27\t 10^{-5}$ & $ 4.56\t 10^{-6}$ [$ 9.80\t 10^{-6}$] \\ \hline
		\end{tabular}
	\end{table}

	\newpage
	\begin{table}[!ht]
		\caption{ Bottom-conserving $B_c \rightarrow A$ transition form factors in CLF QM. Note that Type-II results represent the form factors calculated using $\q2$ dependence given in Eq.~\eqref{eq:q2_ff} with transition pole masses (as listed in Table \ref{tab1}). Type-I (Type-I*) results represent the form factors calculated from $\q2$ dependence given by Eq.~\eqref{eq:q2_t1} with parent pole mass (Eq.~\eqref{eq:q2_ff} with the transition pole masses). The first and second errors correspond to the quark masses and $\beta$ parameters, respectively.}
		\label{tab:ff_bb}
		\setlength{\tabcolsep}{0.5pt}
		\resizebox{\textwidth}{!}{
			\begin{tabular}{|c|cccc|cccc|}
				\hline
				\multirow{3}{*}{F}                     & \multicolumn{4}{c|}{Type-II}                                                                                                                                                                                                  & \multicolumn{4}{c|}{Type-I:}                                                                                      \\  
				&   \multicolumn{4}{c|}{}                                                                                                                                                                                            & \multicolumn{4}{c|}{(Type-I*)}   \\                                                                                \cline{2-9} & \multicolumn{1}{c|}{{$F(0)$}\footnote{We have rounded the errors upto second decimal place, however, in numerical calculations up to three significant digits are considered.}} & \multicolumn{1}{c|}{\text{a}} & \multicolumn{1}{c|}{\text{b}} & \multicolumn{1}{c|}{$F(\q2_{max})$} & \multicolumn{1}{c|}{$F(0)$}    & \multicolumn{1}{c|}{\text{a}}   & \multicolumn{1}{c|}{\text{b}}   &  \multicolumn{1}{c|}{$F(\q2_{max})$} \\ \hline
				
				\multirow{2}{*}{$A^{B_c B_{1A}}$}      & \multicolumn{1}{l|}{${0.17}^{+0.02 +0.00}_{-0.02 -0.01}$}& \multicolumn{1}{l|}{${0.43}^{+0.12 +0.20}_{-0.09-1.23}$} & \multicolumn{1}{l|}{${1.88}^{+0.02 +2.63}_{-0.05 -1.03}$}  & ${0.18}^{+0.02 +0.00}_{-0.02 -0.03}$  & \multicolumn{1}{l|}{$0.15$}    & \multicolumn{1}{l|}{$-7.64$}      & \multicolumn{1}{l|}{$287.67$}      & $0.14$    \\
				& \multicolumn{1}{l|}{}                             & \multicolumn{1}{l|}{}                            & \multicolumn{1}{l|}{}                            &                             & \multicolumn{1}{l|}{($0.17$)}  & \multicolumn{1}{l|}{($0.43$)}   & \multicolumn{1}{l|}{($1.88$)}  & ($0.18$)  \\
				\multirow{2}{*}{$V_{0}^{B_c B_{1A}}$}  & \multicolumn{1}{l|}{${0.19}^{+0.00 +0.01}_{-0.02 -0.02}$}  & \multicolumn{1}{l|}{${0.39}^{+0.01 +0.00}_{-1.18 -0.34}$}   & \multicolumn{1}{l|}{${0.82}^{+0.60 +1.33}_{-0.07 -0.44}$}  & ${0.21}^{+0.00 +0.01}_{-0.02 -0.02}$  & \multicolumn{1}{l|}{$-0.04$}   & \multicolumn{1}{l|}{$-195.33$}    & \multicolumn{1}{l|}{$981.12$}     & $-0.02$   \\
				& \multicolumn{1}{l|}{}                            & \multicolumn{1}{l|}{}                             & \multicolumn{1}{l|}{}                            &                            & \multicolumn{1}{l|}{($-0.09$)} & \multicolumn{1}{l|}{($-28.49$)} & \multicolumn{1}{l|}{($18.66$)}  & ($-0.04$) \\
				\multirow{2}{*}{$V_{1}^{B_c B_{1A}}$}  & \multicolumn{1}{l|}{${6.17}^{+0.13 +0.07}_{-0.24 -0.24}$}  & \multicolumn{1}{l|}{${-0.44}^{+0.04 +0.03}_{-0.04 -0.08}$}  & \multicolumn{1}{l|}{${0.24}^{+0.00 +0.18}_{-0.00 -0.05}$}   & ${6.46}^{+0.15 +0.04}_{-0.27 -0.24}$  & \multicolumn{1}{l|}{$4.93$}    & \multicolumn{1}{l|}{$-1.12$}     & \multicolumn{1}{l|}{$80.33$}      & $4.86$    \\
				& \multicolumn{1}{l|}{}                           & \multicolumn{1}{l|}{}                            & \multicolumn{1}{l|}{}                            &                              & \multicolumn{1}{l|}{($5.18$)}  & \multicolumn{1}{l|}{($-0.32$)}   & \multicolumn{1}{l|}{($0.26$)}   & ($5.47$)  \\
				\multirow{2}{*}{$V_{2}^{B_c B_{1A}}$}  & \multicolumn{1}{l|}{${0.08}^{+0.00 +0.00}_{-0.08 -0.00}$}    & \multicolumn{1}{l|}{${0.20}^{+0.02 +0.00}_{-11.23 -0.43}$}  & \multicolumn{1}{l|}{${0.55}^{+3.76 +0.88}_{-0.03 -0.30}$}  & ${0.09}^{+0.01 +0.00}_{-0.09 -0.00}$    & \multicolumn{1}{l|}{$0.08$}    & \multicolumn{1}{l|}{$-4.09$}      & \multicolumn{1}{l|}{$239.34$}      & $0.07$    \\
				& \multicolumn{1}{l|}{}                            & \multicolumn{1}{l|}{}                            & \multicolumn{1}{l|}{}                            &                             & \multicolumn{1}{l|}{($0.08$)}  & \multicolumn{1}{l|}{($0.20$)}   & \multicolumn{1}{l|}{($0.55$)}  & ($0.09$)  \\ \hline

				\multirow{2}{*}{$A^{B_c B_{1B}}$}      & \multicolumn{1}{l|}{${0.0002}^{+0.00 +0.00}_{-0.00 -0.00}$}   & \multicolumn{1}{l|}{${-19.07}^{+0.03 +0.20}_{-0.02 -0.07}$} & \multicolumn{1}{l|}{${21.16}^{+0.02 +0.07}_{-0.03 -0.20}$} & ${0.0001}^{+0.00 +0.00}_{-0.00 -0.00}$    & \multicolumn{1}{l|}{$0.0001$}   & \multicolumn{1}{l|}{$-154.56$}    & \multicolumn{1}{l|}{$1202.18$}     & $0.00004$   \\
				& \multicolumn{1}{l|}{}                           & \multicolumn{1}{l|}{}                           & \multicolumn{1}{l|}{}                           &                           & \multicolumn{1}{l|}{($0.0002$)} & \multicolumn{1}{l|}{($-19.07$)} & \multicolumn{1}{l|}{($21.16$)} & ($0.0001$) \\
				\multirow{2}{*}{$V_{0}^{B_c B_{1B}}$}  & \multicolumn{1}{l|}{${0.53}^{+0.05 +0.06}_{-0.06 -0.09}$}  & \multicolumn{1}{l|}{${0.41}^{+0.17 +0.27}_{-0.12 -1.23}$}   & \multicolumn{1}{l|}{${1.83}^{+0.25 +2.29}_{-0.21 -0.95}$}  & ${0.57}^{+0.06 +0.08}_{-0.06 -0.13}$  & \multicolumn{1}{l|}{$0.37$}    & \multicolumn{1}{l|}{$-28.59$}     & \multicolumn{1}{l|}{$469.80$}      & $0.30$    \\
				& \multicolumn{1}{l|}{}                           & \multicolumn{1}{l|}{}                            & \multicolumn{1}{l|}{}                            &                             & \multicolumn{1}{l|}{($0.45$)}  & \multicolumn{1}{l|}{($-0.65$)}  & \multicolumn{1}{l|}{($3.17$)}  & ($0.46$)  \\
				\multirow{2}{*}{$V_{1}^{B_c B_{1B}}$}  & \multicolumn{1}{l|}{${7.26}^{+0.91 +1.33}_{-0.89 -1.67}$}  & \multicolumn{1}{l|}{${0.21}^{+0.07 +0.05}_{-0.06 -0.63}$}   & \multicolumn{1}{l|}{${0.70}^{+0.01 +1.04}_{ -0.01 -0.38}$}  & ${7.96}^{+1.05 +1.51}_{-1.01 -2.14}$  & \multicolumn{1}{l|}{$5.42$}    & \multicolumn{1}{l|}{$-9.26$}      & \multicolumn{1}{l|}{$308.75$}      & $4.97$    \\
				& \multicolumn{1}{l|}{}                            & \multicolumn{1}{l|}{}                             & \multicolumn{1}{l|}{}                            &                             & \multicolumn{1}{l|}{($6.10$)}  & \multicolumn{1}{l|}{($0.19$)}   & \multicolumn{1}{l|}{($0.75$)}  & ($6.67$)  \\
				\multirow{2}{*}{$V_{2}^{B_c B_{1B}}$}  & \multicolumn{1}{l|}{${-0.17}^{+0.01 +0.02}_{-0.01 -0.00}$} & \multicolumn{1}{l|}{${-0.08}^{+0.04 +0.42}_{-0.02 -1.52}$}  & \multicolumn{1}{l|}{${1.51}^{+0.15 +1.76}_{-0.14 -0.77}$}  & ${-0.18}^{+0.02 +0.04}_{-0.01 -0.01 }$ & \multicolumn{1}{l|}{$-0.14$}   & \multicolumn{1}{l|}{$-28.32$}     & \multicolumn{1}{l|}{$527.75$}      & $-0.11$   \\
				& \multicolumn{1}{l|}{}                           & \multicolumn{1}{l|}{}                            & \multicolumn{1}{l|}{}                            &                           & \multicolumn{1}{l|}{($-0.17$)} & \multicolumn{1}{l|}{($-0.08$)}  & \multicolumn{1}{l|}{($1.51$)}  & ($-0.18$) \\ \hline

				\multirow{2}{*}{$A^{B_c B_{s1A}}$}     & \multicolumn{1}{l|}{${0.16}^{+0.01 +0.00}_{-0.01 -0.00}$}    & \multicolumn{1}{l|}{${0.61}^{+0.12 +0.02}_{-0.09 -0.69}$}   & \multicolumn{1}{l|}{${1.40}^{+0.00 +1.99}_{-0.03 -0.76}$}  & ${0.17}^{+0.01 +0.00}_{-0.02 -0.01}$   & \multicolumn{1}{l|}{$0.15$}    & \multicolumn{1}{l|}{$-1.69$}      & \multicolumn{1}{l|}{$215.61$}      & $0.15$    \\
				& \multicolumn{1}{l|}{}                            & \multicolumn{1}{l|}{}                            & \multicolumn{1}{l|}{}                            &                            & \multicolumn{1}{l|}{($0.16$)}  & \multicolumn{1}{l|}{($0.61$)}   & \multicolumn{1}{l|}{($1.40$)}  & ($0.17$)  \\
				\multirow{2}{*}{$V_{0}^{B_c B_{s1A}}$} & \multicolumn{1}{l|}{${0.16}^{+0.02 +0.00}_{-0.02 -0.01}$}  & \multicolumn{1}{l|}{${0.18}^{+0.04 +0.00}_{-0.06 -0.16}$}   & \multicolumn{1}{l|}{${0.50}^{+0.03 +0.71}_{-0.04 -0.23}$}  & ${0.17}^{+0.02 +0.00}_{-0.02 -0.01}$  & \multicolumn{1}{l|}{$0.04$}    & \multicolumn{1}{l|}{$-75.31$}     & \multicolumn{1}{l|}{$208.08$}      & $0.03$    \\
				& \multicolumn{1}{l|}{}                            & \multicolumn{1}{l|}{}                             & \multicolumn{1}{l|}{}                            &                             & \multicolumn{1}{l|}{($0.05$)}  & \multicolumn{1}{l|}{($-7.74$)}  & \multicolumn{1}{l|}{($3.32$)}  & ($0.04$)  \\
				\multirow{2}{*}{$V_{1}^{B_c B_{s1A}}$} & \multicolumn{1}{l|}{${6.24}^{+0.44 +0.03}_{-0.53 -0.11}$}  & \multicolumn{1}{l|}{${-0.67}^{+0.06 +0.08}_{-0.06 -0.08}$}  & \multicolumn{1}{l|}{${0.28}^{+0.01 +0.10}_{-0.01 -0.03}$}  & ${6.34}^{+0.46 +0.05}_{-0.55 -0.14}$  & \multicolumn{1}{l|}{$5.00$}    & \multicolumn{1}{l|}{$-1.89$}     & \multicolumn{1}{l|}{$48.47$}      & $4.95$    \\
				& \multicolumn{1}{l|}{}                            & \multicolumn{1}{l|}{}                            & \multicolumn{1}{l|}{}                            &                             & \multicolumn{1}{l|}{($5.21$)}  & \multicolumn{1}{l|}{($-0.52$)}   & \multicolumn{1}{l|}{($0.27$)}   & ($5.34$)  \\
				\multirow{2}{*}{$V_{2}^{B_c B_{s1A}}$} & \multicolumn{1}{l|}{${0.06}^{+0.00 +0.00}_{-0.00 -0.01}$}   & \multicolumn{1}{l|}{${0.21}^{+0.00 +0.00}_{-0.03 -0.14}$}   & \multicolumn{1}{l|}{${0.43}^{+0.02 +0.67}_{-0.02 -0.22}$}  & ${0.06}^{+0.00 +0.00}_{-0.00 -0.01}$   & \multicolumn{1}{l|}{$0.06$}    & \multicolumn{1}{l|}{$1.18$}      & \multicolumn{1}{l|}{$153.43$}      & $0.06$    \\
				& \multicolumn{1}{l|}{}                            & \multicolumn{1}{l|}{}                             & \multicolumn{1}{l|}{}                            &                             & \multicolumn{1}{l|}{($0.06$)}  & \multicolumn{1}{l|}{($0.21$)}    & \multicolumn{1}{l|}{($0.43$)}  & ($0.06$)  \\ \hline

				\multirow{2}{*}{$A^{B_c B_{s1B}}$}     & \multicolumn{1}{l|}{${0.0004}^{+0.00 +0.00}_{-0.00 -0.00}$}   & \multicolumn{1}{l|}{${-21.20}^{+0.07 +0.67}_{-0.06 -0.21}$} & \multicolumn{1}{l|}{${26.07}^{+0.06 +0.23}_{-0.07 -0.77}$} & ${0.0003}^{+0.00 +0.00}_{-0.00 -0.00}$    & \multicolumn{1}{l|}{$0.0002$}   & \multicolumn{1}{l|}{$-153.86$}    & \multicolumn{1}{l|}{$1197.81$}     & $0.0001$   \\
				& \multicolumn{1}{l|}{}                           & \multicolumn{1}{l|}{}                           & \multicolumn{1}{l|}{}                           &                           & \multicolumn{1}{l|}{($0.0004$)} & \multicolumn{1}{l|}{($-21.20$)} & \multicolumn{1}{l|}{($26.07$)} & ($0.0003$) \\
				\multirow{2}{*}{$V_{0}^{B_c B_{s1B}}$} & \multicolumn{1}{l|}{${0.56}^{+0.04 +0.05}_{-0.05 -0.07}$}  & \multicolumn{1}{l|}{${0.55}^{+0.17 +0.10}_{-0.13 -0.71}$}   & \multicolumn{1}{l|}{${1.28}^{+0.20 +1.58}_{-0.16 -0.65}$}  & ${0.59}^{+0.05 +0.05}_{-0.05 -0.09}$  & \multicolumn{1}{l|}{$0.42$}    & \multicolumn{1}{l|}{$-18.09$}     & \multicolumn{1}{l|}{$371.03$}      & $0.38$    \\
				& \multicolumn{1}{l|}{}                            & \multicolumn{1}{l|}{}                            & \multicolumn{1}{l|}{}                            &                            & \multicolumn{1}{l|}{($0.48$)}  & \multicolumn{1}{l|}{($-0.16$)}  & \multicolumn{1}{l|}{($2.16$)}  & ($0.50$)  \\
				\multirow{2}{*}{$V_{1}^{B_c B_{s1B}}$} & \multicolumn{1}{l|}{${9.79}^{+0.95 +1.47}_{-0.95 -1.80}$}  & \multicolumn{1}{l|}{${0.32}^{+0.07 +0.00}_{-0.06 -0.34}$}   & \multicolumn{1}{l|}{${0.60}^{+0.00 +0.88}_{-0.01 -0.32}$}  & ${10.40}^{+1.05 +1.55}_{-1.04 -2.05}$ & \multicolumn{1}{l|}{$7.70$}    & \multicolumn{1}{l|}{$-2.28$}      & \multicolumn{1}{l|}{$220.37$}      & $7.57$    \\
				& \multicolumn{1}{l|}{}                            & \multicolumn{1}{l|}{}                            & \multicolumn{1}{l|}{}                            &                             & \multicolumn{1}{l|}{($8.32$)}  & \multicolumn{1}{l|}{($0.30$)}   & \multicolumn{1}{l|}{($0.64$)}  & ($8.82$)  \\
				\multirow{2}{*}{$V_{2}^{B_c B_{s1B}}$} & \multicolumn{1}{l|}{${-0.17}^{+0.01 +0.01}_{-0.01 -0.00}$} & \multicolumn{1}{l|}{${0.22}^{+0.07 +0.22}_{-0.06 -1.04}$}   & \multicolumn{1}{l|}{${1.31}^{+0.13 +1.58}_{-0.11 -0.66}$}  & ${-0.18}^{+0.01 +0.01}_{-0.01 -0.00}$ & \multicolumn{1}{l|}{$-0.15$}   & \multicolumn{1}{l|}{$-15.34$}      & \multicolumn{1}{l|}{$393.37$}      & $-0.14$   \\
				& \multicolumn{1}{l|}{}                           & \multicolumn{1}{l|}{}                            & \multicolumn{1}{l|}{}                            &                           & \multicolumn{1}{l|}{($-0.17$)} & \multicolumn{1}{l|}{($0.22$)}  & \multicolumn{1}{l|}{($1.31$)}  & ($-0.18$) \\ \hline
		\end{tabular}}
	\end{table}
	\begin{table}[!ht]
		\caption{Bottom-changing $B_c\rightarrow A$ transition form factors in CLF QM. For Type-II, Type-I, and Type-I*, the definitions provided in the caption of Table \ref{tab:ff_bb} apply.}
		\label{tab:ff_bc}
		\setlength{\tabcolsep}{0.5pt}
		\resizebox{\textwidth}{!}{
			\begin{tabular}{|c|cccc|cccc|}
				\hline
				\multirow{3}{*}{F} & \multicolumn{4}{c|}{Type-II} & \multicolumn{4}{c|}{Type-I:} \\  
				& \multicolumn{4}{c|}{} & \multicolumn{4}{c|}{(Type-I*)}   \\                                                                                \cline{2-9} & \multicolumn{1}{c|}{$F(0)$} & \multicolumn{1}{c|}{\text{a}} & \multicolumn{1}{c|}{\text{b}} & \multicolumn{1}{c|}{$F(\q2_{max})$} & \multicolumn{1}{c|}{$F(0)$}    & \multicolumn{1}{c|}{\text{a}}   & \multicolumn{1}{c|}{\text{b}}   &  \multicolumn{1}{c|}{$F(\q2_{max})$} \\ 
				\hline

				\multirow{2}{*}{$A^{B_c D_{1A}} $} & \multicolumn{1}{l|}{${0.12}^{+0.03 +0.04}_{-0.03 -0.04}$}  & \multicolumn{1}{l|}{${1.95}^{+0.04 +0.16}_{-0.08 -0.31}$}    & \multicolumn{1}{l|}{${3.63}^{+1.14 +3.78}_{-1.00 -1.71}$} & ${0.25}^{+0.15 +0.18}_{-0.10 -0.16}$  & \multicolumn{1}{l|}{$0.12$} & \multicolumn{1}{l|}{$2.85$}     & \multicolumn{1}{l|}{$12.66$} & $0.07$    \\
				& \multicolumn{1}{l|}{} & \multicolumn{1}{l|}{} & \multicolumn{1}{l|}{} &  & \multicolumn{1}{l|}{($0.12$)} & \multicolumn{1}{l|}{($1.95$)} & \multicolumn{1}{l|}{($3.63$)}  & ($0.25$)  \\
				\multirow{2}{*}{$V_{0}^{B_c D_{1A}}$} & \multicolumn{1}{l|}{${0.08}^{+0.00 +0.03}_{-0.00 -0.03}$} & \multicolumn{1}{l|}{${1.57}^{+0.15 +0.34}_{-0.20 -0.36}$}    & \multicolumn{1}{l|}{${2.08}^{+0.91 +2.40}_{-0.77 -1.03}$} & ${0.21}^{+0.01 +0.10}_{-0.03 -0.12}$  & \multicolumn{1}{l|}{$0.04$}      & \multicolumn{1}{l|}{$2.90$}     & \multicolumn{1}{l|}{$14.25$}     & $0.02$    \\
				& \multicolumn{1}{l|}{}                            & \multicolumn{1}{l|}{}                              & \multicolumn{1}{l|}{}                              &                            & \multicolumn{1}{l|}{($0.04$  )}  & \multicolumn{1}{l|}{($2.07$)}   & \multicolumn{1}{l|}{($4.11$)}  & ($0.09$)  \\
				\multirow{2}{*}{$V_{1}^{B_c D_{1A}}$}       & \multicolumn{1}{l|}{${0.30}^{+0.05 +0.09}_{-0.05 -0.10}$}  & \multicolumn{1}{l|}{${0.64}^{+0.18 +0.43}_{-0.21 -0.37}$}    & \multicolumn{1}{l|}{${0.92}^{+0.35 +1.06}_{-0.26 -0.38}$}     & ${0.69}^{+0.08 +0.14}_{-0.11 -0.26}$  & \multicolumn{1}{l|}{$0.26$}      & \multicolumn{1}{l|}{$2.06$}     & \multicolumn{1}{l|}{$4.91$}     & $0.28$    \\
				& \multicolumn{1}{l|}{}                            & \multicolumn{1}{l|}{}                              & \multicolumn{1}{l|}{}                              &                             & \multicolumn{1}{l|}{($0.26$)}  & \multicolumn{1}{l|}{($0.69$)}   & \multicolumn{1}{l|}{($0.93$)}   & ($0.61$)  \\
				\multirow{2}{*}{$V_{2}^{B_c D_{1A}}$}       & \multicolumn{1}{l|}{${0.09}^{+0.02 +0.02}_{-0.02 -0.03}$}  & \multicolumn{1}{l|}{${1.49}^{+0.08 +0.22}_{-0.13 -0.32}$}    & \multicolumn{1}{l|}{${2.24}^{+0.77 +2.52}_{-0.65 -1.08}$}     & ${0.23}^{+0.11 +0.11}_{-0.08 -0.13}$  & \multicolumn{1}{l|}{$0.09$}      & \multicolumn{1}{l|}{$2.78$}     & \multicolumn{1}{l|}{$11.43$}     & $0.06$    \\
				& \multicolumn{1}{l|}{}                            & \multicolumn{1}{l|}{}                              & \multicolumn{1}{l|}{}                               &                             & \multicolumn{1}{l|}{($0.09$)}  & \multicolumn{1}{l|}{($1.49$)}   & \multicolumn{1}{l|}{($2.24$)}  & ($0.23$)  \\ \hline

				\multirow{2}{*}{$A^{B_c D_{1B}} $}          & \multicolumn{1}{l|}{${0.001}^{+0.00 +0.03}_{-0.00 -0.00}$}   & \multicolumn{1}{l|}{${-59.96}^{+0.64 +61.94}_{-0.00 -2.45}$} & \multicolumn{1}{l|}{${208.29}^{+0.00 +8.08}_{-2.17 -204.94}$} & ${0.00003}^{+0.00 +0.08}_{-0.00 -0.00}$    & \multicolumn{1}{l|}{$0.001$}     & \multicolumn{1}{l|}{$-92.70$}   & \multicolumn{1}{l|}{$402.04$}   & $0.00001$   \\
				& \multicolumn{1}{l|}{}                           & \multicolumn{1}{l|}{}                            & \multicolumn{1}{l|}{}                            &                           & \multicolumn{1}{l|}{($0.001$)} & \multicolumn{1}{l|}{($-59.96$)} & \multicolumn{1}{l|}{($208.29$)} & ($0.00003$) \\
				\multirow{2}{*}{$V_{0}^{B_c D_{1B}}$}       & \multicolumn{1}{l|}{${0.20}^{+0.04 +0.07}_{-0.04 -0.07}$}  & \multicolumn{1}{l|}{${1.93}^{+0.02 +0.20}_{-0.05 -0.31}$}    & \multicolumn{1}{l|}{${3.23}^{+0.92 +3.43}_{-0.85 -1.54}$}     & ${0.46}^{+0.24 +0.35}_{-0.16 -0.29}$  & \multicolumn{1}{l|}{$0.16$}      & \multicolumn{1}{l|}{$-4.94$}    & \multicolumn{1}{l|}{$45.80$}    & $0.02$    \\
				& \multicolumn{1}{l|}{}                            & \multicolumn{1}{l|}{}                              & \multicolumn{1}{l|}{}                               &                             & \multicolumn{1}{l|}{($0.17$)}  & \multicolumn{1}{l|}{($-2.33$)}  & \multicolumn{1}{l|}{($18.32$)}  & ($0.05$)  \\
				\multirow{2}{*}{$V_{1}^{B_c D_{1B}}$}       & \multicolumn{1}{l|}{${0.11}^{+0.03 +0.06}_{-0.03 -0.05}$}  & \multicolumn{1}{l|}{${1.50}^{+0.08 +0.24}_{-0.12 -0.32}$}    & \multicolumn{1}{l|}{${2.12}^{+0.75 +2.41}_{-0.65 -1.04}$}     & ${0.28}^{+0.15 +0.23}_{-0.11 -0.19}$  & \multicolumn{1}{l|}{$0.004$}     & \multicolumn{1}{l|}{$-92.61$}   & \multicolumn{1}{l|}{$401.38$}   & $0.00004$   \\
				& \multicolumn{1}{l|}{}                            & \multicolumn{1}{l|}{}                              & \multicolumn{1}{l|}{}                              &                             & \multicolumn{1}{l|}{($0.005$)} & \multicolumn{1}{l|}{($-50.04$)} & \multicolumn{1}{l|}{($149.40$)} & ($0.0001$) \\
				\multirow{2}{*}{$V_{2}^{B_c D_{1B}}$}       & \multicolumn{1}{l|}{${-0.06}^{+0.01 +0.02}_{-0.01 -0.01}$} & \multicolumn{1}{l|}{${1.90}^{+0.05 +0.00}_{-0.07 -0.18}$}    & \multicolumn{1}{l|}{${4.82}^{+1.01 +4.29}_{-0.96 -2.02}$}     & ${-0.10}^{+0.03 +0.06}_{-0.05 -0.09}$ & \multicolumn{1}{l|}{$-0.06$}     & \multicolumn{1}{l|}{$2.59$}     & \multicolumn{1}{l|}{$22.00$}     & $-0.02$   \\
				& \multicolumn{1}{l|}{}                           & \multicolumn{1}{l|}{}                              & \multicolumn{1}{l|}{}                              &                           & \multicolumn{1}{l|}{($-0.06$)} & \multicolumn{1}{l|}{($1.90$)}   & \multicolumn{1}{l|}{($4.82$)}  & ($-0.10$) \\ \hline

				\multirow{2}{*}{$A^{B_c \chi_{c1}} $}       & \multicolumn{1}{l|}{${0.21}^{+0.01 +0.01}_{-0.01 -0.01}$}  & \multicolumn{1}{l|}{${2.19}^{+0.08 +0.43}_{-0.10 -0.44}$}    & \multicolumn{1}{l|}{${4.28}^{+0.94 +4.15}_{-0.91 -1.92}$}     & ${0.33}^{+0.01 +0.00}_{-0.02 -0.04}$  & \multicolumn{1}{l|}{$0.21$}      & \multicolumn{1}{l|}{$2.39$}     & \multicolumn{1}{l|}{$6.29$}     & $0.27$    \\
				& \multicolumn{1}{l|}{}                            & \multicolumn{1}{l|}{}                              & \multicolumn{1}{l|}{}                              &                            & \multicolumn{1}{l|}{($0.21$)}  & \multicolumn{1}{l|}{($2.19$)}   & \multicolumn{1}{l|}{($4.28$)}   & ($0.33$)  \\
				\multirow{2}{*}{$V_{0}^{B_c \chi_{c1}}$}    & \multicolumn{1}{l|}{${0.07}^{+0.00 +0.01}_{-0.01 -0.01}$}  & \multicolumn{1}{l|}{${0.98}^{+0.09 +0.48}_{-0.06 -0.36}$}    & \multicolumn{1}{l|}{${1.47}^{+0.42 +1.57}_{-0.42 -0.69}$}     & ${0.09}^{+0.01 +0.01}_{-0.01 -0.01}$  & \multicolumn{1}{l|}{$0.07$}      & \multicolumn{1}{l|}{$2.26$}     & \multicolumn{1}{l|}{$9.98$}     & $0.08$    \\
				& \multicolumn{1}{l|}{}                            & \multicolumn{1}{l|}{}                              & \multicolumn{1}{l|}{}                               &                            & \multicolumn{1}{l|}{($0.07$  )}  & \multicolumn{1}{l|}{($2.26$)}   & \multicolumn{1}{l|}{($7.38$)}   & ($0.10$)  \\
				\multirow{2}{*}{$V_{1}^{B_c \chi_{c1}}$}    & \multicolumn{1}{l|}{${0.47}^{+0.01 +0.02}_{-0.01 -0.03}$}  & \multicolumn{1}{l|}{${-1.20}^{+0.09 +0.52}_{-0.10 -0.39}$}   & \multicolumn{1}{l|}{${2.28}^{+0.09 +0.35}_{-0.09 -0.00}$}      & ${0.45}^{+0.00 +0.00}_{-0.00 -0.01}$   & \multicolumn{1}{l|}{$0.48$}      & \multicolumn{1}{l|}{$-0.01$}    & \multicolumn{1}{l|}{$2.23$}     & $0.44$    \\
				& \multicolumn{1}{l|}{}                            & \multicolumn{1}{l|}{}                             & \multicolumn{1}{l|}{}                               &                             & \multicolumn{1}{l|}{($0.48$)}  & \multicolumn{1}{l|}{($-0.81$)}  & \multicolumn{1}{l|}{($1.98$)}   & ($0.49$)  \\
				\multirow{2}{*}{$V_{2}^{B_c \chi_{c1}}$}    & \multicolumn{1}{l|}{${0.08}^{+0.01 +0.00}_{-0.01 -0.00}$}    & \multicolumn{1}{l|}{${0.79}^{+0.15 +0.57}_{-0.17 -0.51}$}    & \multicolumn{1}{l|}{${2.13}^{+0.35 +1.85}_{-0.31 -0.71}$}     & ${0.11}^{+0.01 +0.00}_{-0.01 -0.01}$  & \multicolumn{1}{l|}{$0.08$}      & \multicolumn{1}{l|}{$1.48$}     & \multicolumn{1}{l|}{$3.84$}     & $0.10$    \\
				& \multicolumn{1}{l|}{}                            & \multicolumn{1}{l|}{}                              & \multicolumn{1}{l|}{}                               &                            & \multicolumn{1}{l|}{($0.08$  )}  & \multicolumn{1}{l|}{($0.79$)}   & \multicolumn{1}{l|}{($2.13$)}   & ($0.11$)  \\ \hline

				\multirow{2}{*}{$A^{B_c \chi_{c1}^\p} $}    & \multicolumn{1}{l|}{${0.04}^{+0.01 +0.01}_{-0.01 -0.03}$}  & \multicolumn{1}{l|}{${2.69}^{+0.00 +0.00}_{-0.01 -97.60}$}    & \multicolumn{1}{l|}{${7.04}^{+1.08 +472.04}_{-1.10 -2.98}$}   & ${0.06}^{+0.02 +0.01}_{-0.01 -0.06}$  & \multicolumn{1}{l|}{$0.04$}      & \multicolumn{1}{l|}{$2.62$}     & \multicolumn{1}{l|}{$9.66$}     & $0.04$    \\
				& \multicolumn{1}{l|}{}                            & \multicolumn{1}{l|}{}                              & \multicolumn{1}{l|}{}                             &                           & \multicolumn{1}{l|}{($0.04$)}  & \multicolumn{1}{l|}{($2.69$)}   & \multicolumn{1}{l|}{($7.04$)}   & ($0.06$)  \\
				\multirow{2}{*}{$V_{0}^{B_c \chi_{c1}^\p}$} & \multicolumn{1}{l|}{${0.35}^{+0.00 +0.03}_{-0.00 -0.04}$}   & \multicolumn{1}{l|}{${2.55}^{+0.04 +0.37}_{-0.06 -0.40}$}    & \multicolumn{1}{l|}{${4.78}^{+1.10 +4.50}_{-1.05 -2.15}$}     & ${0.58}^{+0.02 +0.05}_{-0.02 -0.11}$  & \multicolumn{1}{l|}{$0.33$}      & \multicolumn{1}{l|}{$2.51$}     & \multicolumn{1}{l|}{$8.71$}     & $0.39$    \\
				& \multicolumn{1}{l|}{}                            & \multicolumn{1}{l|}{}                              & \multicolumn{1}{l|}{}                              &                           & \multicolumn{1}{l|}{($0.33$)}  & \multicolumn{1}{l|}{($2.47$)}   & \multicolumn{1}{l|}{($6.13$)}   & ($0.51$)  \\
				\multirow{2}{*}{$V_{1}^{B_c \chi_{c1}^\p}$} & \multicolumn{1}{l|}{${0.30}^{+0.02 +0.08}_{-0.02 -0.07}$}  & \multicolumn{1}{l|}{${1.58}^{+0.12 +0.46}_{-0.15 -0.42}$}    & \multicolumn{1}{l|}{${2.38}^{+0.63 +2.49}_{-0.58 -1.10}$}     & ${0.46}^{+0.03 +0.09}_{-0.03 -0.11}$  & \multicolumn{1}{l|}{$0.25$}      & \multicolumn{1}{l|}{$2.28$}     & \multicolumn{1}{l|}{$5.04$}     & $0.33$    \\
				& \multicolumn{1}{l|}{}                            & \multicolumn{1}{l|}{}                              & \multicolumn{1}{l|}{}                               &                              & \multicolumn{1}{l|}{($0.25$)}  & \multicolumn{1}{l|}{($1.66$)}   & \multicolumn{1}{l|}{($2.42$)}   & ($0.39$)  \\
				\multirow{2}{*}{$V_{2}^{B_c \chi_{c1}^\p}$} & \multicolumn{1}{l|}{${-0.16}^{+0.01 +0.01}_{-0.00 -0.00}$}   & \multicolumn{1}{l|}{${2.40}^{+0.04 +0.17}_{-0.03 -0.28}$}    & \multicolumn{1}{l|}{${6.02}^{+0.76 +4.84}_{-0.75 -2.39}$}     & ${-0.26}^{+0.00 +0.05}_{-0.00 -0.01}$  & \multicolumn{1}{l|}{$-0.16$}     & \multicolumn{1}{l|}{$2.64$}     & \multicolumn{1}{l|}{$10.98$}     & $-0.18$   \\
				& \multicolumn{1}{l|}{}                           & \multicolumn{1}{l|}{}                              & \multicolumn{1}{l|}{}                              &                            & \multicolumn{1}{l|}{($-0.16$  )} & \multicolumn{1}{l|}{($2.40$)}   & \multicolumn{1}{l|}{($6.02$)}  & ($-0.26$) \\ \hline
		\end{tabular}}
	\end{table}
	\begin{table}[!ht]
		\centering
		\caption{Branching ratios of $B_c^{+}\to A^{(\p)}\ell^{+}\nu_{\ell}$ decays in CLF QM. For Type-II, Type-I, and Type-I*, the definitions provided in the caption of Table \ref{tab:ff_bb} apply.}
		\label{semilep_Br}
		\setlength{\tabcolsep}{5pt}
		\begin{tabular}{|l|l|l|}
			\hline
			\multicolumn{1}{|c|}{Decays}                 & \multicolumn{1}{c|}{Type-II}                                & \multicolumn{1}{c|}{Type-I [Type-I*]}     \\ \hline
			$B_c^{+}\to B_{1}^0 e^{+}\nu_{e}$            & $({2.66}^{+1.49 +0.81}_{-0.72 -0.06}) \t 10^{-6}$ & ${2.03} \t 10^{-6}$ [${4.11} \t 10^{-6}$] \\ \hline
			$B_c^{+}\to B_{1}^{0\p} e^{+}\nu_{e}$        & $({8.69}^{+1.17 +2.06}_{-0.79 -2.78}) \t 10^{-5}$ & ${4.46} \t 10^{-5}$ [${6.33} \t 10^{-5}$] \\ \hline
			$B_c^{+}\to B_{1}^0 \mu^{+}\nu_{\mu}$        & $({2.21}^{+1.16 +0.59}_{-0.34 -0.00}) \t 10^{-6}$ & ${3.42} \t 10^{-3}$ [${1.44} \t 10^{-2}$] \\ \hline
			$B_c^{+}\to B_{1}^{0\p} \mu^{+}\nu_{\mu}$    & $({7.20}^{+0.97 +1.72}_{-0.79 -2.32}) \t 10^{-5}$ & ${3.47} \t 10^{-5}$ [${4.99} \t 10^{-5}$] \\ \hline
			$B_c^{+}\to B_{s1}^0 e^{+}\nu_{e}$           & $({2.94}^{+2.26 +1.39}_{-1.28 -1.26}) \t 10^{-5}$ & ${2.21} \t 10^{-5}$ [${2.94} \t 10^{-5}$] \\ \hline
			$B_c^{+}\to B_{s1}^{0\p} e^{+}\nu_{e}$       & $({5.89}^{+0.54 +1.06}_{-0.55 -1.41}) \t 10^{-4}$ & ${3.54} \t 10^{-4}$ [${4.36} \t 10^{-4}$] \\ \hline
			$B_c^{+}\to B_{s1}^0 \mu^{+}\nu_{\mu}$       & $({2.12}^{+1.63 +1.08}_{-0.92 -0.85}) \t 10^{-5}$ & ${1.88} \t 10^{-5}$ [${2.49} \t 10^{-5}$] \\ \hline
			$B_c^{+}\to B_{s1}^{0\p} \mu^{+}\nu_{\mu}$   & $({4.43}^{+0.39 +0.81}_{-0.41 -1.06}) \t 10^{-4}$ & ${2.54} \t 10^{-4}$ [${3.18} \t 10^{-4}$] \\ \hline
			$B_c^{+}\to{D}_{1}^0 e^{+}\nu_{e}$           & $({7.32}^{+3.28 +8.20}_{-2.30 -4.50}) \t 10^{-6}$ & ${2.48} \t 10^{-6}$ [${3.50} \t 10^{-6}$] \\ \hline
			$B_c^{+}\to{D}_{1}^{0\p} e^{+}\nu_{e}$       & $({4.29}^{+1.99 +3.93}_{-1.74 -3.01}) \t 10^{-5}$ & ${2.65} \t 10^{-7}$ [${3.75} \t 10^{-7}$] \\ \hline
			$B_c^{+}\to{D}_{1}^0 \mu^{+}\nu_{\mu}$       & $({7.30}^{+3.28 +8.18}_{-2.30 -4.49}) \t 10^{-6}$ & ${2.47} \t 10^{-6}$ [${3.49} \t 10^{-6}$] \\ \hline
			$B_c^{+}\to{D}_{1}^{0\p} \mu^{+}\nu_{\mu}$   & $({4.28}^{+1.99 +3.92}_{-1.73 -3.00}) \t 10^{-5}$ & ${2.91} \t 10^{-7}$ [${4.12} \t 10^{-7}$] \\ \hline
			$B_c^{+}\to{D}_{1}^0 \tau^{+}\nu_{\tau}$     & $({3.65}^{+1.74 +3.89}_{-1.20 -2.29}) \t 10^{-6}$ & ${2.57} \t 10^{-7}$ [${5.64} \t 10^{-7}$] \\ \hline
			$B_c^{+}\to{D}_{1}^{0\p} \tau^{+}\nu_{\tau}$ & $({1.91}^{+1.00 +1.84}_{-0.84 -1.43}) \t 10^{-5}$ & ${5.57} \t 10^{-8}$ [${1.91} \t 10^{-7}$] \\ \hline
			$B_c^{+}\to\chi_{c1}e^{+}\nu_{e}$            & $({5.22}^{+0.19 +0.27}_{-0.18 -0.41}) \t 10^{-4}$ & ${5.09} \t 10^{-4}$ [${5.59} \t 10^{-4}$] \\ \hline
			$B_c^{+}\to\chi_{c1}^{\p}e^{+}\nu_{e}$       & $({1.05}^{+0.01 +0.27}_{-0.05 -0.29}) \t 10^{-3}$ & ${8.09} \t 10^{-4}$ [${8.91} \t 10^{-4}$] \\ \hline
			$B_c^{+}\to\chi_{c1}\mu^{+}\nu_{\mu}$        & $({5.19}^{+0.19 +0.27}_{-0.18 -0.41}) \t 10^{-4}$ & ${5.06} \t 10^{-4}$ [${5.56} \t 10^{-4}$] \\ \hline
			$B_c^{+}\to\chi_{c1}^{\p}\mu^{+}\nu_{\mu}$   & $({1.04}^{+0.01 +0.27}_{-0.05 -0.29}) \t 10^{-3}$ & ${8.03} \t 10^{-4}$ [${8.85} \t 10^{-4}$] \\ \hline
			$B_c^{+}\to\chi_{c1}\tau^{+}\nu_{\tau}$      & $({5.71}^{+0.23 +0.13}_{-0.25 -0.37}) \t 10^{-5}$ & ${5.34} \t 10^{-5}$ [${6.27} \t 10^{-5}$] \\ \hline
			$B_c^{+}\to\chi_{c1}^{\p}\tau^{+}\nu_{\tau}$ & $({8.77}^{+0.46 +2.38}_{-0.62 -2.78}) \t 10^{-5}$ & ${5.42} \t 10^{-5}$ [${7.04} \t 10^{-5}$] \\ \hline
		\end{tabular}
	\end{table}
	\begin{table}[!ht]
		\centering
		\caption{Predictions of physical observables in semileptonic $B_c^{+}\to A^{(\p)}\ell^{+}\nu_{\ell}$ decays in Type-II CLF QM.\footnote{We did not consider error analysis in other observables.}}
		\label{semilep_obs}
		\setlength{\tabcolsep}{5pt}
		\begin{tabular}{|c|c|c|c|c|c|}
			\hline
			\multicolumn{1}{|c|}{Decays} & \multicolumn{1}{c|}{$\la A_{FB}\ra$} & \multicolumn{1}{c|}{$\la C_{F}^\ell\ra$} & \multicolumn{1}{c|}{$\la P_{L}^\ell\ra$} & \multicolumn{1}{c|}{$\la F_{L}\ra$} & \multicolumn{1}{c|}{$\la \alpha^{*}\ra$} \\ \hline
			$B_c^{+}\to B_{1}^0 e^{+}\nu_{e}$            & ${-0.32}^{+0.27 +0.27}_{-0.33 -0.25}$    & $-0.33$                                  & $1.00$                                   & $0.48$                              & $-0.30$                                \\ \hline
			$B_c^{+}\to B_{1}^{0\p} e^{+}\nu_{e}$        & ${-0.06}^{+0.01 +0.00}_{-0.00 -0.01}$    & $-0.65$                                  & $1.00$                                   & $0.62$                              & $-0.53$                  
			\\ \hline
			$B_c^{+}\to B_{1}^0 \mu^{+}\nu_{\mu}$        & ${-0.19}^{+0.30 +0.29}_{-0.33 -0.29}$    & $-0.01$                                  & $0.54$                                   & $0.49$                              & $-0.31$                                  \\ \hline
			$B_c^{+}\to B_{1}^{0\p} \mu^{+}\nu_{\mu}$    & ${0.02}^{+0.00 +0.00}_{-0.00 -0.01}$     & $-0.40$                                  & $0.78$                                   & $0.60$                              & $-0.50$                                  \\ \hline
			$B_c^{+}\to B_{s1}^0 e^{+}\nu_{e}$           & ${0.09}^{+0.07 +0.08}_{-0.20 -0.25}$      & $-0.95$                                  & $1.00$                                   & $0.76$                              & $-0.72$                                  \\ \hline
			$B_c^{+}\to B_{s1}^{0\p} e^{+}\nu_{e}$       & ${-0.06}^{+0.00 +0.00}_{-0.00 -0.01}$  & $-0.66$                                  & $1.00$                                   & $0.63$                              & $-0.54$                                  \\ \hline
			$B_c^{+}\to B_{s1}^0 \mu^{+}\nu_{\mu}$       & ${0.27}^{+0.05 +0.06}_{-0.22 -0.26}$      & $-0.42$                                  & $0.42$                                   & $0.75$                              & $-0.72$                                  \\ \hline
			$B_c^{+}\to B_{s1}^{0\p} \mu^{+}\nu_{\mu}$   & ${0.04}^{+0.00 +0.00}_{-0.00 -0.00}$  & $-0.36$                                  & $0.72$                                   & $0.60$                              & $-0.50$                                  \\ \hline
			$B_c^{+}\to{D}_{1}^0 e^{+}\nu_{e}$           & ${-0.05}^{+0.00 +0.00}_{-0.00 -0.35}$  & $0.15$                                   & $1.00$                                   & $0.27$                              & $0.16$                                   \\ \hline
			$B_c^{+}\to{D}_{1}^{0\p} e^{+}\nu_{e}$       & ${-0.01}^{+0.00 +0.00}_{-0.00 -0.21}$  & $-0.72$                                  & $1.00$                                   & $0.65$                              & $-0.58$                                  \\ \hline
			$B_c^{+}\to{D}_{1}^0 \mu^{+}\nu_{\mu}$       & ${-0.05}^{+0.00 +0.00}_{-0.00 -0.35}$  & $0.15$                                   & $1.00$                                   & $0.27$                              & $0.16$                                   \\ \hline
			$B_c^{+}\to{D}_{1}^{0\p} \mu^{+}\nu_{\mu}$   & ${-0.002}^{+0.001 +0.000}_{-0.001 -0.205}$ & $-0.70$                                  & $0.99$                                   & $0.65$                              & $-0.58$                                  \\ \hline
			$B_c^{+}\to{D}_{1}^0 \tau^{+}\nu_{\tau}$     & ${-0.05}^{+0.01 +0.00}_{-0.01 -0.26}$    & $0.04$                                   & $0.72$                                   & $0.30$                              & $0.08$                                   \\ \hline
			$B_c^{+}\to{D}_{1}^{0\p} \tau^{+}\nu_{\tau}$ & ${0.18}^{+0.01 +0.01}_{-0.00 -0.18}$     & $-0.20$                                  & $0.44$                                   & $0.62$                              & $-0.52$                                  \\ \hline
			$B_c^{+}\to\chi_{c1}e^{+}\nu_{e}$            & ${-0.58}^{+0.02 +0.01}_{-0.01 -0.00}$    & $0.33$                                   & $1.00$                                   & $0.19$                              & $0.38$                                   \\ \hline
			$B_c^{+}\to\chi_{c1}^{\p}e^{+}\nu_{e}$       & ${-0.05}^{+0.01 +0.04}_{-0.01 -0.01}$     & $-1.24$                                  & $1.00$                                   & $0.89$                              & $-0.88$                                  \\ \hline
			$B_c^{+}\to\chi_{c1}\mu^{+}\nu_{\mu}$        & ${-0.57}^{+0.02 +0.01}_{-0.01 -0.01}$     & $0.34$                                   & $0.99$                                   & $0.18$                              & $0.38$                                   \\ \hline
			$B_c^{+}\to\chi_{c1}^{\p}\mu^{+}\nu_{\mu}$   & ${-0.03}^{+0.01 +0.05}_{-0.02 -0.01}$     & $-1.19$                                  & $0.95$                                   & $0.88$                              & $-0.88$                                  \\ \hline
			$B_c^{+}\to\chi_{c1}\tau^{+}\nu_{\tau}$      & ${-0.41}^{+0.02 +0.01}_{-0.02 -0.00}$     & $0.10$                                   & $0.53$                                   & $0.20$                              & $0.33$                                   \\ \hline
			$B_c^{+}\to\chi_{c1}^{\p}\tau^{+}\nu_{\tau}$ & ${0.28}^{+0.02 +0.06}_{-0.02 -0.03}$      & $-0.13$                                  & $-0.05$                                  & $0.79$                              & $-0.76$                                  \\ \hline
		\end{tabular}
	\end{table}
	\begin{table}[!ht]
		\centering
		\caption{Comparison of our Type-II CLF QM predictions of $B_c\to A^{(\p)}\ell\nu_{\ell}$ branching ratios  with other theoretical works.}
		\label{other_Br}
		\setlength{\tabcolsep}{5pt}
		\begin{tabular}{|c|c|c|c|c|} \hline 
			\text{Decays} & \text{This work} & \text{RQM} \cite{Ebert:2010zu}  & MGI-CLF QM \cite{Li:2023wgq} & \text{CLF QM} \cite{Shi:2016gqt}\footnote{ The $A(A^\p)$ states are in the ${}^{3} P_{1} ({}^{1} P_{1})$ eigenstates \cite{Shi:2016gqt}.} \\ 
			\hline  
			$B_c^{+}\to B_{1}^0 e^{+}\nu_{e}$            & $({2.66}^{+1.49 +0.81}_{-0.72 -0.06}) \t 10^{-6}$ & $3.30\t 10^{-6}$ & $ 1.636\t 10^{-6}$ & $1.52\t 10^{-5}$  \\
			\hline
			$B_c^{+}\to B_{1}^{0\p} e^{+}\nu_{e}$        & $({8.69}^{+1.17 +2.06}_{-0.79 -2.78}) \t 10^{-5}$ & $7.20\t 10^{-6}$ & $ 1.743\t 10^{-5}$ & $7.70\t 10^{-5}$   \\
			\hline
			$B_c^{+}\to B_{1}^0 \mu^{+}\nu_{\mu}$        & $({2.21}^{+1.16 +0.59}_{-0.34 -0.00}) \t 10^{-6}$ & $3.00\t 10^{-6}$ & $ 1.244\t 10^{-6}$ & $1.28\t 10^{-5}$   \\
			\hline 
			$B_c^{+}\to B_{1}^{0\p} \mu^{+}\nu_{\mu}$    & $({7.20}^{+0.97 +1.72}_{-0.79 -2.32}) \t 10^{-5}$ & $5.70\t 10^{-6}$ & $ 1.461\t 10^{-5}$ & $6.28\t 10^{-5}$  \\
			\hline
			$B_c^{+}\to B_{s1}^0 e^{+}\nu_{e}$           & $({2.94}^{+2.26 +1.39}_{-1.28 -1.26}) \t 10^{-5}$ & $2.0\t 10^{-5}$ & $ 3.946\t 10^{-5}$ & $8.31\t 10^{-5}$  \\
			\hline 
			$B_c^{+}\to B_{s1}^{0\p} e^{+}\nu_{e}$       & $({5.89}^{+0.54 +1.06}_{-0.55 -1.41}) \t 10^{-4}$ & $4.5\t 10^{-5}$ & $ 5.824\t 10^{-5}$ & $5.38\t 10^{-4}$   \\
			\hline 
			$B_c^{+}\to B_{s1}^0 \mu^{+}\nu_{\mu}$       & $({2.12}^{+1.63 +1.08}_{-0.92 -0.85}) \t 10^{-5}$ & $1.8\t 10^{-5}$ & $ 2.793\t 10^{-5}$ & $6.33\t 10^{-5}$   \\
			\hline
			$B_c^{+}\to B_{s1}^{0\p} \mu^{+}\nu_{\mu}$   & $({4.43}^{+0.39 +0.81}_{-0.41 -1.06}) \t 10^{-4}$ & $3.1\t 10^{-5}$ & $ 4.230\t 10^{-5}$ & $3.98\t 10^{-4}$   \\
			\hline 		
			\hline 
			$B_c^{+}\to{D}_{1}^0 e^{+}\nu_{e}$           & $({7.32}^{+3.28 +8.20}_{-2.30 -4.50}) \t 10^{-6}$ & $1.1\t 10^{-5}$ & $ 3.510\t 10^{-5}$ & - \\
			\hline 
			$B_c^{+}\to{D}_{1}^{0\p} e^{+}\nu_{e}$       & $({4.29}^{+1.99 +3.93}_{-1.74 -3.01}) \t 10^{-5}$ & $1.9\t 10^{-5}$ & $ 9.84\t 10^{-6}$ & - \\
			\hline
			$B_c^{+}\to{D}_{1}^0 \mu^{+}\nu_{\mu}$       & $({7.30}^{+3.28 +8.18}_{-2.30 -4.49}) \t 10^{-6}$ &       -         & $ 3.491\t 10^{-5}$ & - \\
			\hline 
			$B_c^{+}\to{D}_{1}^{0\p} \mu^{+}\nu_{\mu}$   & $({4.28}^{+1.99 +3.92}_{-1.73 -3.00}) \t 10^{-5}$ &       -         & $ 9.79\t 10^{-6}$ & - \\
			\hline
			$B_c^{+}\to{D}_{1}^0 \tau^{+}\nu_{\tau}$     & $({3.65}^{+1.74 +3.89}_{-1.20 -2.29}) \t 10^{-6}$ & $3.9\t 10^{-6}$ & $ 1.026\t 10^{-5}$ & - \\
			\hline
			$B_c^{+}\to{D}_{1}^{0\p} \tau^{+}\nu_{\tau}$ & $({1.91}^{+1.00 +1.84}_{-0.84 -1.43}) \t 10^{-5}$ & $1.1\t 10^{-5}$ & $ 5.19\t 10^{-6}$ & - \\
			\hline
		\end{tabular}
	\end{table}
	\begin{table}[!ht]
		\centering
		\caption{Comparison of our Type-II CLF QM predictions of $B_c\to \chi_{c1}^{(\p)}\ell\nu_{\ell}$ branching ratios with other theoretical works, along with $A_{FB}$ where available. }
		\label{other_Br_chi}
		\setlength{\tabcolsep}{0.5pt}
		\begin{tabular}{|c|c|c|c|c|c|c|} \hline 
			\text{Decays} & $B_c^{+}\to\chi_{c1}e^{+}\nu_{e}$ & $B_c^{+}\to\chi_{c1}^{\p}e^{+}\nu_{e}$ & $B_c^{+}\to\chi_{c1}\tau^{+}\nu_{\tau}$ & $B_c^{+}\to\chi_{c1}^{\p}\tau^{+}\nu_{\tau}$\\
			\hline
			\text{This work} & $({5.22}^{+0.19 +0.27}_{-0.18 -0.41}) \t 10^{-4}$ & $({1.05}^{+0.01 +0.27}_{-0.05 -0.29}) \t 10^{-3}$ & $({5.71}^{+0.23 +0.13}_{-0.25 -0.37}) \t 10^{-5}$ & $({8.77}^{+0.46 +2.38}_{-0.62 -2.78}) \t 10^{-5}$  \\
			\hline
			\text{RQM} \cite{Ebert:2010zu}      & $8.2\t10^{-4}$                & $9.6\t10^{-4}$                 & $9.2\t10^{-5}$                   & $7.7\t10^{-5}$  \\
			\hline
			\text{pQCD} \cite{Rui:2018kqr}      & $1.53\t10^{-3}$                & $1.06\t10^{-3}$                 & $2.0\t10^{-4}$                    & $1.3\t10^{-4}$  \\
			\hline
			\text{BS} \cite{Wang:2011jt}        & $(1.1\pm 0.3)\t10^{-3}$        & $(2.8\pm 0.8)\t10^{-3}  $       & $(9.7 \pm 6.5)\t10^{-5}  $     & $ (1.9 \pm 1.3)\t10^{-4} $ \\
			\hline
			\text{NRQM} \cite{Hernandez:2006gt} & $6.6\t10^{-4}$                & $1.70\t10^{-3}$                  & $7.2\t10^{-5} $                 & $1.5\t10^{-4} $  \\
			$A_{FB}$&$-0.60_{-0.01}$&$-0.83_{-0.05}\t 10^{-2}$&$-0.46$  &$0.35$  \\
			\hline 
			\text{RQM} \cite{Ivanov:2005fd}     & $9.8\t10^{-4}$               & $3.10\t10^{-3}$                  & $1.2\t10^{-4}$                    & $2.7\t10^{-4}$ \\
			$A_{FB}$&$0.19$&$-3.6 \t 10^{-2}$&$0.34$&$0.31$  \\
			\hline 
			\text{RCQM} \cite{Ivanov:2006ni}    & $9.2\t10^{-4}$                & $2.70\t10^{-3}$                 & $8.9\t10^{-5}$                   & $1.7\t10^{-4}$  \\
			\hline
			\text{QCDSR} \cite{Azizi:2009ny}    & $(1.46\pm 0.42)\t10^{-3}$      & $(1.42 \pm 0.40)\t10^{-3} $     & $(1.47 \pm 0.44)\t10^{-4}$       & $(1.37 \pm 0.38)\t10^{-4} $ \\
			\hline
			\text{CLF QM} \cite{Wang:2009mi}     & $(1.4^{+0.0}_{-0.1})\t10^{-3}$ & $(3.1^{+0.5}_{-0.8})\t10^{-3} $ & $(1.5^{+0.10}_{-0.20})\t10^{-4} $ & $(2.2^{+0.2}_{-0.4})\t10^{-4} $ \\
			\hline
			\text{MGI-CLF QM} \cite{Li:2023wgq}                   & $2.953\t 10^{-4}$              & $1.155\t 10^{-3}$               & $0.256\t 10^{-4}$                  & $0.051\t 10^{-3}$\\
			\hline
		\end{tabular}
	\end{table}	
	\FloatBarrier 
	\begin{center}
		\begin{tabular}{cccc}
			\includegraphics[width=0.52\linewidth]{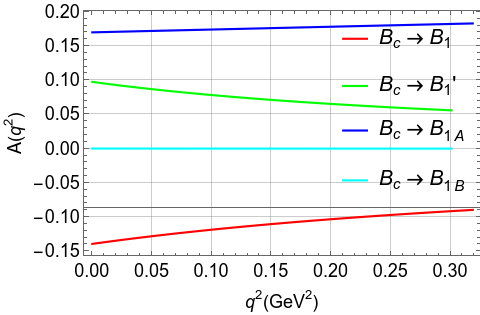} &
			\includegraphics[width=0.50\linewidth]{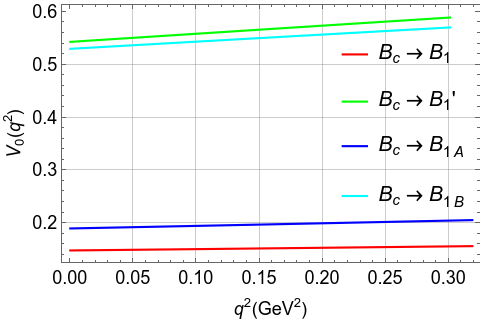} \\
			\includegraphics[width=0.49\linewidth]{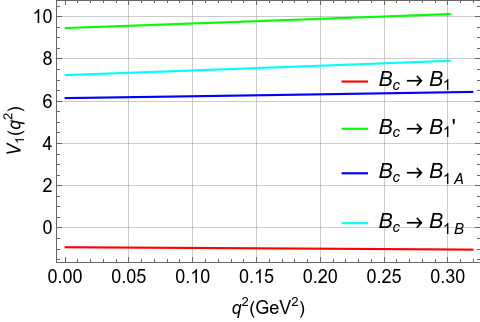} &
			\includegraphics[width=0.52\linewidth]{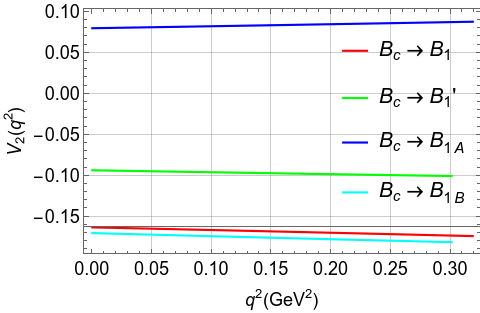} \\
		\end{tabular}
		\captionof{figure}{$\q2$ dependence of Type-II $B_c\rightarrow B_{1}$ form factors using Eq.~\eqref{eq:q2_ff}.}
		\label{fig:B1}
	\end{center}
	
	\begin{center}
		\begin{tabular}{cccc}
			\includegraphics[width=0.52\linewidth]{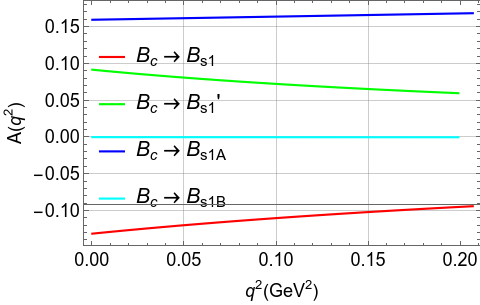} &
			\includegraphics[width=0.50\linewidth]{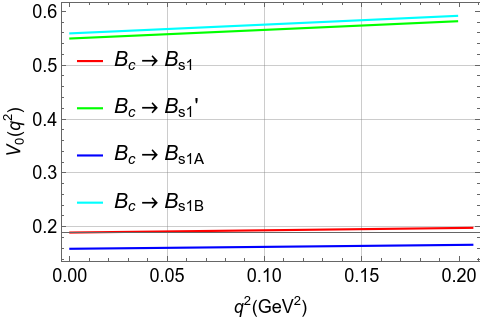} \\
			\includegraphics[width=0.50\linewidth]{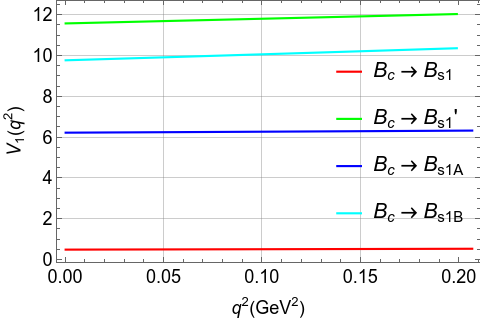} &
			\includegraphics[width=0.53\linewidth]{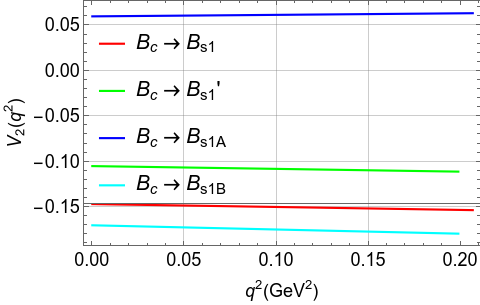} \\
		\end{tabular}
		\captionof{figure}{$\q2$ dependence of Type-II $B_c\rightarrow B_{s1}$ form factors using Eq.~\eqref{eq:q2_ff}.}
		\label{fig:Bs1}
	\end{center}
	
	\newpage
	
	\begin{figure}
		\centering
		\begin{subfigure}{0.49\textwidth}
			\centering
			\includegraphics[width=\linewidth]{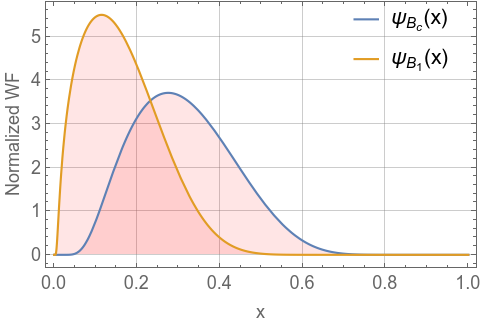}
			\caption{The area under the $B_c \text{~and~} B_1$ overlap is $0.627$ unit${}^2$.}\label{fig:B1_WOL}
		\end{subfigure} %
		\begin{subfigure}{0.49\textwidth}
			\centering
			\includegraphics[width=\linewidth]{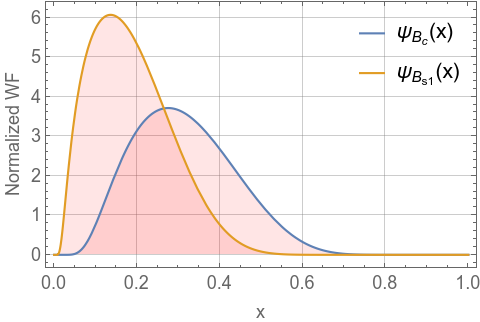}
			\caption{The area under the $B_c \text{~and~} B_{s1}$ overlap is $0.760$ unit${}^2$.}\label{fig:Bs1_WOL}
		\end{subfigure} %
		
		\caption{Overlap plots of normalized $B_c$ and $B_1 /B_{s1}$ LF wave function using Eqs.~\eqref{eq:RWFs} and \eqref{eq:RWFp}, respectively, in Type-II CLF QM.}
		\label{fig:bb_WOL}
	\end{figure}
	
	\begin{figure}
		\centering
		\begin{subfigure}{0.49\textwidth}
			\centering
			\includegraphics[width=\linewidth]{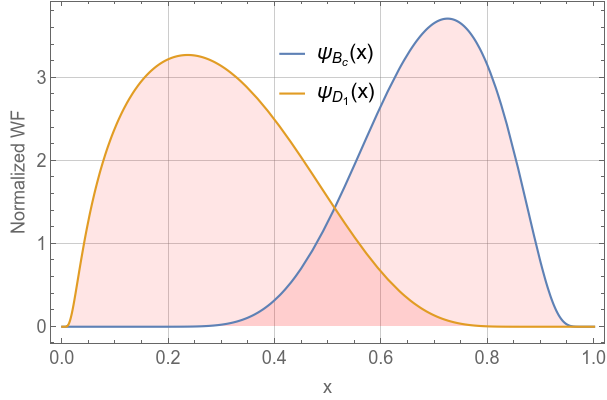}
			\caption{The area under the $B_c \text{~and~} D_{1}$ overlap is $0.235$ unit${}^2$.}\label{fig:D1_WOL}
		\end{subfigure}%
		\begin{subfigure}{0.49\textwidth}
			\centering
			\includegraphics[width=\linewidth]{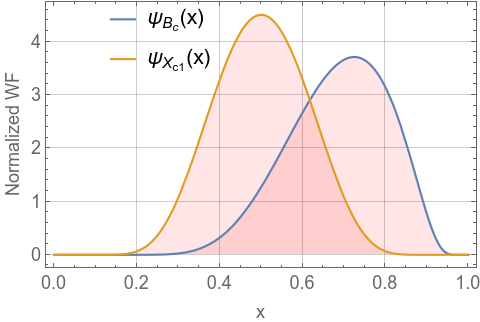}
			\caption{The area under the $B_c \text{~and~} \chi_{c1}$ overlap is $0.540$ unit${}^2$.}\label{fig:Chic1_WOL}
		\end{subfigure}
		\caption{Overlap plots of normalized $B_c$ and $D_1 /\chi_{c1}$ LF wave function using Eqs.~\eqref{eq:RWFs} and \eqref{eq:RWFp}, respectively, in Type-II CLF QM.}
		\label{fig:bc_WOL}
	\end{figure}
	\begin{figure}
		\centering
		\begin{subfigure}{0.49\textwidth}
			\centering
			\includegraphics[width=\linewidth]{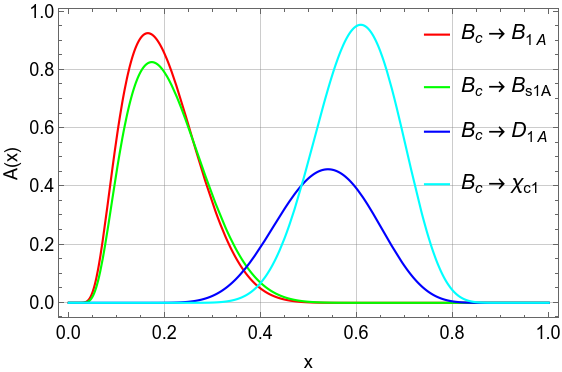}
			\caption{$A(x)$ versus $x$ plot. }\label{fig:Ax}
		\end{subfigure} 
		\begin{subfigure}{0.49\textwidth}
			\centering
			\includegraphics[width=\linewidth]{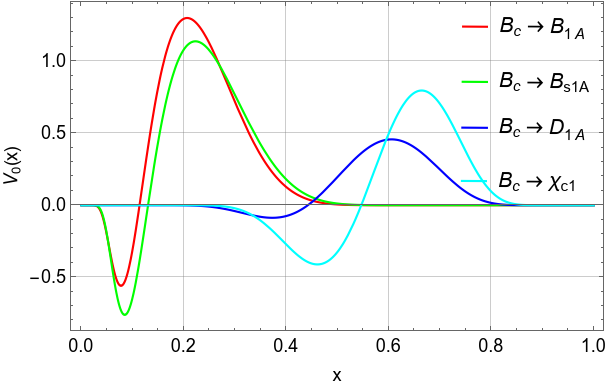}
			\caption{$V_{0}(x)$ versus $x$ plot.}\label{fig:V0x}
		\end{subfigure}
		\begin{subfigure}{0.49\textwidth}
			\centering
			\includegraphics[width=\linewidth]{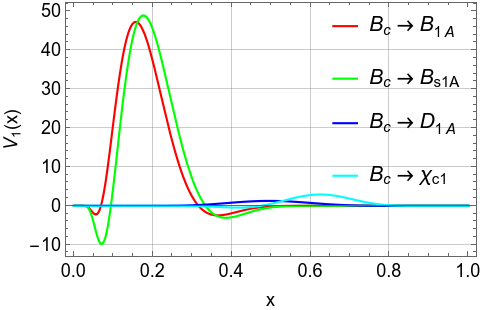}
			\caption{$V_{1}(x)$ versus $x$ plot.}\label{fig:V1x}
		\end{subfigure}
		\begin{subfigure}{0.49\textwidth}
			\centering
			\includegraphics[width=\linewidth]{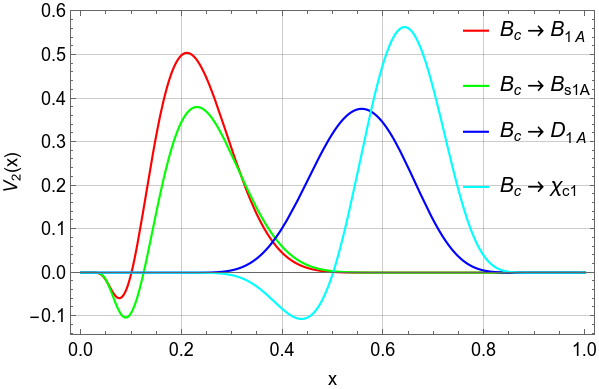}
			\caption{$V_{2}(x)$ versus $x$ plot.}\label{fig:V2x}
		\end{subfigure}
		
		\caption{$x$ dependence of Type-II $F(x)$ form factors ($A_A$-type) at $\q2 \approx 0$ using Eq.~\eqref{eq:Fq2}.}
		\label{fig:Ap_Fx_vs_x}
	\end{figure}
	\begin{figure}
		\centering
		\begin{subfigure}{0.49\textwidth}
			\centering
			\includegraphics[width=\linewidth]{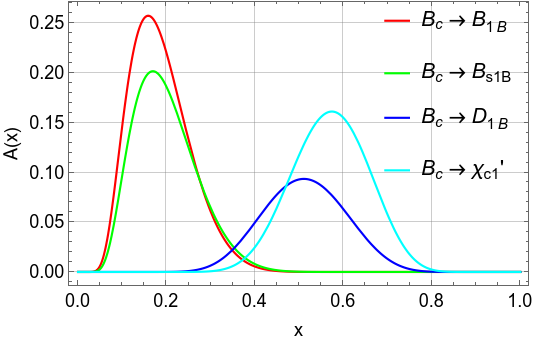}
			\caption{$A(x)$ versus $x$ plot. }\label{fig:Ax_An}
		\end{subfigure} 
		\begin{subfigure}{0.49\textwidth}
			\centering
			\includegraphics[width=\linewidth]{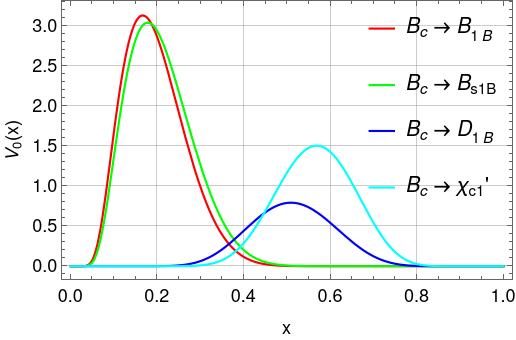}
			\caption{$V_{0}(x)$ versus $x$ plot.}\label{fig:V0x_An}
		\end{subfigure}
		\begin{subfigure}{0.49\textwidth}
			\centering
			\includegraphics[width=\linewidth]{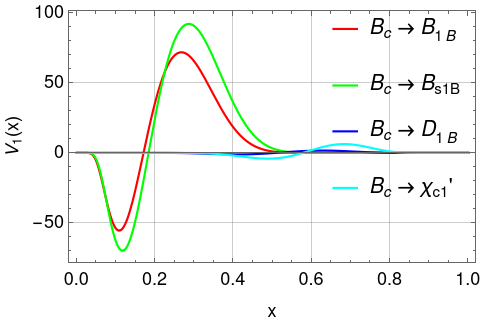}
			\caption{$V_{1}(x)$ versus $x$ plot.}\label{fig:V1x_An}
		\end{subfigure}
		\begin{subfigure}{0.49\textwidth}
			\centering
			\includegraphics[width=\linewidth]{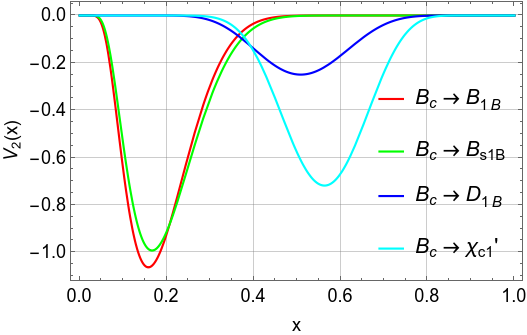}
			\caption{$V_{2}(x)$ versus $x$ plot.}\label{fig:V2x_An}
		\end{subfigure}
		
		\caption{$x$ dependence of Type-II $F(x)$ form factors ($A_B$-type) at $\q2 \approx 0$ using Eq.~\eqref{eq:Fq2}.}
		\label{fig:An_Fx_vs_x}
	\end{figure}
	\begin{center}
		\begin{tabular}{cccc}
			\includegraphics[width=0.52\linewidth]{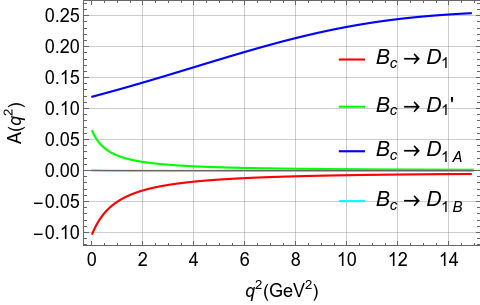} &
			\includegraphics[width=0.50\linewidth]{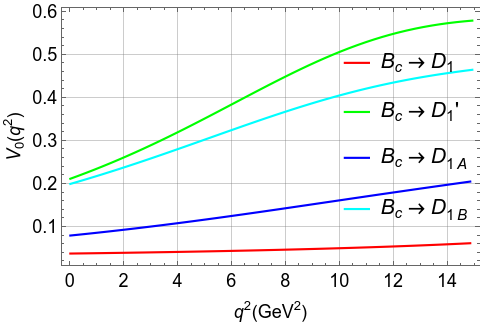} \\
			\includegraphics[width=0.50\linewidth]{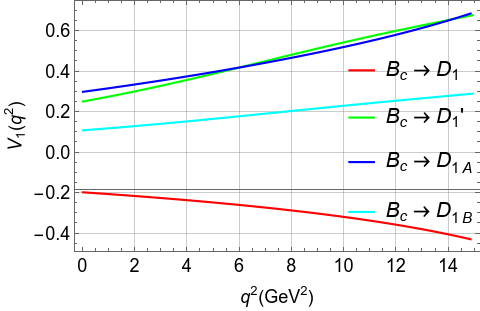} &
			\includegraphics[width=0.50\linewidth]{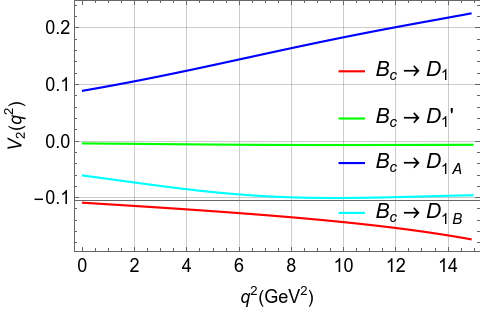} \\
		\end{tabular}
		\captionof{figure}{$\q2$ dependence of Type-II $B_c \to D_{1}$ form factors using Eq.~\eqref{eq:q2_ff}.}
		\label{fig:D1}
	\end{center}
	\begin{center}
		\begin{tabular}{cccc}
			\includegraphics[width=0.51\linewidth]{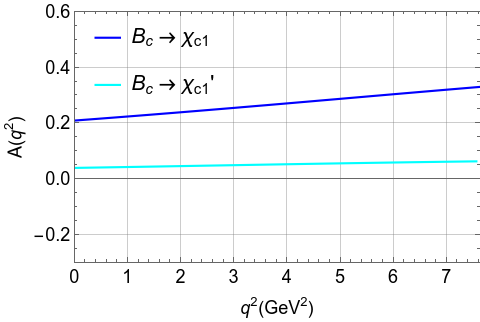} &
			\includegraphics[width=0.50\linewidth]{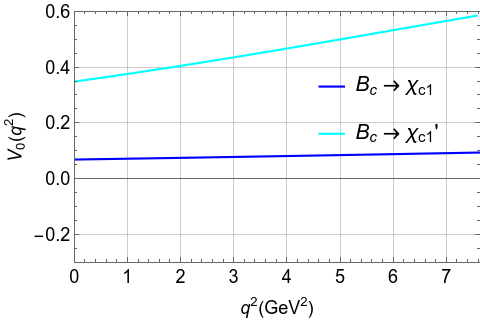} \\
			\includegraphics[width=0.49\linewidth]{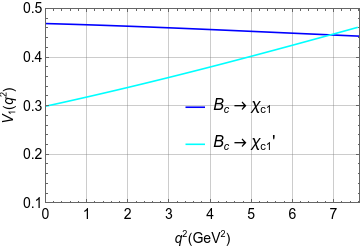} &
			\includegraphics[width=0.51\linewidth]{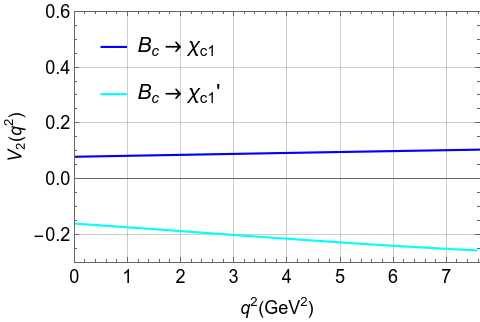} \\
		\end{tabular}
		\captionof{figure}{$\q2$ dependence of Type-II $B_c \to \chi_{c1}$ form factors using Eq.~\eqref{eq:q2_ff}.}
		\label{fig:chi}
	\end{center}
	
	\begin{figure}[!ht]
		\centering
		\begin{subfigure}[b]{0.48\textwidth}
			\centering
			\includegraphics[width=\textwidth]{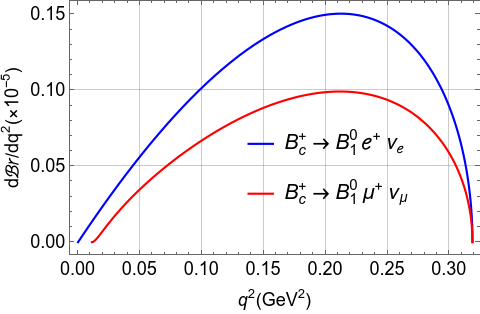}
		\end{subfigure}
		\hfill
		\begin{subfigure}[b]{0.48\textwidth}
			\centering
			\includegraphics[width=\textwidth]{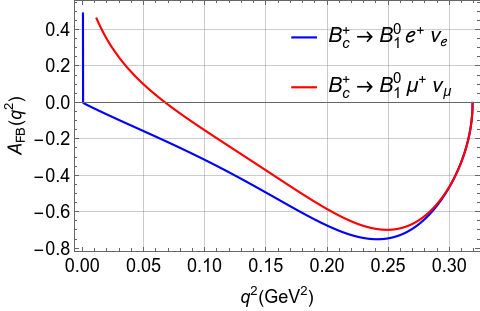}
		\end{subfigure}
		\hfill
		\begin{subfigure}[b]{0.48\textwidth}
			\centering
			\includegraphics[width=\textwidth]{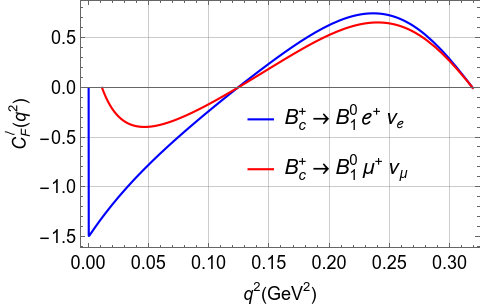}
		\end{subfigure}
		\hfill
		\begin{subfigure}[b]{0.48\textwidth}
			\centering
			\includegraphics[width=\textwidth]{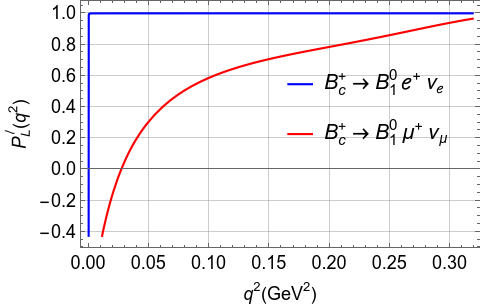}
		\end{subfigure}
		\begin{subfigure}[b]{0.48\textwidth}
			\centering
			\includegraphics[width=\textwidth]{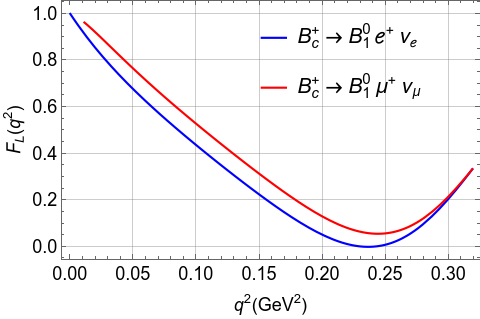}
		\end{subfigure}
		\hfill
		\begin{subfigure}[b]{0.49\textwidth}
			\centering
			\includegraphics[width=\textwidth]{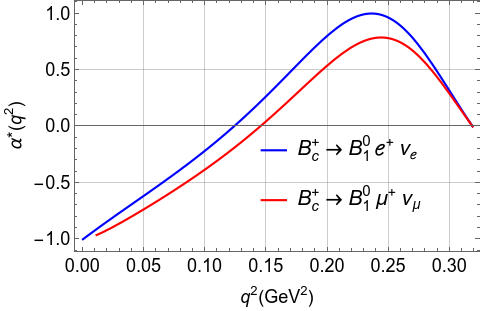}
		\end{subfigure}
		\caption{$\q2$ variation of branching ratios and physical observables of $B_c\to B_{1}\ell\nu_{\ell}$ decays in Type-II CLF QM corresponding to  Eq.~\eqref{eq:q2_ff}.}
		\label{fig:obs_B1}
	\end{figure}
	
	\begin{figure}[!ht]
		\centering
		\begin{subfigure}[b]{0.48\textwidth}
			\centering
			\includegraphics[width=\textwidth]{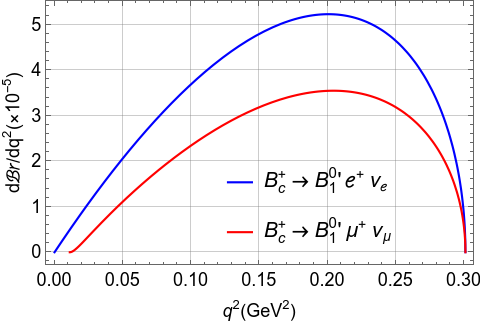}
		\end{subfigure}
		\hfill
		\begin{subfigure}[b]{0.50\textwidth}
			\centering
			\includegraphics[width=\textwidth]{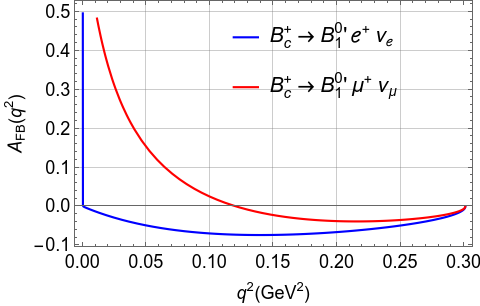}
		\end{subfigure}
		\hfill
		\begin{subfigure}[b]{0.48\textwidth}
			\centering
			\includegraphics[width=\textwidth]{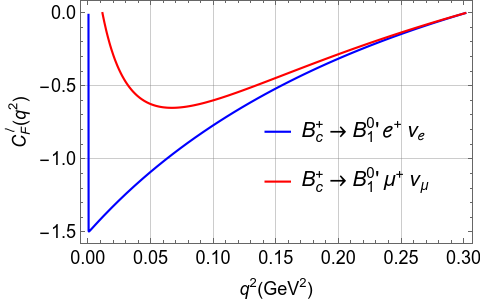}
		\end{subfigure}
		\hfill
		\begin{subfigure}[b]{0.49\textwidth}
			\centering
			\includegraphics[width=\textwidth]{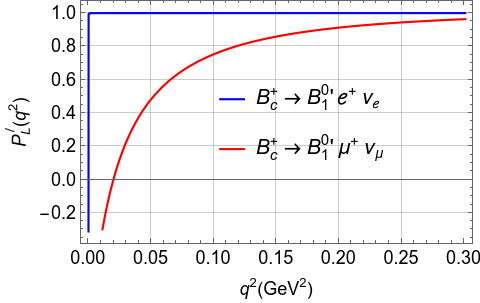}
		\end{subfigure}
		\begin{subfigure}[b]{0.48\textwidth}
			\centering
			\includegraphics[width=\textwidth]{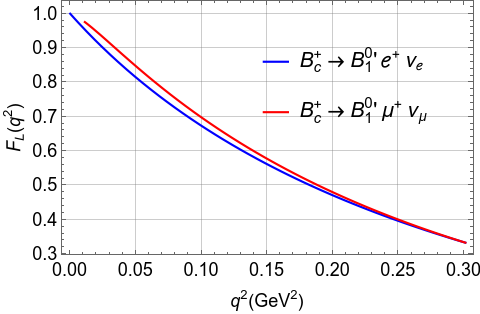}
		\end{subfigure}
		\hfill
		\begin{subfigure}[b]{0.50\textwidth}
			\centering
			\includegraphics[width=\textwidth]{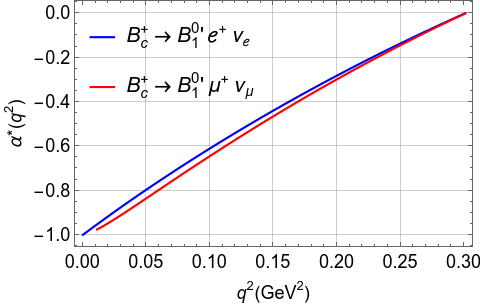}
		\end{subfigure}
		\caption{$\q2$ variation of branching ratios and physical observables of $B_c\to B_{1}^\p\ell\nu_{\ell}$ decays in Type-II CLF QM corresponding to  Eq.~\eqref{eq:q2_ff}.}
		\label{fig:obs_B1p}
	\end{figure}
	
	\begin{figure}[!ht]
		\centering
		\begin{subfigure}[b]{0.48\textwidth}
			\centering
			\includegraphics[width=\textwidth]{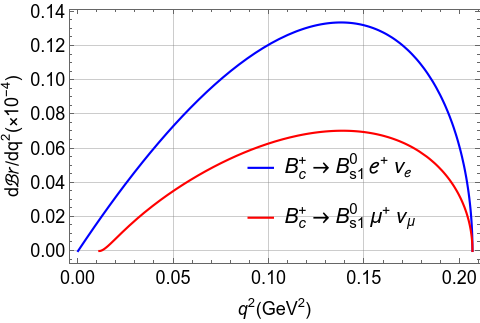}
		\end{subfigure}
		\hfill
		\begin{subfigure}[b]{0.46\textwidth}
			\centering
			\includegraphics[width=\textwidth]{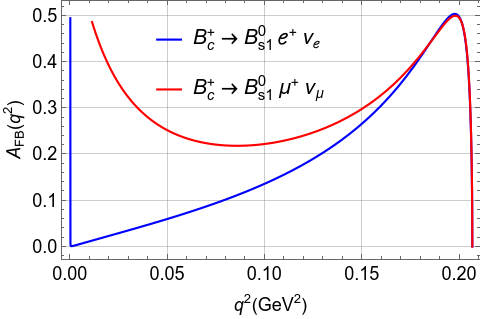}
		\end{subfigure}
		\hfill
		\begin{subfigure}[b]{0.48\textwidth}
			\centering
			\includegraphics[width=\textwidth]{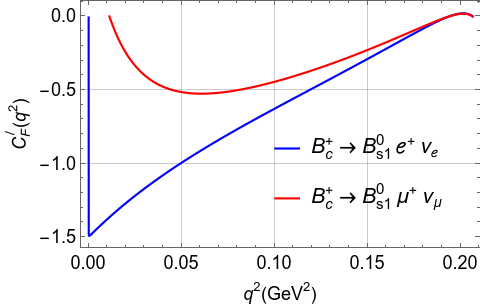}
		\end{subfigure}
		\hfill
		\begin{subfigure}[b]{0.48\textwidth}
			\centering
			\includegraphics[width=\textwidth]{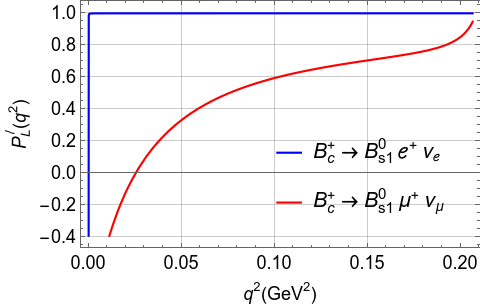}
		\end{subfigure}
		\begin{subfigure}[b]{0.48\textwidth}
			\centering
			\includegraphics[width=\textwidth]{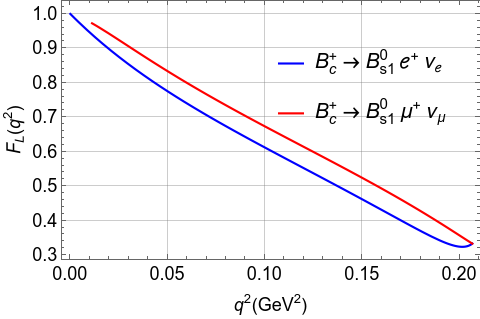}
		\end{subfigure}
		\hfill
		\begin{subfigure}[b]{0.49\textwidth}
			\centering
			\includegraphics[width=\textwidth]{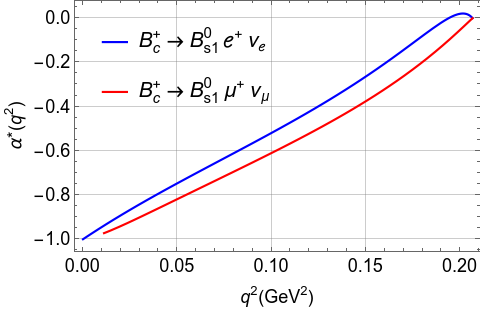}
		\end{subfigure}
		\caption{$\q2$ variation of branching ratios and physical observables of $B_c\to B_{s1}\ell\nu_{\ell}$ decays in Type-II CLF QM corresponding to  Eq.~\eqref{eq:q2_ff}.}
		\label{fig:obs_Bs1}
	\end{figure}
	
	\begin{figure}[!ht]
		\centering
		\begin{subfigure}[b]{0.48\textwidth}
			\centering
			\includegraphics[width=\textwidth]{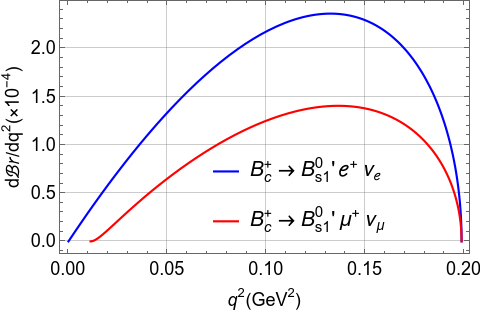}
		\end{subfigure}
		\hfill
		\begin{subfigure}[b]{0.51\textwidth}
			\centering
			\includegraphics[width=\textwidth]{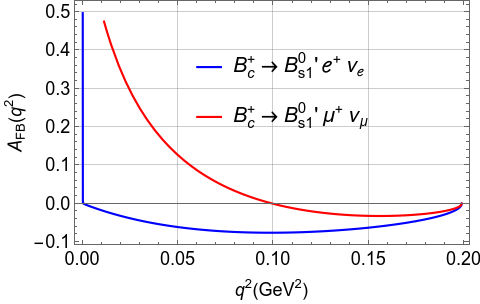}
		\end{subfigure}
		\hfill
		\begin{subfigure}[b]{0.48\textwidth}
			\centering
			\includegraphics[width=\textwidth]{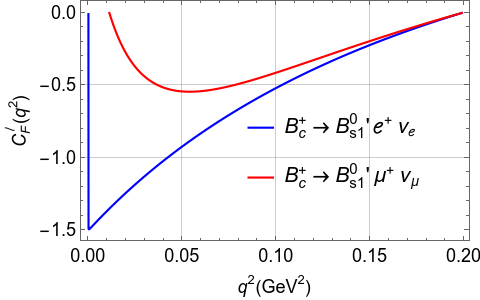}
		\end{subfigure}
		\hfill
		\begin{subfigure}[b]{0.49\textwidth}
			\centering
			\includegraphics[width=\textwidth]{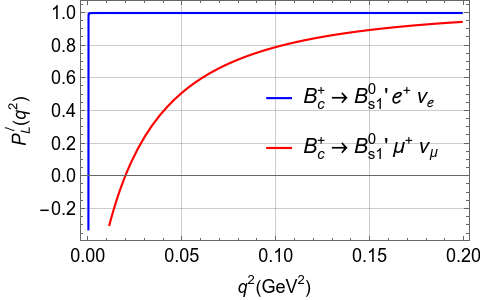}
		\end{subfigure}
		\begin{subfigure}[b]{0.48\textwidth}
			\centering
			\includegraphics[width=\textwidth]{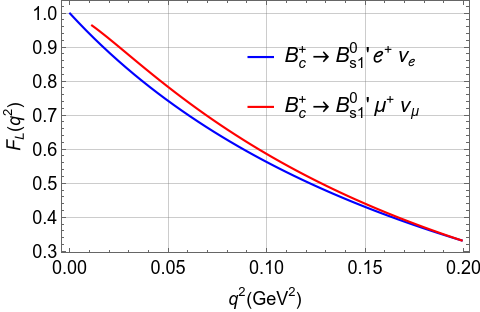}
		\end{subfigure}
		\hfill
		\begin{subfigure}[b]{0.49\textwidth}
			\centering
			\includegraphics[width=\textwidth]{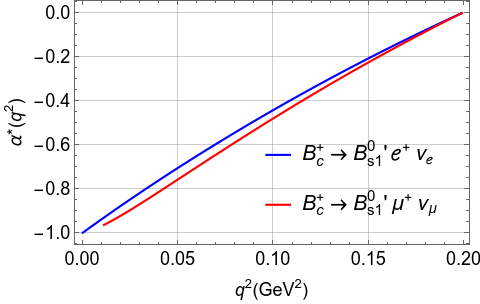}
		\end{subfigure}
		\caption{$\q2$ variation of branching ratios and physical observables of $B_c\to B_{s1}^\p\ell\nu_{\ell}$ decays in Type-II CLF QM corresponding to  Eq.~\eqref{eq:q2_ff}.}
		\label{fig:obs_Bs1p}
	\end{figure}
	
	\begin{figure}[!ht]
		\centering
		\begin{subfigure}[b]{0.48\textwidth}
			\centering
			\includegraphics[width=\textwidth]{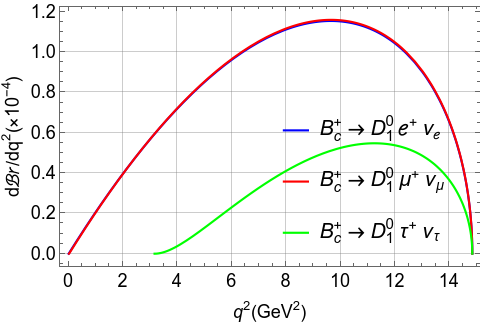}
		\end{subfigure}
		\hfill
		\begin{subfigure}[b]{0.49\textwidth}
			\centering
			\includegraphics[width=\textwidth]{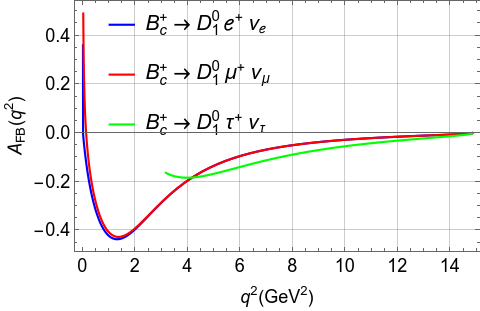}
		\end{subfigure}
		\hfill
		\begin{subfigure}[b]{0.48\textwidth}
			\centering
			\includegraphics[width=\textwidth]{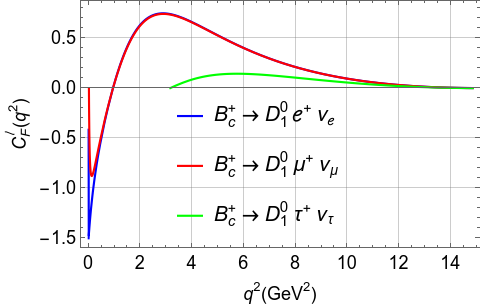}
		\end{subfigure}
		\hfill
		\begin{subfigure}[b]{0.48\textwidth}
			\centering
			\includegraphics[width=\textwidth]{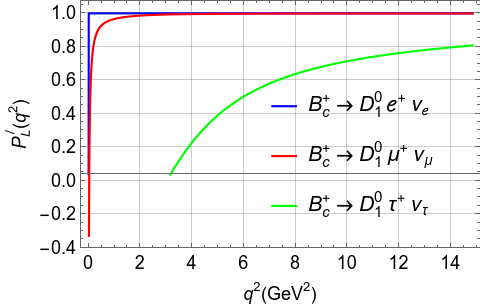}
		\end{subfigure}
		\begin{subfigure}[b]{0.48\textwidth}
			\centering
			\includegraphics[width=\textwidth]{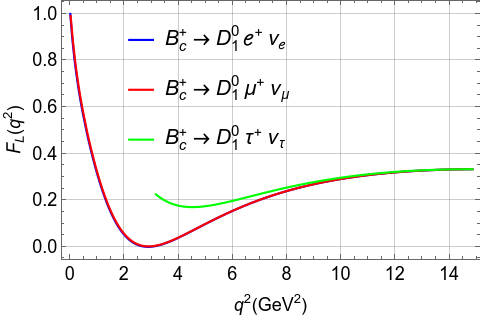}
		\end{subfigure}
		\hfill
		\begin{subfigure}[b]{0.49\textwidth}
			\centering
			\includegraphics[width=\textwidth]{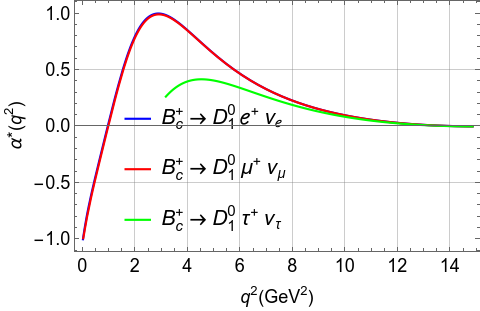}
		\end{subfigure}
		\caption{$\q2$ variation of branching ratios and physical observables of $B_c\to D_{1}\ell\nu_{\ell}$ decays in Type-II CLF QM corresponding to  Eq.~\eqref{eq:q2_ff}.}
		\label{fig:obs_D1}
	\end{figure}
	
	\begin{figure}[!ht]
		\centering
		\begin{subfigure}[b]{0.48\textwidth}
			\centering
			\includegraphics[width=\textwidth]{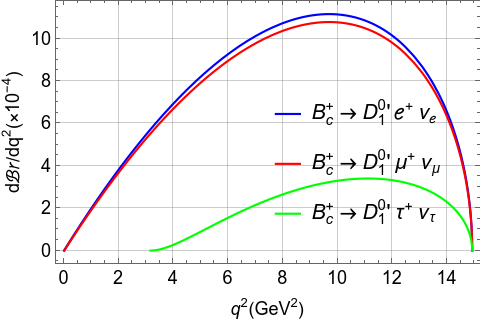}
		\end{subfigure}
		\hfill
		\begin{subfigure}[b]{0.49\textwidth}
			\centering
			\includegraphics[width=\textwidth]{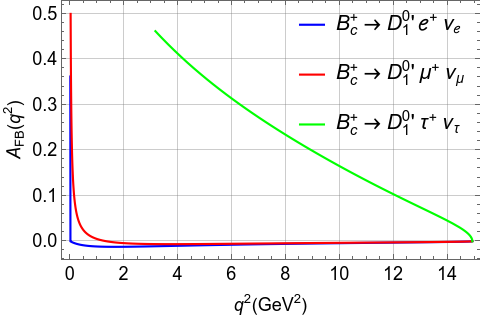}
		\end{subfigure}
		\hfill
		\begin{subfigure}[b]{0.48\textwidth}
			\centering
			\includegraphics[width=\textwidth]{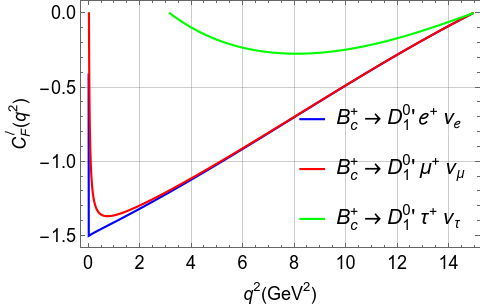}
		\end{subfigure}
		\hfill
		\begin{subfigure}[b]{0.48\textwidth}
			\centering
			\includegraphics[width=\textwidth]{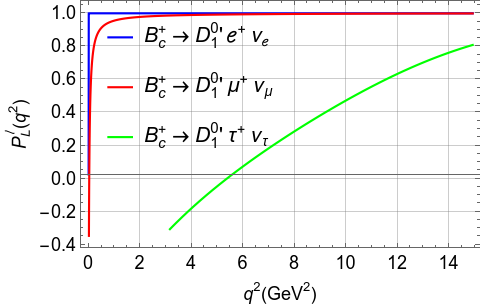}
		\end{subfigure}
		\begin{subfigure}[b]{0.48\textwidth}
			\centering
			\includegraphics[width=\textwidth]{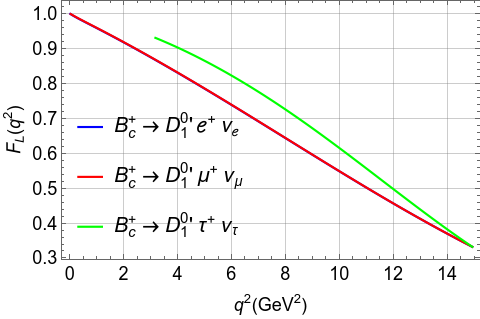}
		\end{subfigure}
		\hfill
		\begin{subfigure}[b]{0.49\textwidth}
			\centering
			\includegraphics[width=\textwidth]{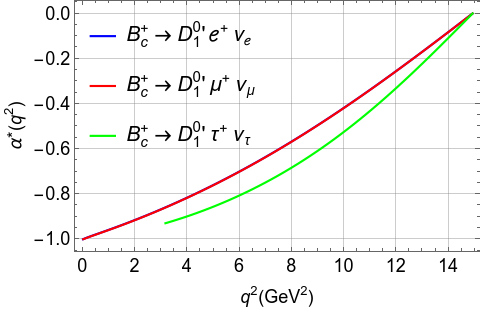}
		\end{subfigure}
		\caption{$\q2$ variation of branching ratios and physical observables of $B_c\to D_{1}^\p\ell\nu_{\ell}$ decays in Type-II CLF QM corresponding to  Eq.~\eqref{eq:q2_ff}.}
		\label{fig:obs_D1p}
	\end{figure}
	
	\begin{figure}[!ht]
		\centering
		\begin{subfigure}[b]{0.48\textwidth}
			\centering
			\includegraphics[width=\textwidth]{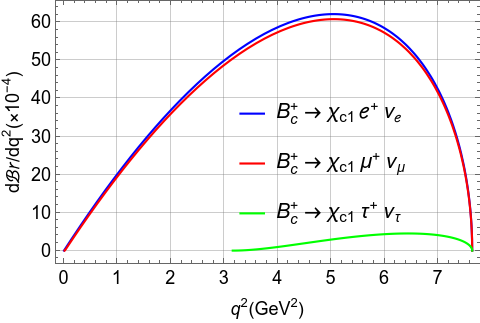}
		\end{subfigure}
		\hfill
		\begin{subfigure}[b]{0.50\textwidth}
			\centering
			\includegraphics[width=\textwidth]{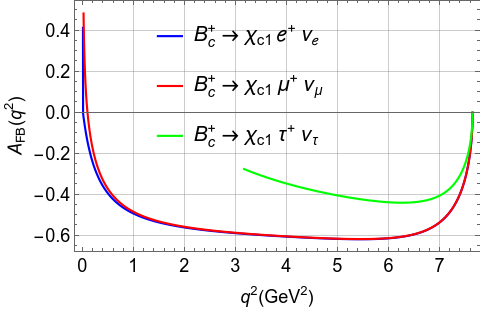}
		\end{subfigure}
		\hfill
		\begin{subfigure}[b]{0.48\textwidth}
			\centering
			\includegraphics[width=\textwidth]{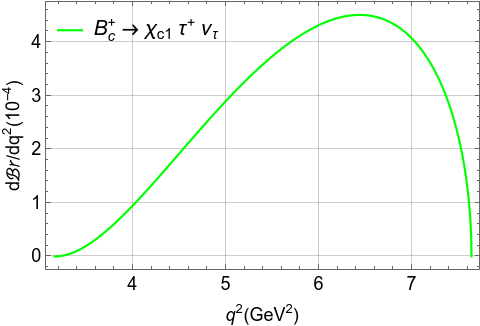}
		\end{subfigure}
		\hfill
		\begin{subfigure}[b]{0.48\textwidth}
			\centering
			\includegraphics[width=\textwidth]{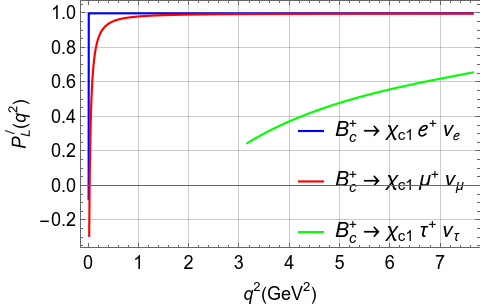}
		\end{subfigure}
		\begin{subfigure}[b]{0.48\textwidth}
			\centering
			\includegraphics[width=\textwidth]{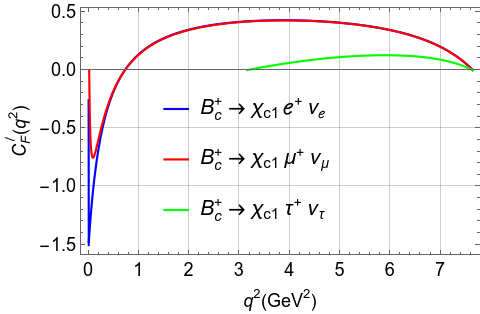}
		\end{subfigure}
		\hfill
		\begin{subfigure}[b]{0.48\textwidth}
			\centering
			\includegraphics[width=\textwidth]{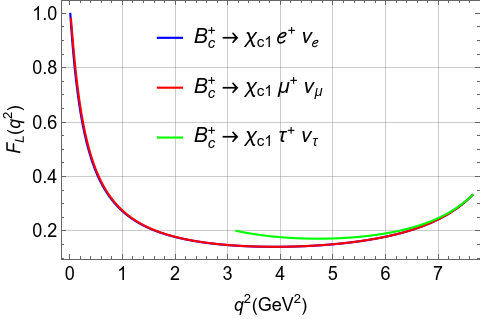}
		\end{subfigure}
		\hfill
		\begin{subfigure}[b]{0.49\textwidth}
			\centering
			\includegraphics[width=\textwidth]{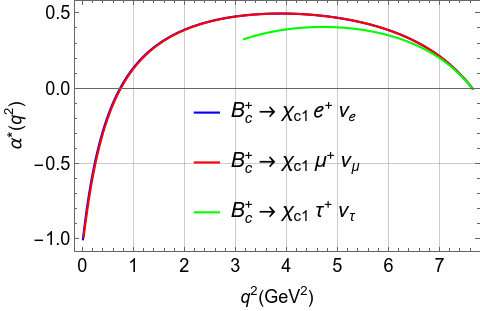}
		\end{subfigure}
		\caption{$\q2$ variation of branching ratios and physical observables of $B_c\to \chi_{c1}\ell\nu_{\ell}$ decays in Type-II CLF QM corresponding to  Eq.~\eqref{eq:q2_ff}.}
		\label{fig:obs_chic1}
	\end{figure}
	
	\begin{figure}[!ht]
		\centering
		\begin{subfigure}[b]{0.48\textwidth}
			\centering
			\includegraphics[width=\textwidth]{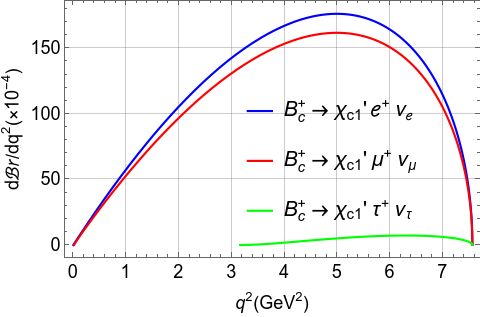}
		\end{subfigure}
		\hfill
		\begin{subfigure}[b]{0.49\textwidth}
			\centering
			\includegraphics[width=\textwidth]{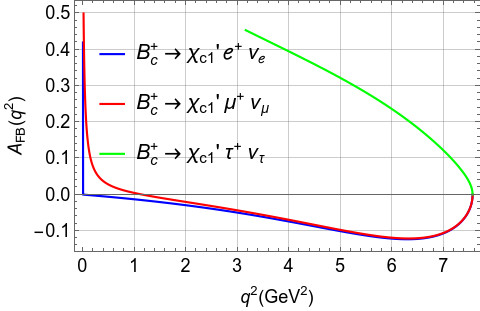}
		\end{subfigure}
		\hfill
		\begin{subfigure}[b]{0.48\textwidth}
			\centering
			\includegraphics[width=\textwidth]{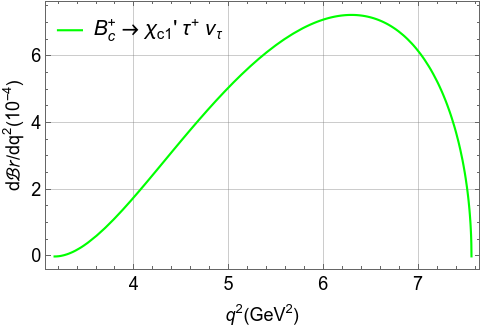}
		\end{subfigure}
		\hfill
		\begin{subfigure}[b]{0.48\textwidth}
			\centering
			\includegraphics[width=\textwidth]{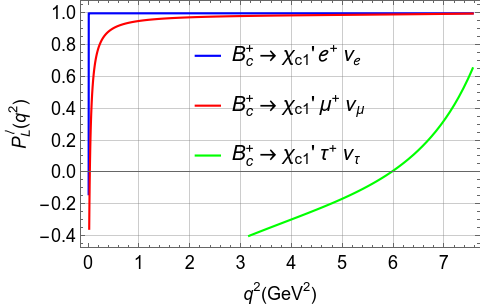}
		\end{subfigure}
		\begin{subfigure}[b]{0.48\textwidth}
			\centering
			\includegraphics[width=\textwidth]{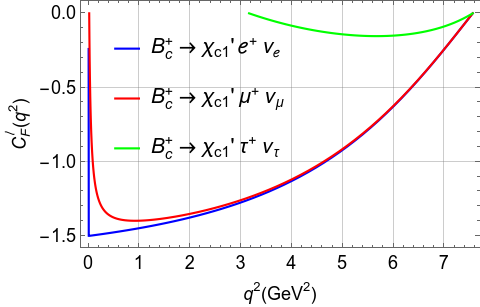}
		\end{subfigure}
		\hfill
		\begin{subfigure}[b]{0.48\textwidth}
			\centering
			\includegraphics[width=\textwidth]{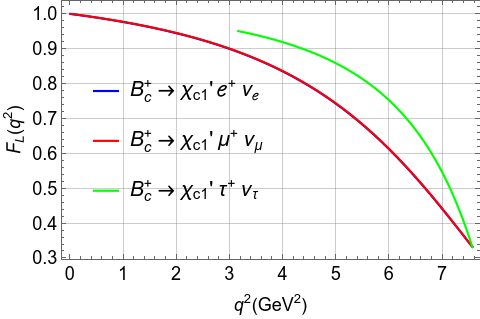}
		\end{subfigure}
		\hfill
		\begin{subfigure}[b]{0.49\textwidth}
			\centering
			\includegraphics[width=\textwidth]{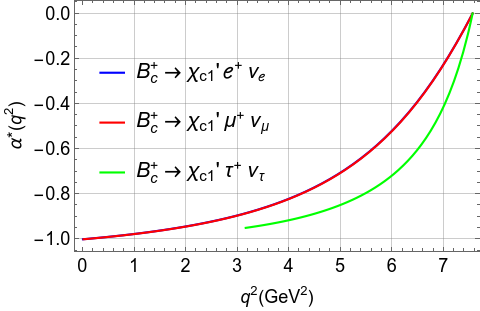}
		\end{subfigure}
		\caption{$\q2$ variation of branching ratios and physical observables of $B_c\to \chi_{c1}^\p\ell\nu_{\ell}$ decays in Type-II CLF QM corresponding to  Eq.~\eqref{eq:q2_ff}.}
		\label{fig:obs_chic1p}
	\end{figure}
	
\end{document}